\documentclass{aa}
\usepackage{txfonts}
\usepackage{graphicx}
\usepackage{multirow}
 \usepackage[colorlinks,linkcolor=magenta,citecolor=blue]{hyperref}

\usepackage{placeins}
\begin{document} 

\title{Amplitude and frequency variations in PG~0101+039 from K2 photometry}
\subtitle{A pulsating hot B subdwarf star in an unsynchronized binary system}

   \author{X.-Y.~Ma\inst{1,2},
   % Xiao-Yu~Ma \inst{1,2},
          W.~Zong\inst{1,2},%\thanks{Corresponding author: Weikai Zong},
          J.-N.~Fu\inst{1,2},%\thanks{Corresponding author: Jian-Ning Fu},
          S.~Charpinet\inst{3},
          J.~Wang\inst{4},
          \and
          K.~Xing\inst{1,2}}
   \institute{Institute for Frontiers in Astronomy and       Astrophysics, Beijing Normal University, Beijing 102206, P.~R.~China \\
              \email{weikai.zong@bnu.edu.cn;jnfu@bnu.edu.cn}\\
         \and
             Department of Astronomy, Beijing Normal University, Beijing 100875, P.~R. China\\
         \and
            Institut de Recherche en Astrophysique et Plan\'{e}tologie,~CNRS,~Universit\'{e} de Toulouse, CNES,~14 Avenue Edouard Belin,~31400 Toulouse,~France \\
         \and 
           College of Science, Chongqing University of Posts and Telecommunications, Chongqing 400065, P.~R.~China
             % \thanks{The University of Heaven temporarily does not accept e-mails}
             }

\titlerunning{Amplitude and frequency variations in PG~0101+039: an unsynchronized binary with a sdB component}
\authorrunning{Ma et al.}

\date{Revised 30 Aug. 2023; accepted 18 Sep. 2023}

% \abstract{}{}{}{}{} 
% 5 {} token are mandatory
 
\abstract
   {{\sl K}2 photometry is suitable for the exploitation of mode variability on short timescales in hot B subdwarf stars and this technique is useful in constraining nonlinear quantities addressed by the stellar theory of high-order perturbation in the future.}
  % aims heading (mandatory)
   {We analyzed high-quality {\sl K}2 data collected for PG~0101+039 over about 80 days and we extracted the frequency content of oscillation. We determined the star's rotational and orbital properties, in addition to characterizing the dynamics of the amplitude and frequency.}
  % methods heading (mandatory)
   {The frequencies were extracted from light curves via a standard prewhitening technique. The binary information was obtained from variations both in brightness and radial velocities. The amplitude and frequency modulations (i.e., AMs and FMs) of the oscillation modes were measured by piece-wise light curves and characterized by a Markov chain Monte Carlo (\texttt{EMCMC}) method.}
  % results heading (mandatory)
   {We extracted 137 independent frequencies in PG~0101+039 and derived period spacing of $\sim 252$~s and 144~s for the dipole and quadruple modes, respectively. We derived a rotation rate of $\sim8.81\pm0.06$~days and $\sim8.60\pm0.16$~days based on {\sl g}- and {\sl p}-mode multiplets, implying a marginally differential rotation with a probability of $\sim 60\%$. We find that the rotation period is much shorter than the orbital period of $\sim0.57$~d, indicating that this system is not synchronized.
   The AMs and FMs were found to be measurable for 44 frequencies with high enough amplitude, including 12 rotational components. We characterized their modulating patterns and found a clear correlation between the amplitude and frequency variation, linked to nonlinear resonant couplings. In general, the modulating scale and timescale are on the order of a few dozen of nanohertz and a few tens of days, respectively. These values can serve as important constraints on future calculations of nonlinear amplitude equations.}
   {PG~0101+039 is an unsynchronized system containing a component whose amplitude and frequency variations are generally found to be on a shorter timescale than previously reported for other sdB pulsators. Those findings are essential to setting observational constraints on the nonlinear dynamics of resonant mode couplings and orbital solutions.}


\maketitle
%
%-------------------------------------------------------------------

 % \input{Sections/introduction}
\section{Introduction}\label{sec:intro}
Hot B subdwarf (sdB) stars are faint blue objects characterized by the following properties: mass of around 0.5\,$M_{\odot}$, effective temperature of $T_{\rm{eff}} \sim 20000-40000$~K, and surface gravity of $\log g \sim 5.2-6.2$~dex \citep[see][for a review]{2016PASP..128h2001H}.
Most of these objects burn helium in the convective core and are covered by a very thin hydrogen-rich envelope. The formation of sdB stars is assumed to trigger some mechanism that accounts for  almost the entire mass of the envelope ending up expelled during the red giant branch. Binary evolution is probably a successful channel for forming sdB stars \citep{2002MNRAS.336..449H},  supported by the fact that more than 50\% sdBs reside in binary systems \citep{2001MNRAS.326.1391M,2013A&A...559A..54V}.

For those intrinsic properties, the sample of sdB stars is relatively small and their candidates only reach a total of up to $\sim6000$ \citep{2020A&A...635A.193G}. A fraction of sdB stars show intrinsic luminosity variations, offering the unique opportunity to probe their internal structure and chemical profiles via asteroseismology \citep[see, e.g.,][]{1996ApJ...471L.103C}. They are typically classified into three pulsating groups, V361~Hya with short-period pressure ($p$-) modes \citep{1997MNRAS.285..640K}, V1093~Her with long-period gravity ($g$)- modes  \citep{2003ApJ...583L..31G}, and DW~Lyn pulsates both in $p$- and $g$-modes \citep{2006A&A...445L..31S}. Those modes are driven by a classical $\kappa$-mechanism as the iron group elements (mostly iron itself) accumulate in the Z-bump region \citep{1997ApJ...483L.123C,2003ApJ...597..518F}.

In the early days, seismic solutions had been successfully obtained only with short-period {\sl p}-mode pulsators \citep[see, e.g.,][]{2008A&A...489..377C,2008A&A...483..875V}, due to the limitations of ground-based observations of {\sl g}-mode pulsators. Seismic studies were performed on the latter ones until when consecutive photometry became available from space, first with the  {\sl MOST} \citep{2005ApJ...633..460R} mission and then {\sl CoRoT} \citep{2010A&A...516L...6C}. The sharp resolution of frequency and the low level of amplitude noise offered an astonishing improvement. Subsequent missions, such as {\sl Kepler} \citep{2010Sci...327..977B,2014PASP..126..398H} and {\sl TESS} \citep{2015JATIS...1a4003R}, certainly shed new light on this field with unprecedented high-quality photometric data delivered on sdB pulsators. More than 100 sdB pulsators have been observed by {\sl Kepler/}2 and {\sl TESS} missions \citep[see, e.g.,][]{2021A&A...650A.205V}. %With broader spatial coverage, {\sl TESS} detected oscillations in more than 100 sdB pulsators \citep[see, e.g.,][]{2022A&A...663A..45K,2022arXiv221109137B}. 
However, only a few of them have been successfully explored with a seismic diagnosis of the interior \citep[see, e.g.,][]{2010ApJ...718L..97V,2011A&A...530A...3C,2019A&A...632A..90C}.

The rich frequency content resolved from space observations has led to many important findings on sdB pulsators. One of the most stringent discoveries is the rotation period in sdB stars, as disclosed by rotational multiplets, with rates on the order of several weeks up to a year \citep[see, e.g.,][]{2018OAst...27..112C,2022MNRAS.511.2201S}. This period distribution has no significant difference from that of the core of red clump stars \citep{2012A&A...548A..10M}. However, for sdB in binary systems, the rotational period is somewhat faster than that of their single counterparts, which may suggest that tidal dynamics had some effect on the redistribution of {their} angular momentum \citep{1989ApJ...342.1079G}. Several claims have been reported that radial differential rotation { is present} in sdB pulsators in such binary systems, based{ on} {\sl p}- and {\sl g}-mode rotational multiplets \citep[see, e.g.,][]{2015ApJ...805...94F,2020MNRAS.492.5202R,2022ApJ...933..211M}. This might suggest that most sdB stars do not rotate synchronously to the orbital rate in close binaries. Interestingly, several studies have also supported evidence of the detection of high degree ($\ell > 3$) modes via rotational multiplets \citep[see, e.g.,][]{2014A&A...570A.129T,2018MNRAS.474.4709K,2019MNRAS.489.4791S}, despite the fact that those modes should not be easily detected due to their much lower visibility \citep{1997A&A...317..919D}.

Investigating the dynamics of oscillation modes in sdB stars is important to the development of nonlinear asteroseismology, a regime that predicts amplitude modulation (AM) and frequency modulation (FM). Pulsating sdB stars show great potential in this forefront as their frequency spectra present various resonance modes \citep[see, e.g.,][]{2016A&A...585A..22Z}. As offered by {\sl Kepler}, a series of works had been concentrated on the analysis of mode variability in sdB pulsators. In the discovery literature, \citet{2016A&A...585A..22Z} found that the behavior of the resonant modes is more complicated than that predicted by the nonlinear theory of amplitude equations. Their later work suggests that most oscillation modes in sdB are probably unstable on timescales of months to years \citep{2018ApJ...853...98Z}. Although {\sl Kepler} collected unprecedentedly high-quality photometry for such research, the modulating patterns of amplitude may exhibit significant differences when choosing different types of flux provided by the standard pipeline \citep{2021ApJ...921...37Z}. Recently, characterizing AM and FM in sdB pulsators were extended to {\sl K}2 photometry but aimed at searching for relatively short-term patterns as limited by the observational duration \citep{2022ApJ...933..211M}. This larger sky coverage indeed provides more suitable targets for such investigation.

% target star
In this paper, we concentrate on the bright sdB star, PG~0101+039 (also known as EPIC~220376019) to study its seismic properties and to characterize the AMs and FMs of its pulsation modes. {PG~0101+039} locates at $ \rm \alpha = 01{^h}04{^m}21.675{^s}$ and $\rm \delta = +04{^d}13{^m}{37.055{^s}}$, with $K_p = 12.113$ (the magnitude in the Kepler band). 
It was originally identified as a sdB star from spectra by \cite{1968ApJ...152..443S}. 
{\cite{2008A&A...477L..13G} provided its effective temperature $T_\mathrm{eff} = 27500\pm500$\,K and surface gravity $\log g = 5.53\pm0.07$\,dex}. With extensive spectroscopic observations, \cite{1999MNRAS.304..535M} concluded that this star is in a close-binary system with an orbital period of $\rm \sim 0.57\,d$. Later, \citet{2002MNRAS.333..231M} suggested that the companion of PG~0101+039 is likely to be a white dwarf star. With about 400-h photometry from {\sl MOST} mission, it was discovered with three {\sl g}-mode oscillations of the low amplitude of less than 1~ppt \citep{2005ApJ...633..460R}. Considering the light variations of ellipsoidal deformation, \citet{2008A&A...477L..13G} concluded that PG~0101+039 ought to be a tidally locked rotation system.
%The distance obtained by GAIA DR3 is $d = 340.781^{+5.470}_{5.300}$\,pc \citep{2021A&A...649A...1G}. 
The present paper is structured as follows: 
We analyzed the {\sl K}2 photometry and extracted the frequencies in PG~0101+039, using these data to investigate the properties of the rotation and period spacing, as described in Sect.\,\ref{sec:FreCon}. In Sect.\,\ref{sec:AFM}, we present our characterization of the amplitude and frequency modulations of the 44 most significant frequencies. W  present our discussion of the orbital information of this binary system and the possible interpretation of the observed modulations in Sect.\,\ref{sec:DisCon}. Finally we present our conclusions in Sect.\,\ref{sec:con}.

\section{Photometry and frequency content} \label{sec:FreCon}

\subsection{Photometry}\label{sec:APT}
 
\begin{figure}
\centering 
\includegraphics[width=0.5\textwidth]{ 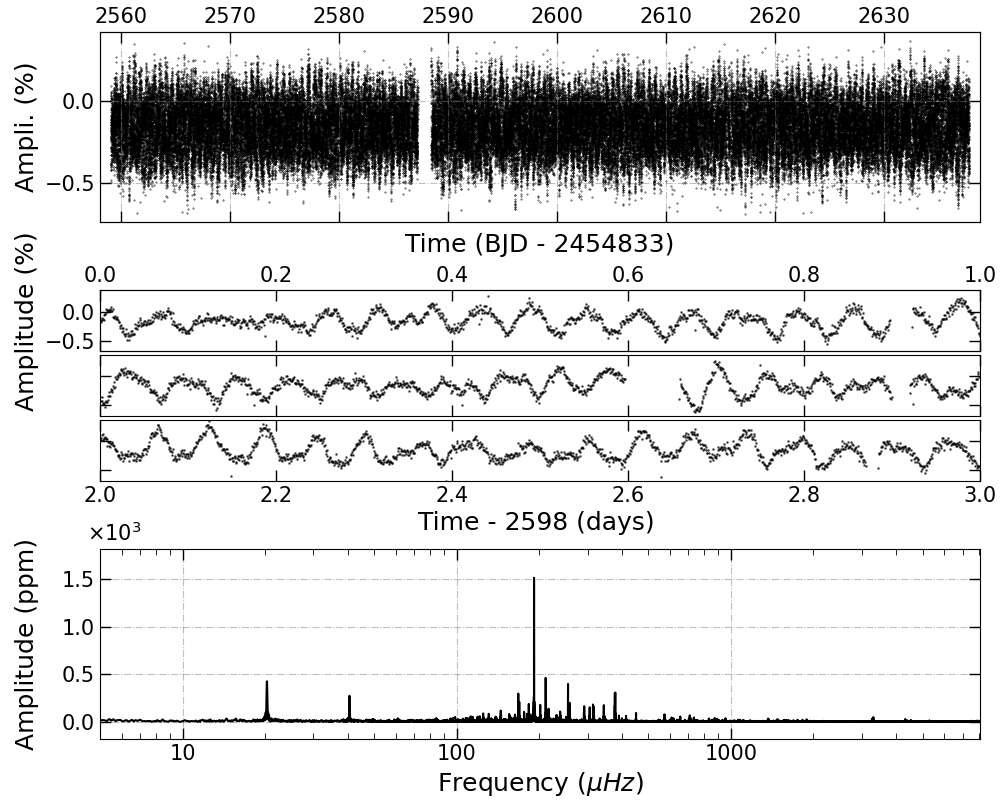}
\caption{{{\em K}2 photometry and frequency signals obtained for {PG~0101+039}}. 
\emph{Top panel}: Complete light curve (amplitude is in percentage, a.k.a. \%, of the mean brightness) with a data sampling of 58.85~s.
\emph{Middle panel}: Close-up view of a 3-d light curve (starting at BJD~2457431) with each panel having a one-day slice.
\emph{Bottom panel}: Lomb-Scargle periodogram of the assembled light curve (amplitude in ppm vs frequency in $\mu$Hz on a logarithmic scale). 
}
\label{fig:light-curve}
\end{figure}

{PG~0101+039}  was observed by {\em K}2 in Campaign~8. As rapid oscillations in sdB pulsators, we completely downloaded the target pixel files (TPFs) in short-cadence (58.85~s) from the Mikulski Archive for Space Telescopes \footnote{https://archive.stsci.edu/}. The \texttt{Lightkurve} \citep{2018ascl.soft12013L} package was used for flux extraction from the available TPFs. To complete the fine pointing during the {\em K}2 observation, a $\sim6.5$~hr thruster firing had to be performed in order to compensate for the solar pressure variation, which led the stamp of the target to move slowly on the CCD module.
We subsequently adopted the \texttt{KEPSFF} routine \citep{2014PASP..126..948V} to correct the systematic photometric variation induced by such a change in the altitude of the craft. 
{To determine the optimal aperture, we tested a series of stamps from the TPFs (see Appendix\,\ref{appecd:lc_stamp}).}
The final light curves cover a duration of 78.7\,days from BJD 2457392.06 to 2457470.78, as shown in the top panel of Fig\,\ref{fig:light-curve}, leaving $111{,}710$ data points. We can clearly see brightness variations with a period of around an hour from the expended light curves as shown in the middle panel of Fig\,\ref{fig:light-curve}. The Lomb-Scargle periodogram (LSP) of {PG~0101+039} is demonstrated at the bottom of Fig\,\ref{fig:light-curve}, with {a Nyquist frequency of approximately 8500\,$\mu$Hz and} significant peaks clustering around 100 to 1000~$\mu$Hz. { This frequency range is suitable for detection of both p- and g-mode in sdB pulsators \citep{2008ASPC..392..231F}.}

\subsection{Frequency extraction and classification}\label{sec:Fre_extra_class}

\begin{table*}
\renewcommand{\arraystretch}{1.22}%设置行高
\caption{List of significant frequencies detected in {PG~0101+039}, by order of increasing frequency.}             
\label{Tab:freq}      
\centering          
\begin{tabular}{l c c c c c c r c c c c}    % 7 columns 
\hline\hline       
{ID} & {Frequency} & {$\sigma$f} & {Period} & {$\sigma$P} & {Amplitude} & {$\sigma$A}  & {S/N} & {$\ell$} & {$m$} & {{Mode}} & {Comments}  \\
{}  & {($\mu$Hz)} & {($\mu$Hz)} & {(s)} & {(s)} 
 & {(ppm)} & {(ppm)} & {}& {} & {} & {}& {}\\
\hline  
\multicolumn{12}{c}{Frequencies in multiplets}  \\
\hline
$f_{44}$   &     144.0292  &     0.0104  &       6943.0373  &     0.5018  &    62.52  &    8.03  &    7.8  &  1  &  0 &  $g$  &  ...  \\ 
$f_{20}$   &     144.7187  &     0.0055  &       6909.9594  &     0.2630  &   117.88  &    8.01  &   14.7  &  1  & 1   & $g$ &  AFM  \\ 
\hline
$f_{30}$   &     155.3015  &     0.0076  &       6439.0879  &     0.3169  &    84.75  &    7.99  &   10.6  &  1  &  0  &  $g$  &  ...  \\ 
$f_{37}$   &     155.9591  &     0.0090  &       6411.9392  &     0.3690  &    72.00  &    7.97  &    9.0  &  1  & 1 & $g$   &  ...  \\ 
\hline
$f_{7}$   &     167.6954  &     0.0021  &       5963.1940  &     0.0739  &   302.10  &    7.74  &   39.0  &  1  &  -1  &  $g$  &  AFM  \\ 
$f_{27}$   &     168.4449  &     0.0070  &       5936.6595  &     0.2469  &    89.20  &    7.71  &   11.6  &  1 & 0    & $g$   &  ...  \\ 
$f_{9}$   &     169.0794  &     0.0030  &       5914.3814  &     0.1045  &   209.27  &    7.71  &   27.1  &   1 & 1 &  $g$  &  AFM  \\ 
\hline
$f_{35}$   &     182.5338  &     0.0082  &       5478.4369  &     0.2454  &    75.49  &    7.62  &    9.9  &  1  &   -1 &  $g$  &  ...  \\ 
$f_{11}$   &     183.1629  &     0.0031  &       5459.6210  &     0.0916  &   200.91  &    7.61  &   26.4  &   1 &   0 &  $g$  &  AFM  \\ 
$f_{58}$   &     183.8808  &     0.0122  &       5438.3060  &     0.3616  &    50.05  &    7.55  &    6.6  &   1 & 1 &  $g$  &  ...  \\ 
\hline
$f_{66}$   &     200.5844  &     0.0130  &       4985.4327  &     0.3234  &    45.26  &    7.26  &    6.2  &   2   &  -1  &  $g$  &  ...  \\ 
$f_{13}$   &     201.6434  &     0.0032  &       4959.2499  &     0.0785  &   181.95  &    7.16  &   25.4  &   2  &  0  &  $g$  &  AFM  \\ 
$f_{40}$   &     202.5375  &     0.0086  &       4937.3562  &     0.2105  &    67.37  &    7.18  &    9.4  &   2   &  1?  &  $g$  &  ...  \\ 
\hline
$f_{43}$   &     231.0496  &     0.0074  &       4328.0757  &     0.1381  &    64.10  &    5.83  &   11.0  &   2   &   -1 &  $g$  &  ...  \\ 
$f_{46}$   &     232.1609  &     0.0076  &       4307.3583  &     0.1407  &    61.37  &    5.74  &   10.7  &   2   &  0  &  $g$  &  AFM  \\ 
$f_{64}$   &     234.3496  &     0.0096  &       4267.1295  &     0.1757  &    47.33  &    5.63  &    8.4  &  2    &   2 &  $g$  & ... \\ 
\hline
$f_{4}$   &     254.6364  &     0.0009  &       3927.1685  &     0.0146  &   404.06  &    4.71  &   85.8  &  1  &  0  &  $g$  &  S  \\ 
$f_{28}$   &     255.2867  &     0.0043  &       3917.1639  &     0.0665  &    88.47  &    4.73  &   18.7  &  1 & 1 &  $g$  &  AFM  \\
\hline
$f_{22}$   &     291.0652  &     0.0031  &       3435.6564  &     0.0362  &   101.26  &    3.84  &   26.4  &  1  &  -1  &  $g$  &  AFM  \\ 
$f_{17}$   &     291.7374  &     0.0021  &       3427.7404  &     0.0244  &   149.76  &    3.83  &   39.1  &  1  &  0  &  $g$  &  AFM  \\ 
\hline
$f_{84}$   &     332.1861  &     0.0079  &       3010.3603  &     0.0711  &    33.87  &    3.28  &   10.3  &   2   &  0  &  $g$  & ...  \\ 
$f_{94}$   &     333.2654  &     0.0091  &       3000.6119  &     0.0822  &    29.11  &    3.28  &    8.9  &    2  & 1 &  $g$  & ...  \\ 
$f_{103}$   &     334.3583  &     0.0107  &       2990.8035  &     0.0960  &    24.70  &    3.27  &    7.5  &   2   &  2  &  $g$  & ... \\ 
\hline
$f_{69}$   &     343.3255  &     0.0061  &       2912.6879  &     0.0515  &    42.30  &    3.17  &   13.3  &   1 &  -1  &  $g$  &  ...  \\ 
$f_{15}$   &     343.9567  &     0.0015  &       2907.3428  &     0.0128  &   169.36  &    3.15  &   53.7  &   $1$ &  $0$  &  $g$  &  AFM  \\ 
\hline
$f_{6}$   &     377.4207  &     0.0008  &       2649.5631  &     0.0053  &   302.17  &    2.84  &  106.4  &  1  &  -1  &  $g$  &  AFM  \\ 
$f_{25}$   &     378.0847  &     0.0026  &       2644.9102  &     0.0179  &    90.28  &    2.84  &   31.8  &  1  &  0  &  $g$  &  AFM  \\ 
$f_{5}$   &     378.7191  &     0.0008  &       2640.4797  &     0.0053  &   305.51  &    2.84  &  107.6  &  1  & 1 &  $g$  &  AFM \\ 
\hline
$f_{111}$   &     450.1190  &     0.0091  &       2221.6350  &     0.0448  &    22.56  &    2.52  &    8.9  & 1   &  -1  &  $g$  & ... \\ 
$f_{24}$   &     450.8067  &     0.0021  &       2218.2457  &     0.0105  &    96.53  &    2.53  &   38.1  &  1  &   0 &  $g$  &  AFM  \\ 
\hline
$f_{33}$   &     570.2222  &     0.0025  &       1753.7022  &     0.0077  &    78.16  &    2.41  &   32.4  &   4   &  -3  &  $g$  &  AFM  \\ 
$f_{60}$  &     571.5085  &     0.0040  &       1749.7552  &     0.0122  &    49.33  &    2.43  &   20.3  &   4   &  -2  &  $g$  &  AFM  \\ 
$f_{102}$   &     572.7683  &     0.0079  &       1745.9066  &     0.0240  &    25.13  &    2.44  &   10.3  &   4   &  -1  &  $g$  & ...  \\ 
$f_{36}$   &     574.0246  &     0.0026  &       1742.0856  &     0.0079  &    75.38  &    2.43  &   31.0  &   4   &  0  &  $g$  &  AFM  \\ 
$f_{99}$   &     576.5143  &     0.0076  &       1734.5624  &     0.0227  &    26.17  &    2.44  &   10.7  &   4   &  2  &  $g$  &  ...  \\ 
\hline
$f_{79}$   &     651.3998  &     0.0056  &       1535.1555  &     0.0132  &    36.24  &    2.51  &   14.4  &   2   &  -1  &  $g$  & ... \\ 
$f_{179}$   &     652.5496  &     0.0148  &       1532.4506  &     0.0348  &    13.83  &    2.53  &    5.5  &  2   &   0 &  $g$ & ... \\ 
$f_{50}$   &     653.5355  &     0.0036  &       1530.1388  &     0.0083  &    57.54  &    2.53  &   22.8  &   2   &  1&  $g$  &  AM  \\
\hline
$f_{87}$   &     702.6498  &     0.0063  &       1423.1840  &     0.0128  &    32.14  &    2.51  &   12.8  &   6   &    &  $g$  &  ...  \\ 
$f_{88}$   &     705.2075  &     0.0066  &       1418.0224  &     0.0133  &    30.93  &    2.53  &   12.2  &   6   &    &  $g$  &  ...  \\ 
$f_{38}$   &     707.7781  &     0.0030  &       1412.8722  &     0.0060  &    69.22  &    2.55  &   27.1  &    6|8  &    &  $g$  &  AFM  \\ 
$f_{67}$   &     710.3852  &     0.0046  &       1407.6870  &     0.0091  &    45.19  &    2.55  &   17.7  &   8   &    &  $g$  &  AFM  \\ 
$f_{80}$   &     711.7116  &     0.0059  &       1405.0636  &     0.0117  &    34.99  &    2.56  &   13.7  &   8   &    &  $g$  &  ...  \\ 
\hline
$f_{132}$   &     884.2697  &     0.0109  &       1130.8767  &     0.0139  &    19.76  &    2.65  &    7.5  &   $\ge 8?$   &    &  $g$  & ... \\ 
$f_{115}$   &     886.9046  &     0.0097  &       1127.5170  &     0.0123  &    22.01  &    2.63  &    8.4  &   $\ge 8$?   &    &  $g$  & ...  \\ 
$f_{96}$   &     889.5773  &     0.0075  &       1124.1294  &     0.0095  &    28.02  &    2.61  &   10.7  &    $\ge 8$?  &    &  $g$  &  ...  \\ 
$f_{86}$   &     890.9649  &     0.0065  &       1122.3786  &     0.0082  &    32.78  &    2.63  &   12.5  &   $\ge 8$?   &    &  $g$  & ...  \\ 
\hline               
\end{tabular}
\end{table*}

\addtocounter{table}{-1} 
\begin{table*} \caption[]{continued.}
\renewcommand{\arraystretch}{1.22}%设置行高
\centering  
\begin{tabular}{l c c c c c c r c c c c} 
\hline\hline
{ID} & {Frequency} & {$\sigma$f} & {Period} & {$\sigma$P} & {Amplitude} & {$\sigma$A}  & {S/N} & {$\ell$} & {$m$} & {{Mode}} & {Comments}  \\
{}  & {($\mu$Hz)} & {($\mu$Hz)} & {(s)} & {(s)} 
& {(ppm)} & {(ppm)} & {}& {} & {} & {}& {}\\
\hline 
$f_{128}$   &     929.4440  &     0.0103  &       1075.9121  &     0.0119  &    20.49  &    2.60  &    7.9  &   $\ge 8$?   &    &  $g$  &  ...  \\ 
$f_{172}$   &     933.3882  &     0.0146  &       1071.3656  &     0.0168  &    14.49  &    2.61  &    5.5  &    $\ge 8$?  &    & $g$   & ...  \\ 
$f_{164}$   &     934.7951  &     0.0142  &       1069.7532  &     0.0162  &    15.08  &    2.64  &    5.7  &   $\ge 8$?   &    & $g$   & ...  \\ 
\hline
$f_{136}$   &    1060.5123  &     0.0107  &        942.9405  &     0.0095  &    18.60  &    2.46  &    7.6  &   8   &    &  $g$  & ...  \\ 
$f_{134}$   &    1063.1768  &     0.0103  &        940.5773  &     0.0091  &    19.34  &    2.45  &    7.9  &    8  &    &  $g$  & ... \\ 
$f_{106}$   &    1065.7602  &     0.0085  &        938.2973  &     0.0075  &    23.46  &    2.45  &    9.6  &   8   &    &  $g$  & ...  \\ 
\hline
$f_{82}$   &    1367.6826  &     0.0060  &        731.1638  &     0.0032  &    34.83  &    2.56  &   13.6  &    4  &    &  mixed &  ...  \\ 
$f_{107}$   &    1370.3923  &     0.0090  &        729.7180  &     0.0048  &    23.21  &    2.57  &    9.0  &   4   &    &  mixed &  ...  \\ 
$f_{101}$   &    1373.1646  &     0.0083  &        728.2448  &     0.0044  &    25.15  &    2.59  &    9.7  &  4    &    & mixed  &  ...  \\ 
\hline
$f_{57}$   &    3306.9347  &     0.0036  &        302.3948  &     0.0003  &    50.25  &    2.25  &   22.3  &  4    &    &  $p$  &  AFM  \\ 
$f_{129}$   &    3310.7620  &     0.0090  &        302.0453  &     0.0008  &    20.33  &    2.26  &    9.0  &   4   &    &  $p$  & ...  \\ 
$f_{180}$   &    3312.2120  &     0.0147  &        301.9130  &     0.0013  &    12.34  &    2.23  &    5.5  &   4   &    &  $p$  & ...  \\ 
$f_{169}$   &    3313.5855  &     0.0123  &        301.7879  &     0.0011  &    14.68  &    2.22  &    6.6  &   4   &    &  $p$  &  ...  \\ 
\hline
\multicolumn{12}{c}{Other independent frequencies}  \\
\hline
$f_{23}$   &     176.0551  &     0.0063  &       5680.0400  &     0.2049  &    99.85  &    7.82  &   12.8  &   $1^*$|$2^*$   &  $0^*$|$-2^*$  &  $g$  &  AFM  \\ 
$f_{1}$   &     191.4365  &     0.0004  &       5223.6643  &     0.0108  &  1532.74  &    7.52  &  203.9  &    $1^*$|$2^*$  &  $0^*$|$2^*$  &  $g$  &  AFM  \\ 
$f_{2}$   &     211.0751  &     0.0012  &       4737.6492  &     0.0264  &   466.68  &    6.77  &   69.0  &    $2^*$ &  $0^*$ &  $g$  &  AFM  \\ 
$f_{19}$   &     212.3919  &     0.0041  &       4708.2776  &     0.0918  &   130.76  &    6.68  &   19.6  &   $1^*$  &  $0^*$ &  $g$  &  AFM  \\ 
$f_{18}$   &     216.2481  &     0.0036  &       4624.3181  &     0.0761  &   146.22  &    6.42  &   22.8  &      &    &  $g$  &  AFM  \\ 
$f_{21}$   &     238.2989  &     0.0040  &       4196.4107  &     0.0698  &   110.22  &    5.39  &   20.4  &   $1^*$   &  $0^*$  &  $g$  &  AFM  \\ 
$f_{10}$   &     258.4635  &     0.0018  &       3869.0180  &     0.0273  &   207.41  &    4.67  &   44.4  &      &    &  $g$  &  AFM  \\ 
$f_{16}$   &     305.5117  &     0.0019  &       3273.1966  &     0.0202  &   152.56  &    3.55  &   43.0  &   $2^*$   &  $-2^*$  &  $g$  &  AFM  \\  
$f_{12}$   &     314.0817  &     0.0015  &       3183.8846  &     0.0150  &   187.03  &    3.42  &   54.7  &   $1^*$  &  $0^* $ &  $g$  &  AFM  \\ 
$f_{14}$   &     315.3790  &     0.0016  &       3170.7879  &     0.0159  &   174.72  &    3.41  &   51.3  &      &    &  $g$  &  AFM  \\ 
$f_{31}$   &     316.7995  &     0.0034  &       3156.5707  &     0.0341  &    80.68  &    3.41  &   23.7  &     &    & $g$   &  AFM  \\
$f_{39}$   &     390.5330  &     0.0034  &       2560.6031  &     0.0220  &    67.72  &    2.80  &   24.2  &   $2^*$   &  $-2^*$  &  $g$  &  FM  \\ 
$f_{41}$   &     414.4970  &     0.0033  &       2412.5626  &     0.0192  &    66.65  &    2.71  &   24.6  &   $1^*$   &  $0^*$  &  $g$  &  AFM  \\ 
$f_{65}$   &     607.7807  &     0.0044  &       1645.3304  &     0.0120  &    45.86  &    2.51  &   18.3  &      &    &  $g$  &  AFM  \\ 
$f_{73}$   &     617.1960  &     0.0052  &       1620.2307  &     0.0136  &    39.34  &    2.51  &   15.7  &      &    &  $g$  &  AFM  \\ 
$f_{68}$   &     732.8429  &     0.0047  &       1364.5489  &     0.0088  &    44.95  &    2.61  &   17.2  &      &    & $g$   &  AFM  \\ 
$f_{85}$   &    3269.8990  &     0.0056  &        305.8198  &     0.0005  &    32.85  &    2.29  &   14.3  &      &    &  $p$  &  AFM  \\ 
$f_{90}$   &    4318.7076  &     0.0067  &        231.5508  &     0.0004  &    30.30  &    2.52  &   12.0  &      &    &  $p$  & AFM  \\ 
\hline 
\multicolumn{12}{c}{Orbitial Information}  \\
\hline 
$f_{3}$   &      20.3085  &     0.0014  &      49240.4255  &     3.4977  &   438.78   &    7.81  &   56.2  &      &    &    &  AFM  \\ 
$f_{8}$   &      40.6194  &     0.0022  &      24618.7763  &     1.3508  &   272.22  &    7.49  &   36.4  &      &    &    & AFM  \\ 
\hline 
\multicolumn{12}{c}{Combination Frequencies}  \\
\hline 
$f_{48}$   &     346.2998  &     0.0042  &       2887.6715  &     0.0354  &    59.60  &    3.12  &   19.1  &      &    &    & AFM  \\ 
$f_{89}$   &    4320.7351  &     0.0068  &        231.4421  &     0.0004  &    30.48  &    2.56  &   11.9  &      &    &    & AFM  \\ 
\hline
\end{tabular}
\tablefoot{Column 1 gives the identification ID in order of decreasing amplitude); columns 2 and 3 give the frequencies in $\mu$Hz and errors; columns 4 and 5 give the periods in seconds and errors; columns 6 and 7 show the amplitudes in ppm (parts per million) and errors; column 8 gives the signal-to-noise ratio (S/N) level; columns 9 and 10 show the quantum number identified by the asymptotic regime and rotation (see Sect.~\ref{sec:RoMul} and {Appendix~\ref{sec:Period_Spacing}});{ column11 gives the preliminary classification of modes (see Sect.~\ref{sec:Fre_extra_class});} and column12  comments on whether amplitude or frequency modulations were measured or not.
The AM/FM/AFM labels indicate whether the frequency has modulations of amplitude~(AM), frequency~(FM), or both~(AFM) and 'S' stands for stable.'*' in $\ell$ and $m$ means they are identified by period spacing or the \'Echelle diagram {(see details in Appendix \ref{sec:Period_Spacing})}. 
}
\end{table*}

Computing the Fourier transforms (FT) of the light curves is helpful in examining the periodic signals presented in the data. We used the dedicated software \texttt{FELIX}\footnote{Frequency Extraction for Lightcurve exploitation, developed by S.~Charpinet, greatly optimizes the algorithm and accelerates the speed of calculation when performing frequency extraction from dedicated consecutive light curves. See details in \citet{2010A&A...516L...6C,2019A&A...632A..90C} and \citet{2016A&A...585A..22Z,2016A&A...594A..46Z}.} to perform frequency extraction from the coordinated light curves. The significant frequencies were prewhitened in order of decreasing amplitude until it goes down to the adopted threshold of 5.2 times the local noise level, which is defined by the median value of the amplitude in the LSP. % \citep[see, e.g.,][]{2021ApJ...921...37Z}. 
This signal-to-noise ratio limit of $S/N = 5.2$  was adopted as a compromise between the testing results from 2-yr {\em Kepler} and 27-d TESS photometry \citep{2016A&A...585A..22Z,2019A&A...632A..90C}. 
The highest peak was meant to be extracted when several close frequencies were encompassed within $0.4\,\mu$Hz, namely, about $3\times \Delta f$, where $\Delta f = 1/T$ {is the frequency resolution $\sim0.147\,\mu$Hz} and T $\sim78.7$ days. {Table \ref{Tab:freq} lists 83 significant frequencies, which includes} 44 frequencies whose amplitude modulations (AM) and frequency modulations (FM) are characterized in the following section. Another {39} significant frequencies were identified as rotational components, but their amplitudes were not high enough to characterize the AM or FM.  The entire frequency data set is listed in Table \ref{Tab:freq_all}, containing 137 independent frequencies, 2 orbital frequencies, and 51 linear combination frequencies. Compared to the three pulsation frequencies listed in \citet{2005ApJ...633..460R}, the frequency around 7235~s was not detected in our list, whereas the other two were found, but with a different significant amplitude, regardless of the observation band.

\begin{figure*}[t]
\centering 
\includegraphics[width=\textwidth]{ 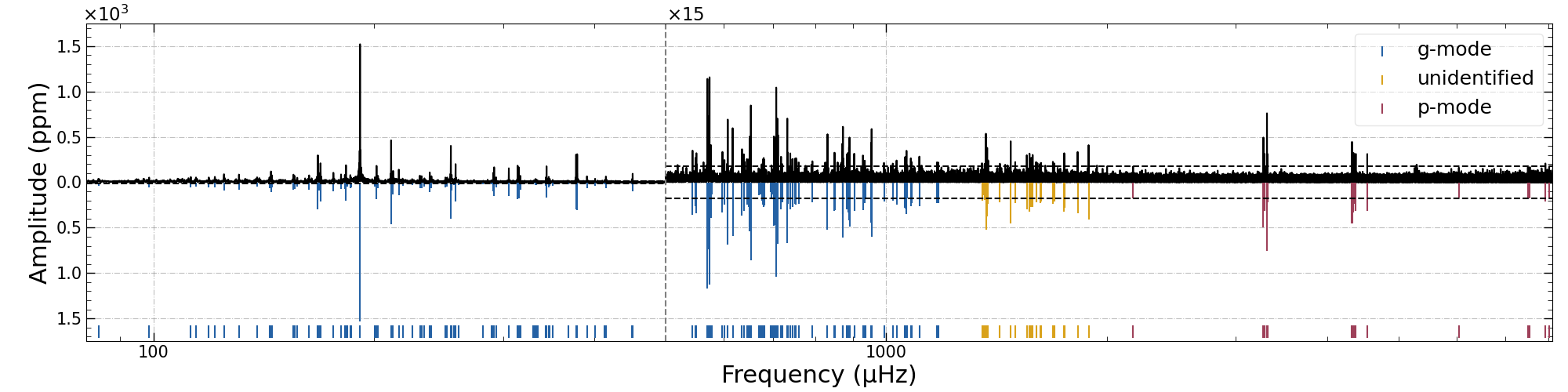}
\caption{LSP of the {\sl K}2 photometry collected on {PG~0101+039}, covering the frequency range where the pulsation signals are found. The entire periodogram is divided into two different ranges: the low-frequency {\sl g}-mode region (blue lines) and the high-frequency {\sl p}-mode region (yellow and red lines).
The dashed horizontal lines are the 5.2$\sigma$ threshold of the local noise and the colorful vertical segments at the bottom of the LSP indicate the frequencies we extracted.
}
\label{fig:MD}
\end{figure*}

As rich frequencies are presented in {PG~0101+039}, it is wise to give a preliminary classification of acoustic {\sl p}-modes and gravity {\sl g}-modes, since their seismic properties are different. We followed a similar treatment of this classification to \citet{2022ApJ...933..211M}, barely by their pulsation period. 
Theoretical calculations of sdB pulsators suggest that dipole {\sl p}-modes typically exhibit periods of $<400$\,s (or $f > 2500\,\mu$Hz), whereas their {\sl g}-modes are $>1000$\,s (or $f < 1000\,\mu$Hz) \citep[see, e.g.,][]{2003ApJ...597..518F,2005A&A...443..251C,2011A&A...530A...3C}. However, p-mode periods can increase beyond 600~s when $T_\mathrm{eff}$ and $\log g$ decreases \citep{1999PhDT........26C,2001PASP..113..775C,2002ApJS..139..487C}. { This brings on difficulties in terms of the direct classification of {\sl p}- and {\sl g}-mode when their periods are around 600~s, even though PG~0101+039 is clearly located in the {\sl g}-mode dominating pulsator}.
Figure\,\ref{fig:MD} shows the part of the LSP where the pulsation frequencies are extracted and labeled with preliminary classification. We detected 94 independent frequencies in the range of [$\sim80-1200$]~$\mu$Hz that are highly probable {\sl g}-modes (the blue vertical segments). Another 16 independent frequencies were found in the high-frequency {\sl p}-mode region, [$\sim2600-8000$]~$\mu$Hz (the red vertical segments); there are another {9} independent frequencies in the region of [$\sim1200-2000$]~$\mu$Hz (the yellow segments), which might be low-order high-degree ( $\ell>3$) {\sl g}-modes or mixed modes. This latter assumption requires further classification \citep[see, e.g.,][]{2011A&A...530A...3C,2019A&A...632A..90C}.  
In addition, within the frequency range of 1200 and 2000~$\mu$Hz, we resolved {ten} linear combinations that could be intrinsic resonant modes \citep{2016A&A...594A..46Z} or nonlinear effects from the linear eigenfrequencies \citep{1995ApJS...96..545B}. In the following, we discuss a more sophisticated determination of quantum numbers of those pulsation modes by their seismic properties.

\subsection{Rotational multiplets}\label{sec:RoMul}

\begin{figure*}
\centering
    \includegraphics[width=0.98\textwidth]{ 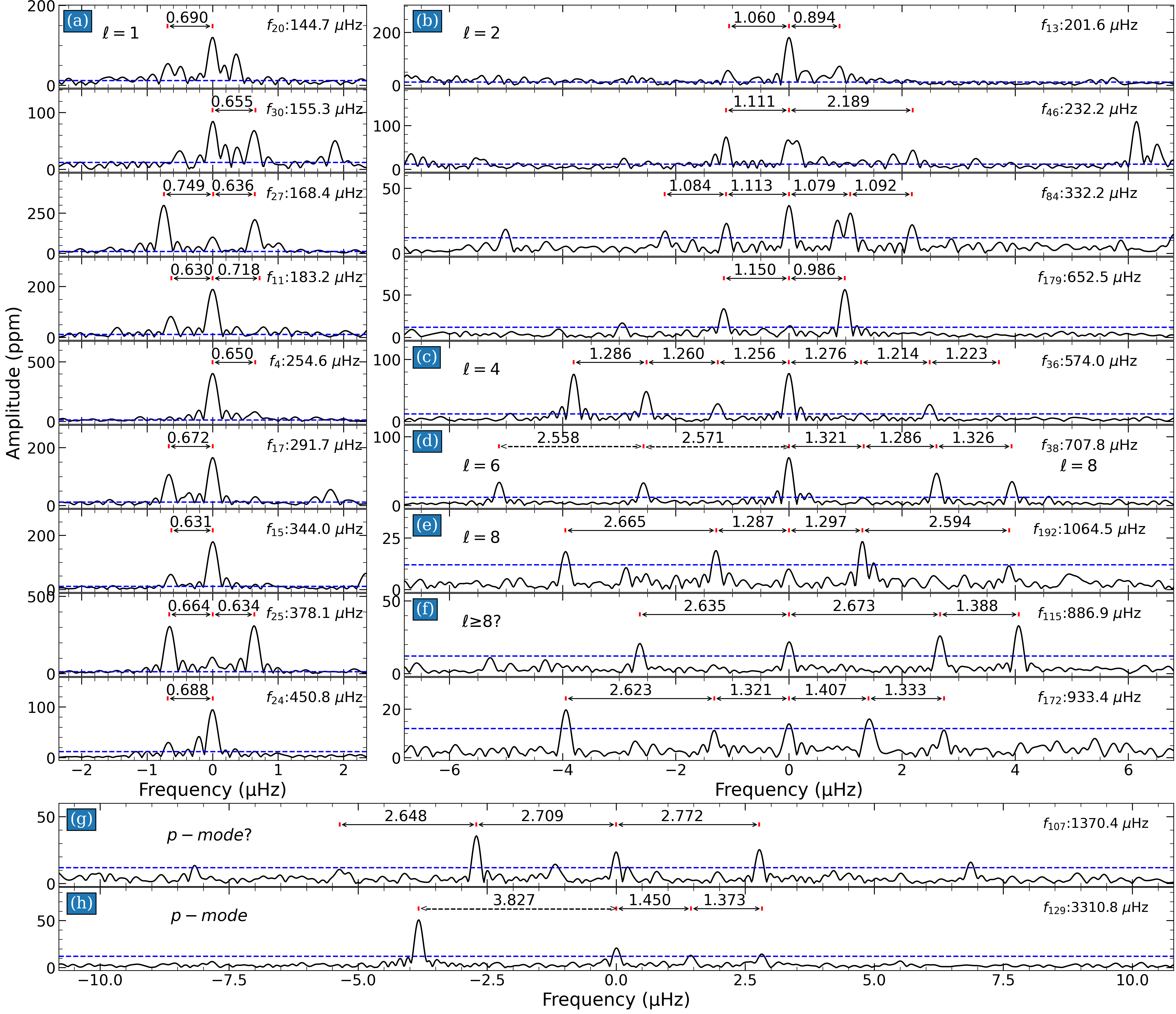}
    \caption{Rotational multiplets identified in {PG~0101+039} by the common frequency spacing. All the components, labeled by vertical segments, are shifted to the frequencies as given in the text. The value of splitting is given in the text and indicated by the annotation of double arrows between the segments, whereas two types of arrow lines indicate the different interpretations of rotation multiplets (see text for details). The blue horizontal lines are the $5.2\sigma$ detection threshold.
    (a)-(e) Different degree $\ell$ multiplets in {\sl g}-mode region, (f) suspected high degree multiplets with $\ell \ge 8$ in {\sl g}-mode region, (g) mixed mode region to be identified of {\sl p}-modes? and (h) {\sl p}-mode region.}
    \label{fig:RotaMult}
\end{figure*}

When a star rotates slowly, the non-radial degenerate frequencies will be split into $2\ell+1$ components. According to the first { order of approximation of the effect of rigid rotation treated on a} perturbation,
the frequencies { of the components} follow the formulae \citep{1992ApJ...394..670D}:
\begin{equation}
    \label{eq:fre_splitting}
    \nu_{n,l,m} = \nu_{n,l,0} + m\Omega(1-C_{n,\ell}),
\end{equation}
where $\nu_{n,l,0}$ is the frequency of the central ($m=0$) component, $\Omega$ is the solid rotational frequency, and the Ledoux constant, $C_{n,\ell} \ll 1,$ is valid for acoustic {\sl p}-mode and $C_{n,\ell} \sim 1/ \ell(\ell +1)$ for high-radial order gravity {\sl g}-mode. According to the frequency spacing of rotational multiplets, it is helpful to determine the rotation period and give precise identification of oscillation modes in pulsating stars.

In {PG~0101+039}, we detected a common frequency spacing of $\sim 0.6 - 1.3\,\mu$Hz, as per the frequencies listed in Table~\ref{Tab:freq}. In Fig.~\ref{fig:RotaMult}, we present all the groups of rotational components identified by frequency spacing. In the low-frequency {\sl g}-mode region, we detected nine groups of frequency spacing of $\sim 0.66\,\mu$Hz, from $0.63$ to $0.75\,\mu$Hz, i.e., $f_{20}\sim$ 144.72~$\mu$Hz, $f_{30}\sim$ 155.30~$\mu$Hz, $f_{27}\sim$ 168.44~$\mu$Hz, $f_{11}\sim$ 183.16~$\mu$Hz, $f_{4}\sim$ 254.64~$\mu$Hz, $f_{17}\sim$ 291.74~$\mu$Hz, $f_{15}\sim$ 343.96~$\mu$Hz, $f_{25}\sim$ 378.08~$\mu$Hz, and $f_{24}\sim$ 450.80~$\mu$Hz, including three triplets and six doublets with one missing component.
We note that the worst precision of frequency in those components is around $0.02\,\mu$Hz. We interpreted the frequency spacing of $\sim {0.66}\,\mu$Hz as being the minimum common spacing for dipole mode rotation, since frequency spacing increases with the degree $\ell$. {According to Eq.\,\ref{eq:fre_splitting}, the frequency spacings of $\ell = 2,4,6,8$ are 1.10\,$\mu$Hz, 1.26\,$\mu$Hz, 1.29\,$\mu$Hz, and 1.30\,$\mu$Hz, respectively}. Roughly considering that $C_{n,1} = 1/2$, we can deduce a rotation period of $P = 8.75 \pm 0.13$~d from the weighted average value $\Delta \nu = 0.66\pm 0.01~\mu$Hz based on the above nine triplets (where the weight is set according to their S/N). We also found four groups of frequencies that can be associated with quadrupole modes with frequency spacing of $\sim 1.08\,\mu$Hz, namely, $f_{13}\sim$ 201.64~$\mu$Hz, $f_{46}\sim$ 232.16~$\mu$Hz, $f_{84}\sim$ 332.18~$\mu$Hz, and $f_{179}\sim$ 652.55~$\mu$Hz including one full quintuplet and three incomplete ones. Similarly to $\ell =1$ (but considering $C_{n,2} = 1/6$), the rotation period can be derived to be $P = 8.78\pm 0.10 $~d from the above four quintuplets. Thus, both the $\ell = 1$ and $\ell = 2$ modes suggest that the rotation period should be around 8.8~days.

With the rotation period known, we can further identify with high confidence several high-degree ($\ell > 3$) frequencies in { the} {\sl g}-mode region. With a frequency spacing of $\sim 1.26 \pm 0.01\,\mu$Hz, the five significant frequencies and two suspected ones around {$f_{36}\sim$ 574.03~$\mu$Hz would correspond} to an $\ell = 4$ multiplet with two missing components, { since the $\ell = 4$ modes should exhibit a frequency splitting of $1.26\,\mu$Hz according to Eq.~\ref{eq:fre_splitting}}. If we only count the five significant frequencies, the splitting has the same value, which still satisfies the $\ell=4$ identification. We note that the $\ell =5$ modes with $\Delta \nu \sim 1.27\,\mu$Hz could meet the measured splitting too, but high odd-degree modes suffer a much higher geometric cancellation effect \citep{2010aste.book.....A}. We thus do not consider any high odd-degree modes in {PG~0101+039} anymore and classified the frequency group around $f_{36}\sim$ 574.03~$\mu$Hz as an $\ell = 4$ multiplet. The frequency spacing found around $f_{38} \sim 707.78\,\mu$Hz shows a distinct feature, namely, the most left and right three components can be derived with $\sim 1.28\,\mu$Hz and $\sim 1.31\,\mu$Hz, respectively. One explanation is that they belong to two different multiplets with $\ell=6$ ($\Delta \nu \sim 1.283\pm0.003\,\mu$Hz) and $\ell =8$ ($\Delta \nu \sim 1.314\pm0.012\,\mu$Hz), respectively. Otherwise, we measured an average frequency splitting of $\sim 1.30 \pm 0.01\,\mu$Hz from all six components, which could be attributed to an $\ell = 8$ multiplet. One clearly incomplete $\ell = 8$ multiplet is probably found around $f_{134} \sim 1063.17\,\mu$Hz, where five components were identified by a frequency spacing of $\Delta \nu \sim 1.30\,\mu$Hz. We also detected two incomplete multiplets near $f_{115} \sim 886.90\,\mu$Hz and $f_{172} \sim 933.38\,\mu$Hz, possibly  having a degree of $\ell \ge 8$, based on  a frequency spacing of $\Delta \nu \sim 1.34\pm 0.01\,\mu$Hz. If we exclude the two relatively large spacing with $\Delta \nu = 1.404$ and $1.388~\mu$Hz in $f_{172}$ and $f_{115}$, respectively, the frequency spacing is $\sim 1.325\pm 0.005\,\mu$Hz; this value is close to the $\ell = 8$ splitting. 
We recall here that a high even-degree, for instance, $\ell = 12, 14, \cdots$, can also satisfy the splitting of $\Delta \nu \sim 1.32\,\mu$Hz but the geometric cancellation effect increases with the value of $\ell$. Thus, it is wise to keep  $\ell$ at the minimum value.

For the multiplets above the low-frequency {\sl g}-modes, we give the preliminary identification of degree number as the Ledoux constant $C_{n,l}$ is near to 0 for {\sl p}-modes. 
In the mixed region near $f_{107} \sim 1370.39\,\mu$Hz, the frequency spacing is derived to be $1.355 \pm 0.017\,\mu$Hz if we count all three significant frequencies and one suspected frequency of $f_{190}^* \sim 1365.03\,\mu$Hz. This measurement may be a bit larger, namely, $\sim1.37\,\mu$Hz, when derived from only the three significant frequencies. Whether this multiplet is high degree {\sl g}-modes or {\sl p}-modes, this splitting is somewhat larger than the previous multiplets with a rational frequency splitting of $\sim 1.315\,\mu$Hz. We prefer this multiplet to be in {\sl p}-modes, in line with a similar frequency splitting found in the {\sl p}-mode multiplet around $\sim 3310.8\,\mu$Hz. We measured a frequency spacing of $\sim 1.34\,\mu$Hz for all four components or $\sim 1.41\,\mu$Hz for the  rightmost three components around $f_{129} \sim 3310.8\,\mu$Hz. Consistently with the chosen minimum degree number, the preliminary classifications are both $\ell=4$ for the multiplets around $1370.4\,\mu$Hz and $3310.8\,\mu$Hz.
{It would be very surprising to find that both two multiplets are identified with degrees, not of $\ell = 1 $ or 2. The low-degree modes seem to have a higher chance of being observed. To reman cautious, these {\sl p}-mode multiplets may also be mixed with dipole or quadrupole modes if we discard the frequencies with an amplitude of S/N$<5.2$. However, to give an exact discriminant mode, it is necessary to wait for the future seismic model on PG~0101+039}.

Finally, we note that the frequency spacing between different components within the same multiplet may display a frequency mismatch up to $0.1\,\mu$Hz, for instance, with respect to the triplet of $f_{7} \sim 168\,\mu$Hz. This is not a defect in rotational components considering the measuring uncertainty and variation of frequency. {In general, the frequency variation is $\sim1.6\times$ the frequency mismatch, even though the measuring uncertainty is merely 17\% of the frequency mismatch.} Therefore, we derived a rotational period of $\sim 8.81 \pm 0.06$ days based on the weighted average splitting from the 14 {\sl g}-mode multiplets (see Fig.\,\ref{fig:RotaMult} a-e). With the two {\sl p}-mode multiplets (Fig.\,\ref{fig:RotaMult}, g-h), we derived a rotational period of $8.60 \pm 0.16$~days, which is slightly faster than that of the {\sl g}-modes. Our result suggests that {PG~0101+039} has a differential rotation with a faster rotating envelope than the core. To remain cautious,
{we assume that PG~0101+039} may still be a rigid object with a probability up to $\sim40\%$ if the uncertainties
of rotational periods are fully considered. {Moreover, we also tried to analyze the period spacing of g-modes in {PG~0101+039}, but without any clear-cut patterns found (see Appendix\,\ref{sec:Period_Spacing} for details).}

 \section{Amplitude and frequency modulations}\label{sec:AFM} 

This section is devoted to characterizing amplitude modulation (AM) and frequency modulation (FM) occurring in significant frequencies with high-enough amplitudes. Recent studies indicate that most oscillation modes are unstable in pulsating sdB stars \citep[see, e.g.,][]{2016A&A...594A..46Z,2022ApJ...933..211M}.
To obtain enough amplitude and frequency measuring points, we have to choose the frequencies with $\rm{S/N}>8.8$ for our aims. {The frequency and amplitude errors derived from \texttt{Felix} had been estimated by \cite{2016A&A...594A..46Z} and \citet{2021ApJ...921...37Z}, whose results demonstrate that those errors are well determined and robust.}
In {PG~0101+039}, there are 44 frequencies that met this criterion, including 19 rotational components. 
After comparing the results of AM and FM from photometry that are extracted via stamps with various pixels (see Appendix\,\ref{sec:AFM_T} for details), the AMs and FMs were characterized  (as given in Sections\,\ref{sec:AFM_Mul} and\,\ref{sec:Other_AFM}).
% We first present comparing results of AM and FM from photometry that is extracted via stamps with various pixels in Appendix\,\ref{sec:AFM_T}. The 

As supported by the comparing results in Appendix\,\ref{sec:AFM_T}, we ultimately selected a conservative stamp ($N=16$) for measuring AMs and FMs of 44 frequencies that are commented in the last column of Table~\ref{Tab:freq}. As proposed in the previous work by  \citet{2022ApJ...933..211M}, the modulating patterns of AMs and FMs can be quantitatively characterized with several simple fitting functions. Here, we follow the same strategy and characterize the discovered AMs and FMs occurring in {PG~0101+039}, via the linear combinations of three simple types of fittings (linear, parabolic, and sinusoidal waves), as defined by a matrix function of: 
\begin{equation}
\begin{aligned}
&G_{k}(t)=\left[g_1(t),g_2(t),g_3(t),g_1(t)+g_3(t),g_2(t)+g_3(t)\right],\\
&\left\{
\begin{array}{l}
    g_1(t) = bt+c,\\
    g_2(t) = at^2+bt+c,\\
    g_3(t) = A\sin(\omega t + \phi)+A_0,
\end{array} 
\right.
\end{aligned}    
\end{equation}
where $k = 1, 2, 3, 4, 5$. We finally chose the fitting function by increasing the index of $k$ after a quantitative evaluation. Then the {Markov chain Monte Carlo (\texttt{MCMC}) method} was applied to derive the uncertainties for the optimal fittings. {\texttt{MCMC} is implemented by the \texttt{EMCEE} code \citep{2013PASP..125..306F} and sample posterior distributions of parameters. We can see the detailed processes for such analysis in Sect.~3 of \citet{2022ApJ...933..211M}}. Table\,\ref{Tab:AFM_D} lists the information of each fitting function for the 45 frequencies with AM and FM, including the fitting coefficients and the correlation between AM and FM.

\begin{figure*}
\centering
\includegraphics[width=0.2693\textwidth]{ 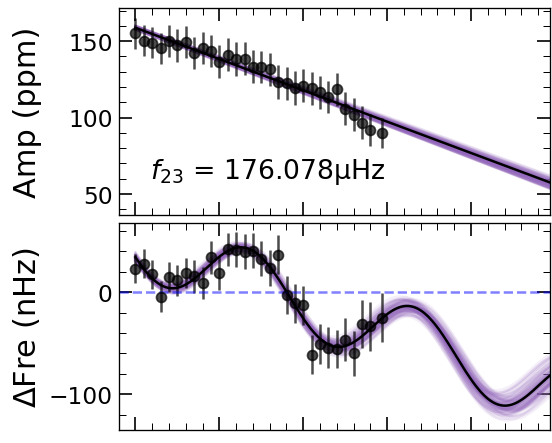}
\includegraphics[width=0.234\textwidth]{ 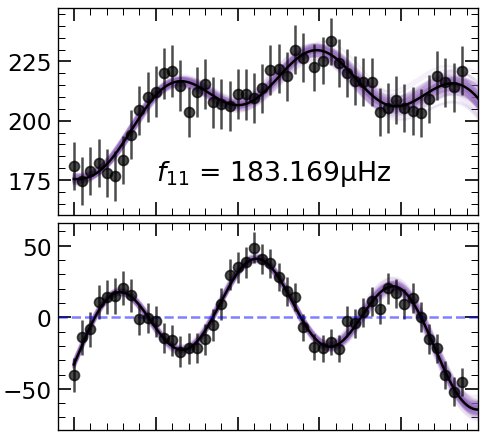}
\includegraphics[width=0.234\textwidth]{ 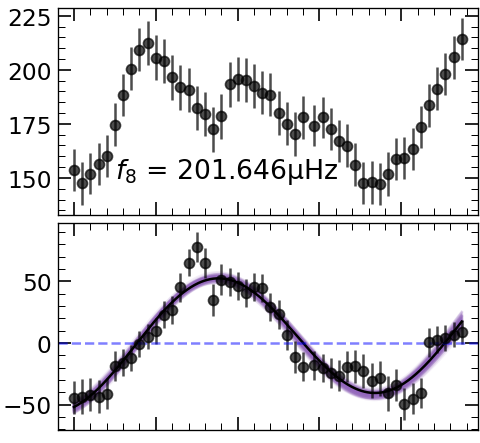}
\includegraphics[width=0.234\textwidth]{ 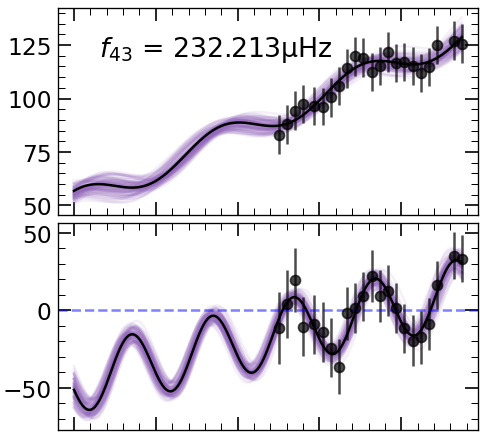}

\includegraphics[width=0.2693\textwidth]{ 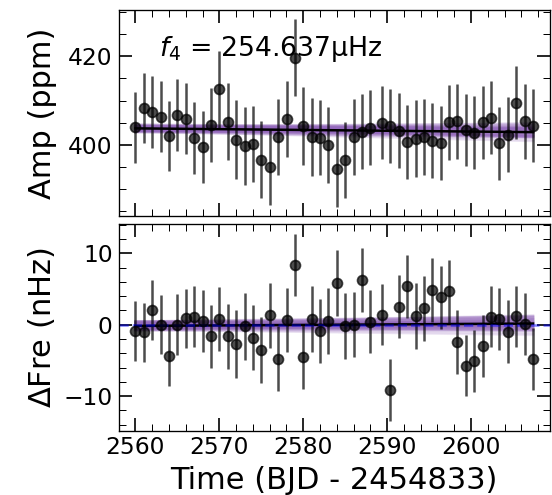}
\includegraphics[width=0.234\textwidth]{ 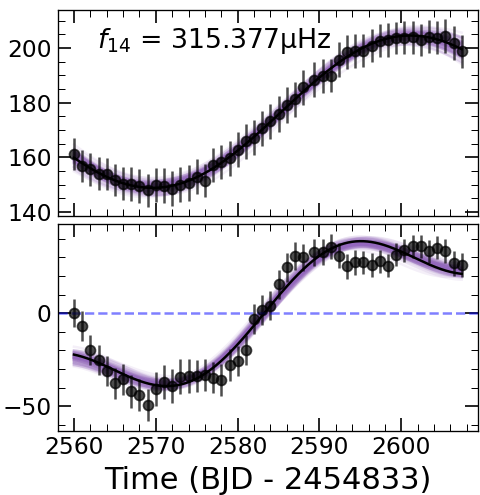}
\includegraphics[width=0.234\textwidth]{ 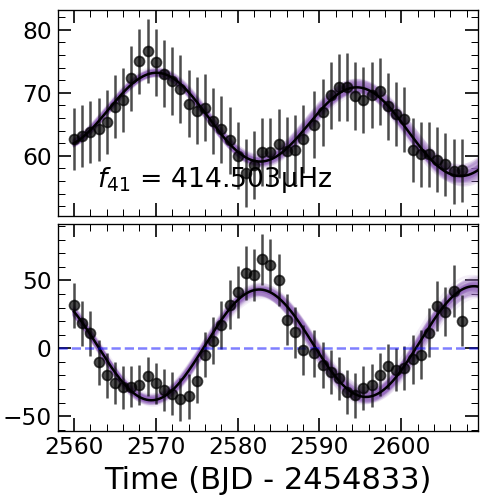}
\includegraphics[width=0.234\textwidth]{ 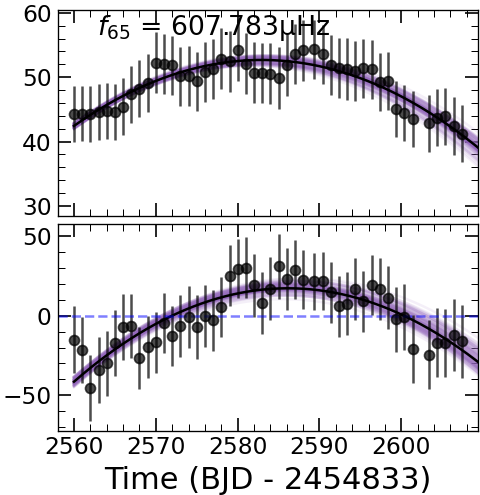}

\caption{Gallery of eight typical frequencies with AM/FM variations. Each module contains AM (top panel) and FM (bottom panel) occurring in one single frequency. The frequencies are shifted to their averages as represented by the horizontal lines in each bottom panel. The solid curves in purple and black represent the fitting results from the \texttt{MCMC} method and the optimal model, respectively. \label{fig:AFM}
}
\end{figure*}

Figure\,\ref{fig:AFM} is a gallery composed of eight portraits that depict AM and FM of each frequency through various fitting functions, $G_k(t)$. Except for the modulating patterns of $f_{4}\sim 254.6\,\mu$Hz, they clearly exhibit evident AM and FM in various patterns, with a majority of them containing a sinusoidal component. Most of those frequencies and amplitudes have undergone an oscillatory behavior back and forth around their average values, for instance, modulations of $f_{44}\sim 414.5\,\mu$Hz fitted with $G_4$ containing a slight linear trend. However, two frequencies, $f_{23}\sim 176.1\,\mu$Hz and $f_{46}\sim 232.1\,\mu$Hz, show a significant linear decreasing and increasing trend, which makes them detectable in only half of the observation. We note that some correlation or anti-correlation happens between the AMs and FMs, for instance, a significant anti-phase of $\rho_\mathrm{am,fm} = -0.82$ is derived for $f_{44}\sim 414.5\,\mu$Hz.

\subsection{AM and FM in Multiplets}\label{sec:AFM_Mul}

\begin{figure*}
\centering
\includegraphics[width=\textwidth]{ 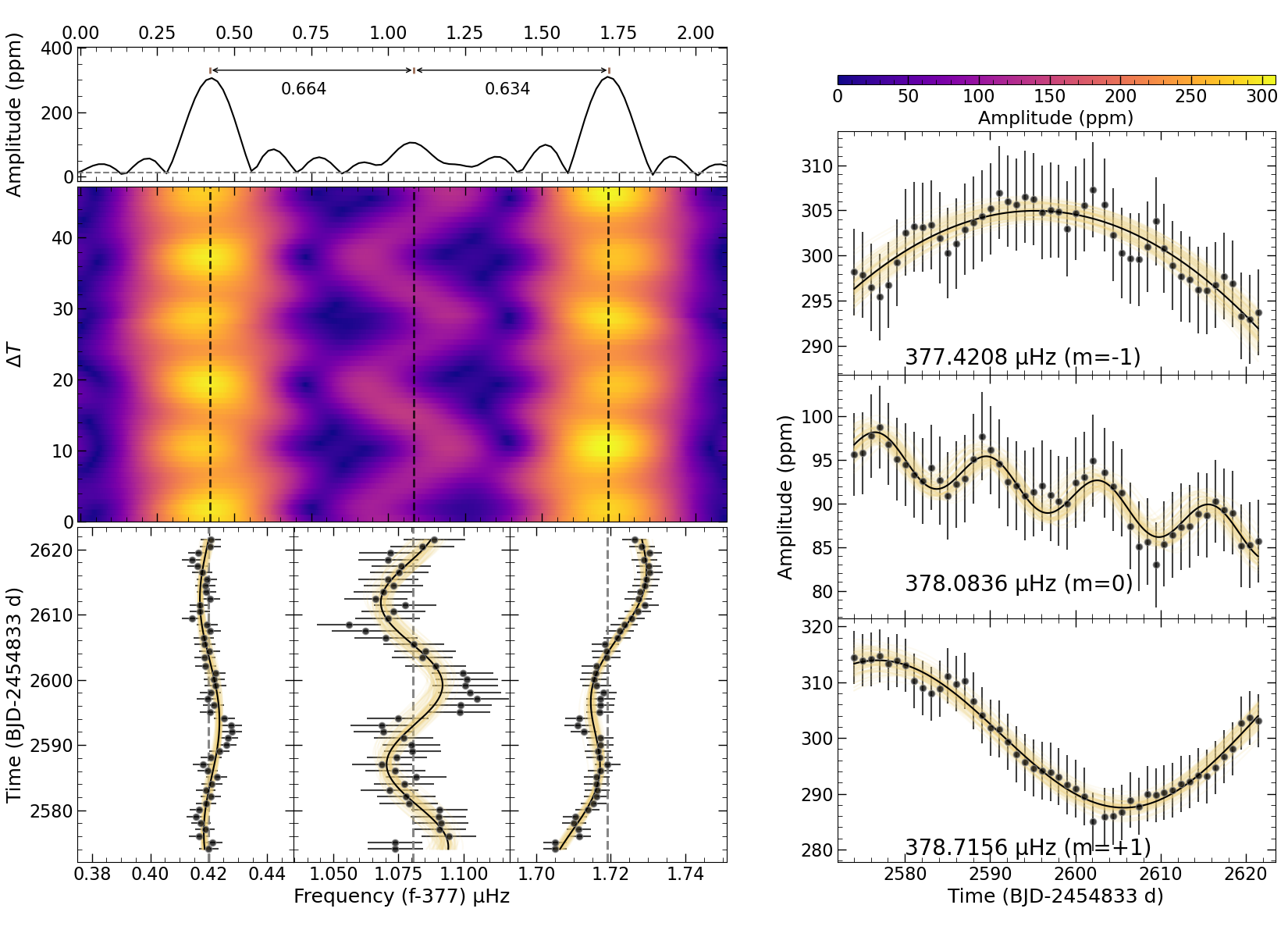}
\caption{Frequency and amplitude modulation in the $g-$mode triplet near 377\,$\mu$Hz. \emph{Top-left:}  Fine profile of the triplet in LSP, with near-symmetric frequency spacing. The dashed horizontal line represents the 5.2\,$\sigma$ detection threshold. \emph{Middle-left:} Sliding LSP giving a 3D view of the dynamics of the concerning triplet in general. \emph{Bottom-left:} Expanded view of FM around the average frequency (vertical lines, also in the \emph{middle-left} panel) of each component. \emph{Right:}  Measured amplitudes of the three components as a function of time obtained from each piece of light curves. The solid curves are the optimal fitting models of the AM and FM patterns, covering with \texttt{MCMC} uncertainties (see text for details). }
\label{fig:Mul_377}
\end{figure*}

\begin{figure*}
\centering
\includegraphics[width=\textwidth]
{ 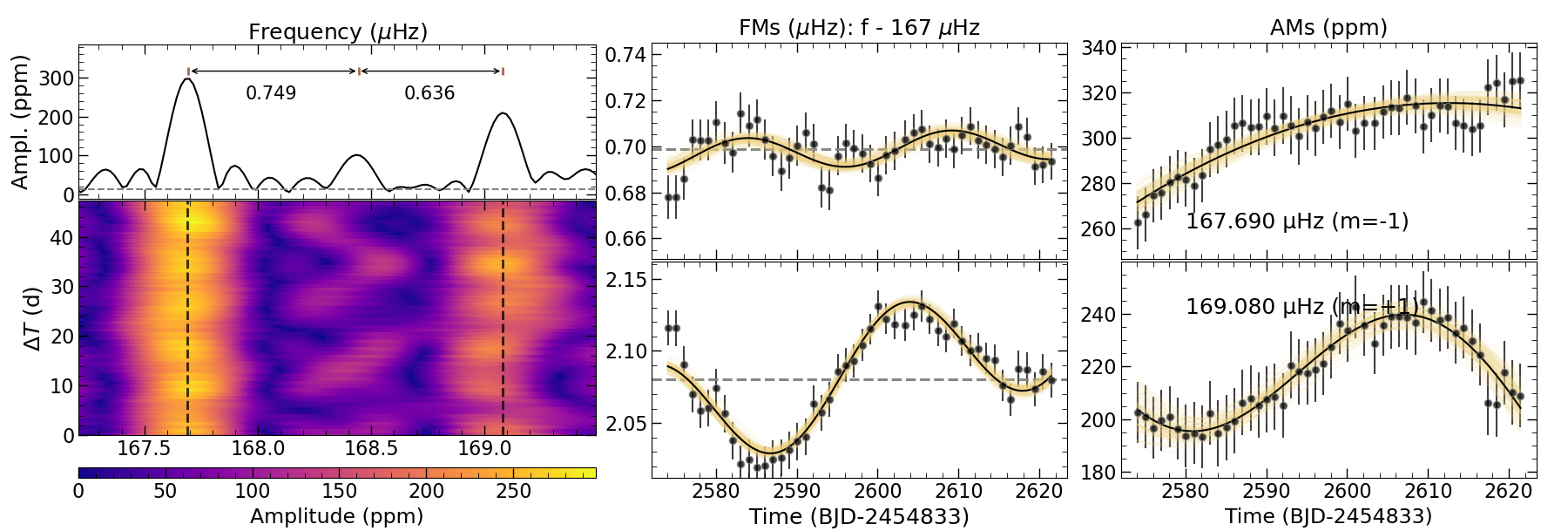}
\caption{Similar to Fig.\,\ref{fig:Mul_377} but for the g-mode triplet near 167\,$\mu$Hz, in a different layout of the sub-panels. Note: the AM and FM of the central component are not presented, as these measurements are limited by its low amplitude and large scale of the FM.}
\label{fig:Mul_167}
\end{figure*}

In this subsection, we focus on the AMs and FMs occurring in five multiplets as themselves are under the resonant condition, including  one triplet near 377\,$\mu$Hz (Fig.\,\ref{fig:Mul_377}), two doublets near 167\,$\mu$Hz and 291\,$\mu$Hz (Fig.\,\ref{fig:Mul_167} and \ref{fig:Mul_291}), one incomplete $\ell=4$ multiplet { near 570\,$\mu$Hz} (Fig.\,\ref{fig:Mul_570}), and one multiplet possibly containing $\ell=6$ and $\ell=8$ components near 707\,$\mu$Hz (Fig.\,\ref{fig:Mul_707}). Here, we describe the details of AMs and FMs occurring in the complete triplet near 377\,$\mu$Hz and one doublet near 167\,$\mu$Hz. For simplicity, the detailed behaviors of AMs and FMs of the other mentioned multiplets are given in Appendix\,\ref{sec:AFM_ap}.

% \subsection{Characterization of modulation patterns}\label{sec:AFM_chara}
% \clearpage

% \input{AFM_Details}
\begin{table*} 
\renewcommand{\arraystretch}{1.31}%设置行高
\caption[]{\label{Tab:AFM_D} AM/FM detected in \textbf{PG~0101+039}, sorted by order of increasing frequency.}
\small
\begin{center}
\begin{tabular}{lcrcrrrrrrrc}
\hline
\hline
\multicolumn{1}{c}{ID} &  \multicolumn{1}{c}{Fre} &  \multicolumn{1}{c}{Corr} &  \multicolumn{1}{c}{AM/FM} &  \multicolumn{1}{c}{A} &  \multicolumn{1}{c}{T=2 $\pi$/$\omega$} &  \multicolumn{1}{c}{$\phi$} &  \multicolumn{1}{c}{a} &  \multicolumn{1}{c}{b} &  \multicolumn{1}{c}{c} &  \multicolumn{1}{c}{Fitting}  \\
\multicolumn{1}{c}{} &  \multicolumn{1}{c}{($\mu$Hz)} &  \multicolumn{1}{c}{} &  \multicolumn{1}{c}{} &  \multicolumn{1}{c}{(ppm/nHz)} &  \multicolumn{1}{c}{(d)} &  \multicolumn{1}{c}{[0, 2$\pi$)} &  \multicolumn{1}{c}{(10$^{-3}$)} &  \multicolumn{1}{c}{(10$^{-2}$)} &  \multicolumn{1}{c}{} &  \multicolumn{1}{c}{}  \\

\hline
\multicolumn{11}{c}{AFMs in multiplets}  \\
\hline 
\multirow{2}{*}{$f_{7}$} & \multirow{2}{*}{167.6896} & \multirow{2}{*}{0.34} & AM & --   & --  &  -- &  $ {29_{-4}^{+4}}$ &  ${226_{-16}^{+16}}$ & $ {271_{-2}^{+2}}$ &  $G_{2}$   \\ 
 &  & & FM &  $12.7_{-1.2}^{+1.2}$   & $16.8_{-0.3}^{+0.3}$  &  $3.8_{-0.2}^{+0.2}$ & -- & $-13_{-6}^{+7}$ &  ${3.3_{-1.7}^{+1.7}}$ &  $G_{4}$  \\
\multirow{2}{*}{$f_{9}$} & \multirow{2}{*}{$169.0805$} & \multirow{2}{*}{$0.33$} & AM & $38.5_{-1.7}^{+1.7}$ & $29.1_{-0.4}^{+0.4}$  &  $1.8_{-0.1}^{+0.1}$ &  -- & $150_{-10}^{+10}$ & $-30.6_{-2.7}^{+2.6}$ & $G_{4}$  \\ 
 & & & FM  & $17.0_{-1.6}^{+1.6}$ & $18.9_{-0.3}^{+0.3}$ & $2.6_{-0.2}^{+0.2}$ & -- & $70_{-10}^{+10}$ &  $204_{-2}^{+2}$ & $G_{4}$    \\ 
 \multirow{2}{*}{$f_{22}$}   & \multirow{2}{*}{291.0679}   & \multirow{2}{*}{$0.62$} & AM  &  $8.9_{-1.5}^{+1.5}$  & $41.2_{-3.5}^{+3.9}$ &  $2.9_{-0.4}^{+0.3}$ &  -- &  $6_{-12}^{+12}$ &  $104_{-3}^{+3}$  &  $G_{4}$  \\ 
 & & & FM    &  $31.4_{-2.0}^{+2.0}$   & $39.2_{-1.4}^{+1.3}$  & $4.3_{-0.2}^{+0.2}$ &  -- &  $-11_{-16}^{+16}$ &  $0.19_{-4.14}^{+4.10}$  &  $G_{4}$  \\
\multirow{2}{*}{$f_{17}$}   & \multirow{2}{*}{291.7364}  &  \multirow{2}{*}{-0.13} & AM  & $4.8_{-0.8}^{+0.8}$   & $33.9_{-2.5}^{+2.9}$  &  $1.14_{-0.35}^{+0.36}$ & -- &  $-8_{-6}^{+6}$ &   $151_{-1}^{+1}$  &  $G_{4}$\\ 
 & & & FM & $20.4_{-1.2}^{+1.3}$ & $26.8_{-0.4}^{+0.4}$ & $2.7_{-0.1}^{+0.1}$ & --   &  $15_{-8}^{+8}$  &  $-5.3_{-1.7}^{+1.7}$ &  $G_{4}$  \\
\multirow{2}{*}{$f_{6}$}   & \multirow{2}{*}{377.4208}   & \multirow{2}{*}{0.49}& AM   &  --   & -- &  -- &  $-18_{-1}^{+1}$ &  $77_{-4}^{+5}$ &  $297_{-1}^{+1}$  &  $G_{2}$   \\ 
 & & & FM  &  $2.7_{-0.4}^{+0.4}$ & $35.7_{-2.2}^{+2.5}$  &  $1.1_{-0.3}^{+0.3}$&  -- &  $-4_{-3}^{+3}$ &  $1.36_{-0.65}^{+0.63}$ &  $G_{4}$   \\ 
 \multirow{2}{*}{$f_{25}$}   & \multirow{2}{*}{378.0836}   & \multirow{2}{*}{$0.26$} & AM  &  $2.5_{-0.3}^{+0.3}$  & $13.0_{-0.2}^{+0.2}$ &  $0.15_{-0.21}^{+0.21}$ &  -- &  $-24_{-3}^{+3}$ &  $96.8_{-0.4}^{+0.4}$  &  $G_{4}$   \\ 
 & & & FM    &  $12.9_{-1.6}^{+1.6}$   & $27.1_{-0.8}^{+0.8}$  & $5.1_{-0.2}^{+0.2}$ &  -- &  $-1_{-10}^{+10}$ &  $1.6_{-2.1}^{+2.1}$  &  $G_{4}$  \\
\multirow{2}{*}{$f_{5}$}   & \multirow{2}{*}{378.7156}  &  \multirow{2}{*}{-0.51} & AM  & $15.5_{-1.6}^{+1.6}$   & $60.1_{-3.0}^{+3.2}$  &  $60_{-3}^{+3}$ & -- &  $13_{-10}^{+10}$ &   $298_{-2}^{+2}$  &  $G_{4}$  \\ 
 & & & FM & $3.8_{-0.3}^{+0.3}$ & $28.8_{-0.6}^{+0.7}$ & $5.6_{-0.1}^{+0.1}$ & --   &  $40_{-3}^{+3}$  &  $-10.0_{-0.5}^{+0.5}$ &  $G_{4}$ \\
\multirow{2}{*}{$f_{33}$}   & \multirow{2}{*}{$570.2229$}  &  \multirow{2}{*}{-0.07} & AM &  --   & --  &  -- & $14_{-1}^{+1}$ & $-59_{-5}^{+5}$ & $79_{-1}^{+1}$ & $G_{2}$ \\ 
 & & & FM  & $9.7_{-1.3}^{+1.3}$  & $31.2_{-1.3}^{+1.4}$ & $5.4_{-0.2}^{+0.2}$ &  -- &  $-40_{-9}^{+8}$ &  $8.0_{-1.9}^{+1.9}$ & $G_{4}$ \\
\multirow{2}{*}{$f_{60}$}   & \multirow{2}{*}{$571.5070$}   & \multirow{2}{*}{-0.48} &AM   &  -- & -- & -- & $-14_{-1}^{+1}$ &  $72_{-4}^{+4}$ &  $45.8_{-0.5}^{+0.5}$  & $G_{2}$   \\ 
 & & & FM  &  -- & -- & -- &  -- &  $-87_{-9}^{+9}$ &  $20.3_{-2.4}^{+2.3}$ &  $G_{1}$    \\
\multirow{2}{*}{$f_{36}$}   & \multirow{2}{*}{$574.0266$}  &  \multirow{2}{*}{0.53} & AM  & $1.85_{-0.25}^{+0.26}$ & $8.8_{-0.1}^{+0.1}$ & $1.65_{-0.26}^{+0.26}$ & $10_{-1}^{+1}$ &  $-60_{-4}^{+4}$ &   $82.7_{-0.5}^{+0.5}$  &  $G_{5}$   \\ 
 & & & FM & $5.3_{-1.1}^{+1.1}$ & $26.8_{-1.6}^{+1.8}$ & $3.9_{-0.4}^{+0.4}$ & --   &  $-34_{-8}^{+8}$  &  $8.2_{-1.5}^{+1.5}$ &  $G_{4}$   \\
\multirow{2}{*}{$f_{38}$}   & \multirow{2}{*}{$707.7771$}  &  \multirow{2}{*}{-0.49} & AM &  $5.2_{-0.7}^{+0.8}$   & $51.7_{-3.7}^{+4.2}$  &  $2.2_{-0.2}^{+0.2}$ & -- & -- & $70.6_{-1.1}^{+1.1}$ & $G_{3}$  \\ 
 & & & FM  & $11.3_{-1.5}^{+1.5}$   & $32.2_{-1.3}^{+1.3}$  &  $-2.9_{-0.2}^{+0.2}$ & -- & $14_{-1}^{+1}$ & $-1.2_{-2.0}^{+2.0}$ & $G_{4}$  \\
\multirow{2}{*}{$f_{67}$}   & \multirow{2}{*}{$710.3810$}  &  \multirow{2}{*}{0.20} & AM &  --   & -- &  -- & -- & $1_{-1}^{+1}$ & $45.6_{-0.4}^{+0.6}$ & $G_{1}$ \\ 
 & & & FM  & $17_{-2}^{+2}$   & $29.8_{-0.1.0}^{+0.9}$  &  $-3.4_{-0.2}^{+0.2}$ & -- & $78_{-14}^{+14}$ & $-13.8_{-2.7}^{+2.7}$ & $G_{4}$  \\ 
\hline
\multicolumn{11}{c}{Other AFMs}  \\
\hline  
\multirow{2}{*}{$f_{20}$}  & \multirow{2}{*}{$144.7211$} &  \multirow{2}{*}{$0.3$} & AM  &  $6_{-1}^{+1}$  &$15.3_{-0.4}^{+0.5}$  &  $5.5_{-0.3}^{+0.3}$ &  -- &  $20_{-10}^{+10}$ &  $110_{-2}^{+2}$ &  $G_{4}$   \\ 
& & & FM & $52^{+3}_{-3} $  & $27.5^{+0.4}_{-0.4}$  &  $2.4^{+0.1}_{-0.1}$ &  -- &  $230^{+20}_{-20}$ &  $-52^{+4}_{-4}$ &  $G_{4}$   \\
\multirow{2}{*}{$f_{23}$} & \multirow{2}{*}{$176.0526$} & \multirow{2}{*}{$0.72$} & AM & -- & -- & -- & -- & $-205_{-8}^{+8}$ &  $158.7_{-1.3}^{+1.4}$ & $G_{1}$  \\ 
 & & & FM & $33.0_{-2.4}^{+2.3}$ & $19.9_{-0.4}^{+0.5}$  &  $-2.7_{-0.12}^{+0.12}$ &  -- & $-290_{-20}^{+20}$ & $-48.7_{-2.9}^{+3.0}$ & $G_{4}$  \\ 
\multirow{2}{*}{$f_{11}$} & \multirow{2}{*}{$183.1638$} & \multirow{2}{*}{-0.13} & AM & $10_{-1}^{+1}$   & $17.9_{-0.3}^{+0.3}$  &  $3.7_{-0.2}^{+0.2}$ &  $-46_{-3}^{+3}$ &  $270_{-10}^{+10}$ &  $181_{-2}^{+2}$ & $G_{5}$ \\        
 & & & FM & $28_{-1}^{+1}$   & $17.3_{-0.1}^{+0.1}$  &  $6.1_{-0.1}^{+0.1}$ &  $-74_{-3}^{+4}$ &  $340_{-20}^{+20}$ & $-27_{-2}^{+2}$ &  $G_{5}$   \\ 
\multirow{2}{*}{$f_{1}$}   & \multirow{2}{*}{191.4365}   & \multirow{2}{*}{0.12} & AM &  $8.2_{-1.2}^{+1.3}$   & $13.9_{-0.3}^{+0.3}$  &  $1.7_{-0.3}^{+0.3}$ &  $68_{-4}^{+4}$ &  $-460_{-20}^{+20}$ &  $1577_{-2}^{+2}$   &  $G_{4}$   \\ 
 & & & FM & -- & --  & -- &  -- &  -- &  --  &  --  \\ 
\multirow{2}{*}{$f_{13}$}   & \multirow{2}{*}{201.6427}   & \multirow{2}{*}{0.62} & AM &  --   & --  &  -- &  -- &  -- &  --  & -- \\ 
 & & & FM &  $51_{-2}^{+2}$   & $39_{-1}^{+1}$  &  $5.1_{-0.1}^{+0.1}$ &  -- &  $38_{-17}^{+17}$ &  $-4.9_{-4.7}^{+4.5}$  &  $G_{4}$ \\ 
\multirow{2}{*}{$f_{2}$}   & \multirow{2}{*}{211.0748}   & \multirow{2}{*}{-0.03} & AM &  --   & --  &  -- &  -- &  $-31_{-3}^{+3}$ &  $476_{-1}^{+1}$  &  $G_{1}$  \\ 
 & & & FM & $2.9_{-0.5}^{+0.5}$   & $30.1_{-1.6}^{+1.9}$  &  $5.1_{-0.3}^{+0.3}$ &  -- &  $-1.2_{-3.2}^{+3.2}$ &  $0.6_{-0.7}^{+0.7}$  &  $G_{4}$  \\ 
\multirow{2}{*}{$f_{19}$}   & \multirow{2}{*}{212.3933}  &  \multirow{2}{*}{$0.53$} & AM  &  $5.5_{-0.9}^{+0.9}$   & $23.8_{-0.8}^{+0.9}$  &  $6.1_{-0.3}^{+0.3}$ &  -- &  $44_{-6}^{+6}$ &  $124_{-1}^{+1}$  &  $G_{4}$  \\ 
 &  & & FM & $19.2_{-2.5}^{+2.5} $  & $39_{-3}^{+3}$  &  $5.2_{-0.3}^{+0.3}$ &  -- &  $25_{-19.9}^{+18.8}$ &  $-1.4_{-4.9}^{+5.1}$  &   $G_{4}$    \\ 
\multirow{2}{*}{$f_{18}$}   & \multirow{2}{*}{216.2493}   & \multirow{2}{*}{0.75} & AM   &  --   & --  &  -- &  $-100_{-3}^{+3}$ &  $300_{-10}^{+10}$ &  $163_{-1}^{+1}$  &  $G_{2}$  \\ 
 & & & FM & --   & -- &  -- & $-120_{-6}^{+6}$ &  $510_{-30}^{+30}$ &  $-31_{-3}^{+3}$  &  $G_{2}$  \\ 
\multirow{2}{*}{$f_{46}$}   & \multirow{2}{*}{232.1455}   & \multirow{2}{*}{0.45} & AM   &  --   & --  &  -- &  -- &  $167_{-14}^{+14}$ &  $91.7_{-1.8}^{+1.7}$  &  $G_{1}$  \\ 
 & & & FM & $21_{-2}^{+2}$   & $9.9_{-0.2}^{+0.2}$ &  $0.45_{-0.18}^{+0.19}$ & -- &  $120_{-30}^{+30}$ &  $-13_{-3}^{+3}$  &  $G_{4}$ \\ 
\multirow{2}{*}{$f_{21}$}   & \multirow{2}{*}{238.3032}   &  \multirow{2}{*}{$0.06$}  & AM & --   & --  &  -- &  -- &  $15_{-4}^{+3}$ &  $104_{-1}^{+1}$   &  $G_{1}$  \\ 
 &  & & FM & $38_{-2}^{+2}$   & $30.4_{-0.5}^{+0.6}$  & $1.1_{-0.1}^{+0.1}$ &  -- &  -- & --   &  $G_{3}$  \\

\hline
\end{tabular}
\end{center}
% \tablefoot{B and I denote the beating effect and intrinsic modulations of amplitude and frequency, respectively.}
\end{table*}

\addtocounter{table}{-1} 
\begin{table*} \caption[]{continued.}
\renewcommand{\arraystretch}{1.31}%设置行高
\small
\begin{center}
\begin{tabular}{lcrcrrrrrrrc}
\hline
\hline
 \multicolumn{1}{c}{ID} &  \multicolumn{1}{c}{Fre} &  \multicolumn{1}{c}{Corr} &  \multicolumn{1}{c}{AM/FM} &  \multicolumn{1}{c}{A} &  \multicolumn{1}{c}{T=2 $\pi$/$\omega$} &  \multicolumn{1}{c}{$\phi$} &  \multicolumn{1}{c}{a} &  \multicolumn{1}{c}{b} &  \multicolumn{1}{c}{c} &  \multicolumn{1}{c}{Fitting}  \\
 \multicolumn{1}{c}{} &  \multicolumn{1}{c}{($\mu$Hz)} &  \multicolumn{1}{c}{} &  \multicolumn{1}{c}{} &  \multicolumn{1}{c}{(ppm/nHz)} &  \multicolumn{1}{c}{(d)} &  \multicolumn{1}{c}{[0, 2$\pi$)} &  \multicolumn{1}{c}{(10$^{-3}$)} &  \multicolumn{1}{c}{(10$^{-2}$)} &  \multicolumn{1}{c}{} &  \multicolumn{1}{c}{} \\               
\hline

\multirow{2}{*}{$f_{4}$}  &  \multirow{2}{*}{254.6373}  & \multirow{2}{*}{$0.12$} & AM  & --  & --  &  -- &  -- &  $-2_{-3}^{+4}$ &  $404_{-1}^{+1}$  &  $G_{1}$   \\ 
& & & FM & -- & -- & -- &  -- &  $0.6_{-2.4}^{+2.4}$ &  $-0.17_{-0.64}^{+0.67}$   &   $G_{1}$  \\ 
\multirow{2}{*}{$f_{28}$}   & \multirow{2}{*}{255.2858}  & \multirow{2}{*}{$0.33$} & AM  &  $4.4_{-0.5}^{+0.6}$   & $20.1_{-0.6}^{+0.6}$  &  $1.87_{-0.23}^{+0.23}$ &  -- &  $-16_{-5}^{+5}$ &  $93.6_{-0.8}^{+0.8}$   &  $G_{4}$ \\ 
 & & & FM  &  $13.1_{-2.6}^{+2.5}$   & $16.2_{-0.5}^{+0.5}$  &  $4.1_{-0.3}^{+0.3}$ &  -- &  $-41_{-14}^{+15}$ &  $9.7_{-3.4}^{+3.4}$  &  $G_{4}$   \\
\multirow{2}{*}{$f_{10}$}   & \multirow{2}{*}{258.4617}   & \multirow{2}{*}{$-0.53$}& AM   &  --   & -- &  -- &  -- &  $12_{-3}^{+3}$ &  $206_{-1}^{+1}$   &  $G_{1}$  \\ 
 & & & FM  &  $7.3_{-1.2}^{+1.2}$ & $37.2_{-3.1}^{+3.7}$  &  $5.8_{-0.4}^{+0.4}$&  -- &  $-10_{-10}^{+10}$ &  $1.9_{-2.4}^{+2.4}$ &  $G_{4}$   \\ 

\multirow{2}{*}{$f_{16}$}   & \multirow{2}{*}{305.5128}   & \multirow{2}{*}{0.4} & AM   &  --   & --  &  -- &  -- &  $62.9_{-2.8}^{+2.7}$ &  $140_{-1}^{+1}$   &  $G_{1}$ \\ 
 & & & FM  & --   & --  &  -- &  $39_{-3}^{+4}$ &  $-140_{-20}^{+20}$ &  $4.4_{-2.0}^{+2.0}$  &  $G_{2}$   \\ 
\multirow{2}{*}{$f_{12}$}   &  \multirow{2}{*}{314.0823}  &  \multirow{2}{*}{$-0.05$} & AM  &  $4.3_{-0.4}^{+0.4}$   & $23.0_{-0.5}^{+0.6}$  &  $4.2_{-0.2}^{+0.2}$ &  -- &  $14_{-4}^{+4}$ &  $181_{-1}^{+1}$   &  $G_{4}$  \\
 & & & FM & -- & -- & -- & $-25_{-3}^{+3}$ &  $140_{-10}^{+10}$ &  $-13.7_{-1.5}^{+1.5}$   & $G_{2}$  \\ 
 \multirow{2}{*}{$f_{14}$}   & \multirow{2}{*}{315.3790}  & \multirow{2}{*}{$0.95$} & AM  &  $24.1_{-2.5}^{+2.7}$   & $59.8_{-2.9}^{+3.0}$  &  $2.67_{-0.13}^{+0.13}$ &  -- &  $25_{-13}^{+12}$ &  $170.5_{-3.9}^{+4.1}$   &  $G_{4}$   \\
 & & & FM & $20.4_{-1.5}^{+1.6}$ & $36.6_{-2.1}^{+2.2}$ & $0.86_{-0.2}^{+0.2}$ & -- &  $170_{-20}^{+20}$ &  $-38.7_{-3.8}^{+3.9}$   & $G_{4}$  \\
 
\multirow{2}{*}{$f_{31}$}   & \multirow{2}{*}{$316.7981$}  &  \multirow{2}{*}{-0.28} & AM &  --   & --  &  -- &  -- & $-15_{-2}^{+2}$ & $87.0_{-0.5}^{+0.5}$ & $G_{1}$ \\ 
 & & & FM  & $6.4_{-2.4}^{+2.3}$  & $17_{-1}^{+1}$ & $0.05_{-0.51}^{+0.53}$ &  -- &  $26_{-13}^{+13}$ &  $-6.5_{-3.0}^{+3.0}$ & $G_{4}$  \\
\multirow{2}{*}{$f_{15}$}   & \multirow{2}{*}{343.9562}   & \multirow{2}{*}{-0.46} &AM   &  $8.9_{-0.6}^{+0.6}$ & $18.7_{-0.2}^{+0.2}$ & $0.52_{-0.12}^{+0.12}$ &  -- &  $-24_{-4}^{+4}$ &  $178_{-1}^{+1}$  & $G_{4}$\\ 
 & & & FM  &  $12.7_{-1.0}^{+1.0}$   & $21.2_{-0.3}^{+0.4}$  &  $1.7_{-0.2}^{+0.2}$ &  -- &  $-44_{-6}^{+6}$ &  $9.4_{-1.4}^{+1.4}$ &  $G_{4}$ \\
\multirow{2}{*}{$f_{39}$}   & \multirow{2}{*}{390.5333}  &  \multirow{2}{*}{-0.06} & AM  & $1.8_{-0.3}^{+0.3}$  & $18.4_{-0.7}^{+0.8}$  & $2.2_{-0.4}^{+0.4}$  & -- &  $-6_{-4}^{+4}$ &   $66.1_{-0.6}^{+0.6}$  &  $G_{4}$  \\ 
 & & & FM & $ 9.8_{-1.3}^{+1.3}$ & $ 33.7_{-1.6}^{+1.8}$ & $0.46_{-0.24}^{+0.25}$ & --   &  $-90_{-10}^{+10}$  &  $ 18.7_{-1.8}^{+1.8}$ &  $G_{4}$   \\
\multirow{2}{*}{$f_{41}$}   & \multirow{2}{*}{414.4996}   & \multirow{2}{*}{-0.82} & AM   & $6.5_{-0.3}^{+0.3}$  & $24.5_{-0.3}^{+0.3}$  & $5.2_{-0.1}^{+0.1}$ &  -- &  $-9_{-3}^{+3}$ &  $67.7_{-0.4}^{+0.4}$   &  $G_{4}$ \\ 
 & & & FM  & $40_{-2}^{+2}$  & $26.2_{-0.3}^{+0.3}$ & $5.6_{-0.1}^{+0.1}$ & -- &  $10_{-12}^{+12}$ &  $1.0_{-2.7}^{+2.6}$  &  $G_{4}$  \\ 
\multirow{2}{*}{$f_{24}$}   &  \multirow{2}{*}{450.8051}  &  \multirow{2}{*}{$-0.88$} & AM  &  $3.8_{-0.3}^{+0.3}$   & $20.3_{-0.3}^{+0.3}$  &  $3.80_{-0.13}^{+0.14}$ & $55_{-1}^{+1}$ &  $-292_{-4}^{+4}$ &  $118_{-1}^{+1}$   &  $G_{5}$ \\
 & & & FM & $14.0_{-1.3}^{+1.3}$ & $18.0_{-0.3}^{+0.3}$ & $0.20_{-0.15}^{+0.15}$ & $-112_{-4}^{+4}$ &  $560_{-20}^{+20}$ &  $-50_{-2}^{+2}$   & $G_{5}$  \\ 

\multirow{2}{*}{$f_{65}$}   & \multirow{2}{*}{$607.7826$}   & \multirow{2}{*}{0.74} & AM   & -- & -- & -- & $-19_{-1}^{+1}$ &  $90_{-4}^{+4}$ &  $42.4_{-0.6}^{+0.6}$   &  $G_{2}$   \\ 
 & & & FM  & -- & -- & -- & $-86_{-6}^{+6}$ &  $450_{-26}^{+26}$ &  $-42_{-3}^{+3}$  &  $G_{2}$   \\ 
\multirow{2}{*}{$f_{73}$}   & \multirow{2}{*}{$617.1965$}   & \multirow{2}{*}{-0.42} & AM   & $1.5_{-0.3}^{+0.3}$ & $29.3_{-2.3}^{+2.5}$ & $-0.2_{-0.3}^{+0.3}$ & -- &  $-13_{-5}^{+5}$ &  $45.1_{-0.4}^{+0.4}$   &  $G_{4}$   \\ 
 & & & FM  & $9.3_{-1.8}^{+01.8}$ & $20.3_{-1.2}^{+1.4}$ & $-0.98_{-0.34}^{+0.36}$ & -- &  $130_{-20}^{+20}$ &  $-19.5_{-2.5}^{+2.5}$   &  $G_{4}$   \\ 
\multirow{2}{*}{$f_{50}$}   &  \multirow{2}{*}{$653.5360$}  &  \multirow{2}{*}{$-0.32$} & AM  &  $1.6^{+0.4}_{-0.4}$ & $11.8_{-0.16}^{+0.29}$ & $0.75_{-0.43}^{+0.45}$ & -- &  $11_{-3}^{+3}$ &  $56.2_{-0.6}^{+0.6}$ & $G_{4}$   \\
 & & & FM & -- & -- & -- & -- &  $-18_{-7}^{+7}$ &  $4.2_{-1.9}^{+2.0}$   & $G_{1}$   \\ 
\multirow{2}{*}{$f_{68}$}   & \multirow{2}{*}{$732.8429$}  &  \multirow{2}{*}{0.30} & AM &  $1.2_{-0.3}^{+0.3}$   & $29.3_{-2.3}^{+2.9}$  &  $-1.8_{-0.6}^{+0.6}$ & -- & $12_{-3}^{+3}$ & $42.6_{-0.5}^{+0.5}$ & $G_{4}$  \\ 
 & & & FM  & --  & --  &  -- & $75_{-10}^{+9}$ & $-390_{-50}^{+50}$ & $40_{-7}^{+7}$ & $G_{2}$ \\ 
\multirow{2}{*}{$f_{85}$}   & \multirow{2}{*}{$3269.8987$}  &  \multirow{2}{*}{-0.79} & AM &  --  &-- & -- & -- &  $-45_{-1}^{+1}$ & $45.8_{-0.2}^{+0.2}$ & $G_{1}$  \\ 
 & & & FM  & $6.2_{-1.5}^{+1.4}$   & $9.5_{-0.4}^{+0.4}$  &  $-0.2_{-0.4}^{+0.4}$ & -- & $110_{-20}^{+20}$ & $-17_{-2}^{+2}$ & $G_{4}$ \\
\multirow{2}{*}{$f_{57}$}   & \multirow{2}{*}{$3306.9341$}   & \multirow{2}{*}{-0.18} &AM   & $4.1_{-0.2}^{+0.2}$ & $44.6_{-1.4}^{+1.5}$ & $4.97_{-0.12}^{+0.12}$ & -- &  $-43_{-2}^{+2}$ &  $63.2_{-0.4}^{+0.4}$  & $G_{4}$  \\ 
 & & & FM  & $21.3_{-10.6}^{+16.1}$ & $69.0_{-19.7}^{+33.9}$ & $0.35_{-0.92}^{+0.75}$ &  -- &  $120_{-50}^{+70}$ &  $-34.5_{-21.1}^{+16.6}$ &  $G_{4}$  \\
\multirow{2}{*}{$f_{90}$}   & \multirow{2}{*}{$4318.7053$}   & \multirow{2}{*}{0.49} &AM   & -- & -- & -- & $58_{-9}^{+9}$ &  $250_{-15}^{+16}$ &  $30.9_{-0.9}^{+0.9}$  & $G_{2}$  \\ 
 & & & FM  & -- & -- & -- &  $539_{-43}^{+44}$ &  $1150_{-80}^{+80}$ &  $-43.0_{-4.6}^{+4.6}$ &  $G_{2}$ \\

\\

\hline 
\multicolumn{11}{c}{Combination frequencies}  \\
\hline 
\multirow{2}{*}{$f_{48}$}   & \multirow{2}{*}{$346.4111$}   &  \multirow{2}{*}{-0.32} & AM & --  & -- & -- &  -- &  $25_{-2}^{+2}$ &  $54_{-1}^{+1}$  &   $G_{1}$   \\ 
& & & FM &  $25_{-2}^{+2}$  & $10.5_{-0.1}^{+0.1}$  &  $1.5_{-0.2}^{+0.2}$ &  -- &  $-71_{-13}^{+13}$ &  $15.7_{-2.9}^{+2.9}$   &  $G_{4}$ \\ 
\multirow{2}{*}{$f_{89}$}   & \multirow{2}{*}{$4320.7365$}   & \multirow{2}{*}{-0.1} &AM   & -- & -- & -- & $54_{-3}^{+3}$ &  $178_{-7}^{+8}$ &  $30.2_{-0.7}^{+0.7}$  & $G_{2}$  \\ 
 & & & FM  & $24.1_{-3.3}^{+3.3}$ & $17.8_{-0.7}^{+0.8}$ & $-1.5_{-0.2}^{+0.2}$ &  -- &  $-400_{-30}^{+30}$ &  $-61.3_{-4.7}^{+4.6}$ &  $G_{4}$ \\
\hline 
\multicolumn{11}{c}{Orbitial information}  \\
\hline 
\multirow{2}{*}{$f_{3}$}  & \multirow{2}{*}{$20.3107$}  & \multirow{2}{*}{$-0.58$} & AM  &  $19_{-7}^{+9}$  & $72_{-15}^{+26}$  &  $3.5_{-0.6}^{+0.5}$ &  -- & -- & $ 432_{-12}^{+13}$ &  $G_{3}$   \\ 
& & &  FM & $6.7_{-0.5}^{+0.5}$   & $35_{-1}^{+1}$  &  $3.98_{-0.16}^{+0.15}$ & -- &  -- &  $2.9_{-0.7}^{+0.7}$ &  $G_{3}$   \\
\multirow{2}{*}{$f_{8}$}  & \multirow{2}{*}
{$40.6194$}  & \multirow{2}{*}{$-0.21$} & AM  &  $4.9_{-0.6}^{+0.7}$  & $18.1_{-0.5}^{+0.5}$  &  $5.8_{-0.3}^{+0.3}$ & $32_{-3}^{+3}$ &  $186_{-10}^{+10}$ &  $258.3_{-1.3}^{+1.3}$ &  $G_{5}$   \\
& & &  FM &   --   &  -- &  -- &  $19_{-3}^{+3}$&  $80_{-10}^{+11}$ &  $5.2_{-1.4}^{+1.4}$ &  $G_{2}$   \\
\hline
\end{tabular}
\end{center}
\tablefoot{Columns 1 and 2 give the ID and frequency taken from Table\,\ref{Tab:freq}; column 3 gives the value of correlation between the amplitude and frequency modulations; column 4 shows the label of amplitude modulation (AM) or frequency modulation (FM); columns 5, 6, and 7 give the fitting coefficients of amplitude (A), period ($T=2\pi/\omega$) and phase ($\phi$) of sinusoidal function if periodic patterns were identified; columns 8, 9, and 10 give the coefficients of polynomial fitting up to the second order; column 11 gives the matrix index of fitting function to the modulation pattern.
% B and I denote the beating effect and intrinsic modulations of amplitude and frequency, respectively. 
}
\end{table*}

\FloatBarrier

Figure \ref{fig:Mul_377} displays the AMs and FMs for the three components that form the {\sl g}-mode triplet near $377\,\mu$Hz. A close view of the profile of the triplet shows that the three components are not strictly systematically spaced in frequency. We measured a significant frequency mismatch of $\Delta f \sim 0.031\,\mu$Hz that is one order higher than the frequency uncertainty.
The middle-left panel illustrates the dynamic patterns of frequency and amplitude over time through the sLSP diagram in a general way. A clear FM has been observed in the central component. Thus, we provide an expanded view centered on the average frequency of each component in the bottom-left panels, while their corresponding amplitude behaviors are presented in the right panels. With the current measuring precision in amplitude and frequency, we disclose the various modulating patterns in AMs and FMs of these three components. The optimal fitting and the associated \texttt{MCMC} curves suggest that the AMs and FMs follow simple patterns. During the $\sim80$~day observations, the central component is found to exhibit modulation periods of $\sim27$\,days and 13\,days for FM and AM, respectively, whereas the AM values show a slightly decreasing trend. Both the two side components exhibit FM with a similar period of $\sim30$~d and vary in a small scale of a few nano Hz. We have observed some clear (anti-) correlation between the three resonant components. For instance, the two side components evolve somewhat in anti-phase both in amplitude and frequency with a similar timescale.

Figure\,\ref{fig:Mul_167} shows the behavior of AMs and FMs occurring in the two side components forming the triplet near 167\,$\mu$Hz. From the LSP, a very large asymmetry is found in the frequency spacing of the three components, with a frequency mismatch of 0.105~$\mu$Hz. This could be related to the large FM of the central component (with a zigzag pattern), as revealed from the sliding LSP. However, it failed to explore piece-wise measurement of frequency for the central component due to its low amplitude. The retrograde component exhibits relatively stable frequency and amplitude from the sLPS. With piece-wise measurement, FM varies in a scale of a few nano Hz and AM exhibits a slightly increasing pattern from $\sim270$ to 320 ppm. We note that the frequency uncertainty is comparable to the FM scale. 
The prograde component shows periodic patterns of FM and AM as indicated by the fitting models. The period of FM is comparable in both side components, $\sim 17$~days, if we ignore the slight linear trends.

\subsection{Other AFMs}\label{sec:Other_AFM}
Apart from the above AMs and FMs, the others are presented in Fig.\,\ref{fig:AFM} and Fig.\,\ref{fig:AF_L}. Overall, these modulations reveal rich and various patterns in {PG~0101+039}. 
Most of the observed modulations exhibit quasi-sinusoidal patterns with additional linear or parabolic fittings, namely, by fitting with $G_4$ or $G_5$. For instance, the AM and FM of $f_{24}\sim450.8~\mu$Hz have evolved in anti-phase and exhibit a sinusoidal plus a parabolic pattern of $G_5$. Its amplitude decreases from $\sim115$~ppm to the local minimum of $\sim 75$~ppm and rises afterward; correspondingly, the frequency varies from -50\,nHz up to +50\,nHz and returns back to -50\,nHz relative to its average. Both the AM and FM of $f_{15}\sim344.0~\mu$Hz exhibit a sinusoidal pattern with an additional linear fitting. There are a few frequencies that merely demonstrate somewhat sinusoidal patterns such as the FM of $f_{21}\sim238.3~\mu$Hz and the AM of $f_{38}\sim707.8~\mu$Hz. Some completely exhibit simple parabolic patterns, for instance, $f_{18}\sim216.3~\mu$Hz whose AM and FM change in a scale of 100~ppm and 100~nano~Hz, respectively. 
Several frequencies can be characterized by simple linear fitting of a decreasing or increasing amplitude and frequency. For instance, the frequency of $f_{59}\sim571.5~\mu$Hz decreases nearly 70\,nHz over an interval of $\sim50$~days.
We note that several modes exhibit visible amplitudes just over a fraction of the period during the observation such as $f_{90}\sim4318.7~\mu$Hz and $f_{89}\sim4320.7~\mu$Hz and there are also several modes whose amplitudes decrease down to detect threshold and vanish in the noise; for instance, $f_{85}\sim3269.9~\mu$Hz.

There are only a few stable AMs and FMs, characterized for example, by a frequency and amplitude of $f_4\sim254.6~\mu$Hz, or amplitudes of $f_{21}\sim238.3~\mu$Hz and $f_{10}\sim258.5~\mu$Hz modulating around 10~ppm. The majority of our discovered FMs are characterized within the range of $20-100$~nHz. One exceptional case is that $f_{20}\sim144.7~\mu$Hz exhibits a large FM up to 200~nHz. We find a general correlation that the AMs and FMs are proportional to each other, that is: a mode that modulates on a large scale both in amplitude and frequency. Most of the concerning frequencies are on a modulating timescale of $\sim30$~days and can be well resolved by the light curves. Interestingly, we observe several AMs with a relatively short period on a timescale down to $\sim 10$\,days, for instance: AMs of $f_{5}\sim378.1~\mu$Hz, $f_{36}\sim574.0~\mu$Hz, and $f_{50}\sim653.5~\mu$Hz. However, these quick patterns of AMs are very shallow (i.e., on the order of a few ppm, comparable to the amplitude uncertainties). 

A notable feature of the (anti-) correlation is found between AM and FM among several frequencies. As listed in Table~\ref{Tab:AFM_D}, the strongest correlation is found for $f_{14}\sim315.4~\mu$Hz with a value up to 0.95. The strong anti-correlations are derived for the frequencies of $f_{41}\sim 414.5~\mu$Hz, $f_{24}\sim 450.8~\mu$Hz, and $f_{85}\sim 3269.9~\mu$Hz, with values of -0.82, -0.88, and -0.79, respectively. Several other frequencies also show strong correlation of $|\rho_\mathrm{am,fm}|>0.7$, for instance, $f_{23}\sim 176.1~\mu$Hz, $f_{18}\sim 216.3~\mu$Hz, $f_{12}\sim 314.1~\mu$Hz, and $f_{65}\sim 607.8~\mu$Hz. Additionally, we find that the FM of $f_{1}\sim191.4~\mu$Hz and the AM of $f_{13}\sim201.6~\mu$Hz exhibit quasi-regular behavior. However, their patterns cannot be fitted by our simple $G_k$ functions. As a result, we reserve not to perform fitting or apply \texttt{EMCEE} on these relatively complicated AMs and FMs. Finally, we note that most of these AMs and FMs change in scales in a similar order like a few tens of ppm and micro Hz, respectively.

 \section{Discussion}\label{sec:DisCon}

In the previous sections, we describe our analysis of the {\sl K}2 photometry of {PG~0101+039} and our detections of rich pulsation signals for the first time, which offer an opportunity to investigate the properties of those pulsation modes from linear to nonlinear way. With splitting frequencies, we derived a rotation period of $\sim8.81\pm0.06$ and $8.60\pm0.16$~days by the {\sl g}-modes and {\sl p}-modes, respectively, { implying a marginally differential rotation with a probability of $\sim 60\%$}. Thus, this suggests that {PG~0101+039} is now clearly an unsynchronized system with the orbital period of $0.570$~days \citep{2008A&A...477L..13G}. In this section, we discuss the rotational and orbital properties, as well as the dynamics of amplitude and frequency of oscillation modes.

\subsection{Binary information}\label{sec:Inclination}

As reported in the literature \citep{1999MNRAS.304..535M,2005ApJ...633..460R,2008A&A...477L..13G}, {PG~0101+039} has been considered a highly probable synchronized system due to its close orbit of about half a day, requiring the tidal force to be strong enough for the synchronization process to begin. Recently, \citet{
2023A&A...673A..90S} analyzed {\sl K}2 data collected on {PG~0101+039} and concluded that the orbital light variation is dominated by the beaming effect and accompanied by tiny ellipsoidal deformation. These authors derived a very high orbital inclination that is close to $89.4^\circ\pm0.06^\circ$, which, in turn, suggests that the companion is probably a helium-core white dwarf with a mass of $0.34\pm0.04\,M_{\odot}$. In addition, they derived the rotation period of $0.85\pm0.09$~days for {PG~0101+039} based on $v \sin i$ from spectroscopy, which is slightly slower than its orbital period. This is the first claim that {PG~0101+039} should be in a non-synchronized system.

From our thorough analysis of the pulsation signals, we firmly derived a rotation period of $\sim9$~days based on frequency splitting. This rotation rate is about $\times10$ slower than the preliminary result by \citet{2023A&A...673A..90S}, which means that the rotational velocity measured by \citet{2008A&A...477L..13G} is overestimated heavily. Interestingly, we find that the rotation period derived from {\sl p}-modes, $8.60\pm0.16$~days, is slightly faster than that coming from {\sl g}-modes, namely, $8.81\pm0.06$~days, although this differential is marginally and comes with a probability of $\sim 60\%$. In general, {\sl p}-modes propagate outer parts than {\sl g}-modes, which means that PG~0101+039 has a radial differential rotation with a relatively faster envelope. From the {\sl Kepler/K}2 photometry, two sdB pulsators have been reported to have radial differential rotations via {\sl p}- and {\sl g}-multiplets, KIC~3527751 \citep{2015ApJ...805...94F} and EPIC~220422705 \citep{2022ApJ...933..211M}, even the former were challenged by \citet{2018ApJ...853...98Z}. 
We stress that {PG~0101+039} is among them with the fastest rotation and orbit. If we consider the tidal effect from the companion, it would be a natural case that the envelope rotates faster than the inner part. The tidal force introduces internal gravity modes, which are located between the radiative envelope and the convective core, then it transports angular momentum to the out layer, leading to the envelope becoming first synchronized and then gradually
proceeding to the inner part \citep{1989ApJ...342.1079G}. The radial differential sdB pulsators are on the way to being synchronized and could be used for the stellar chronology of binary systems. Therefore, we paid attention to the binary information of {PG~0101+039} from both photometry and spectroscopy.

\begin{figure}
\centering
\includegraphics[width=0.49\textwidth]{ 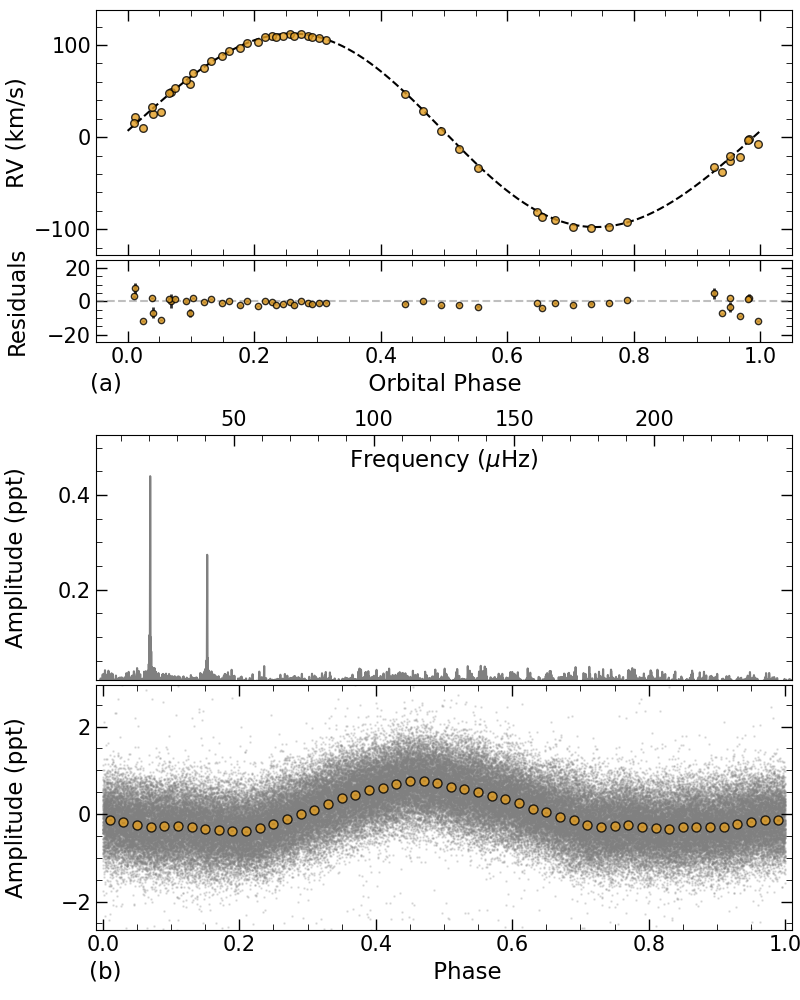}
\caption{Orbital signals of radial velocity and photometry. (a) Phase-folded radial velocities with a Keplerian orbit fitting for the binary system and residuals. Uncertainties are included but are smaller than the symbols themselves for most measurements. (b) LSP of the orbital signals with the removal of oscillation signals (top panel) and the corresponding phase-folded light curve (bottom panel). Note: the {\sl K}2 photometry is averaged into 50 measurements by their phase (filled cycles).}
\label{fig:rv-phase}
\end{figure}

For {PG~0101+039}, we collected 50 high-quality spectra with S/N~>~10 from the LAMOST\footnote{The Large Sky Area Multi-object Fiber Spectroscopic Telescope is located at the Xinglong Observatory, China. The diameter of its field of view is 5$^\circ$, and it is equipped with 4000 fibers at the focus.} DR8. They were observed under the LK-MRS projects during 2018-2019 \citep[see details in][]{2020ApJS..251...15Z} and exhibit a medium resolution of $R\sim7500$, covering the wavelength of 495-535~nm and 630-680~nm in two bands \citep{2022Innov...300224Y}. As we only analyzed the binary signals, we only considered the radial velocity (RV) of {PG~0101+039} derived from these spectra. We note that all spectra are found to be single-line dominated as a consequence of the high contrast of light fraction between the sdB and the companion. Among several pipelines to measure the RVs of LAMOST spectra, we chose the RVs from \citet{2021ApJS..256...14Z}, who adopted the cross-correlation function method. In addition, they have proposed a robust method for a self-consistent determination that aims to correct the systematic zero-point RV offsets, combined with the RVs from the Gaia DR2 \citep{2019A&A...622A.205K}. We list the information of those RVs in Table~\ref{AP:Spectrum}, with the attributes of observation time, RVs, and associated errors, as well as the S/N values. We first adopted the method of phase dispersion minimization (PDM) to derive the period of the RV curves. Then we used the Keplerian orbit to fit the radial velocities around the PDM period $\sim0.5699$\,d (see Fig.\,\ref{fig:rv-phase}, a). The fitting result of the orbital solution is obtained through the \texttt{EMCEE} code \citep[see,][for a review]{2013PASP..125..306F} and their parameters are provided in Table\,\ref{table:rv}. We then derived the mass limit, $M_2 \geq 0.36$\,M$_\odot$, for the companion based on the mass function and assumed the sdB star with a canonical mass of $M_1 = 0.47$\,M$_\odot$.
Our results are overall consistent with the orbital parameters derived from previous literature \citep{1999MNRAS.304..535M,2008A&A...477L..13G,2023A&A...673A..90S}, as compared in Table\,\ref{table:rv}

\begin{table*}
\renewcommand{\arraystretch}{1.2}%设置行高
\caption{Orbital parameters derived by the \texttt{EMCEE} code v.s. the results from the literature.} 
\label{table:rv}     
\centering         
\begin{tabular}{ccccccccc}     
\hline
\hline
$P$  & $\gamma$ & $K$  & $\omega$  & $e$ & $a$ &  $i$  & $M_\mathrm{2,min}$  &  Reference  \\ 
(d) & (km\,s$^{-1}$) & (km\,s$^{-1}$) & ($^{\circ}$)  &  & ($\mathrm{R_{\odot}}$) &  ($^{\circ}$)  & ($\mathrm{M_{\odot}}$)  &  \\
\hline
$0.5699079\pm0.0000068$ & $8.47\pm0.34$ & $104.29\pm0.47$ & -- & -- & -- & -- & $0.37\pm0.02$ & [1] \\
$0.569899\pm0.000001$ & $7.3\pm0.2$ & $104.5\pm0.3$ & -- & -- & $3.1 \pm 0.4$ & $40\pm 6$& $0.45$ & [2] \\
-- & -- & -- & -- & -- & $2.53 \pm 0.01$ & $89.4\pm 0.6$& $0.33$ & [3] \\
$0.5698990^{+0.0000003}_{-0.0000004}$ &  $6.9^{+\,0.2}_{-\,0.3} $ &  $105.0^{+\,0.2}_{-\,0.4}$ & $82.4^{+\,3.8}_{-\,8.5}$ & $0.032^{+\,0.002}_{-\,0.004}$ &  --  &  -- & $0.36$ & This work \\
\hline
\hline
\end{tabular}
\tablefoot{The parameters $P$, $\gamma$, $K$, $\omega$, $e$, $a,$ and $i$ denote the orbital period, system velocity, the semi-amplitude for the primary star, the angle of the pericenter, the orbital eccentricity, the orbital semi-major axis, and the orbital inclination of the binary system, respectively.
References: (1) \cite{1999MNRAS.304..535M}; (2) \cite{2008A&A...477L..13G}; (3) \cite{2023A&A...673A..90S}} %  (2) \cite{2015A&A...576A..44K};
\end{table*}

\begin{figure}
    \centering
    \includegraphics[width=0.259\textwidth]{ 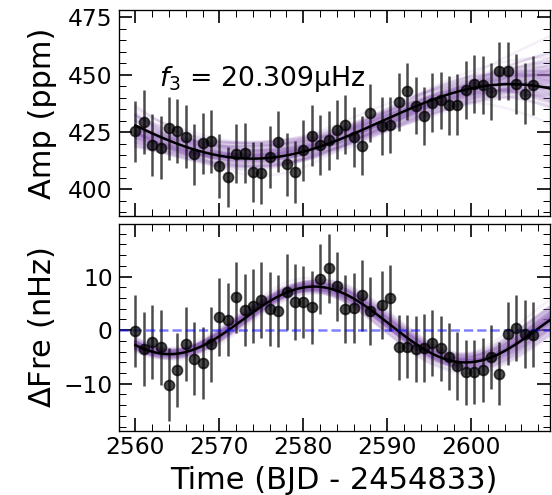}
    \includegraphics[width=0.224\textwidth]{ 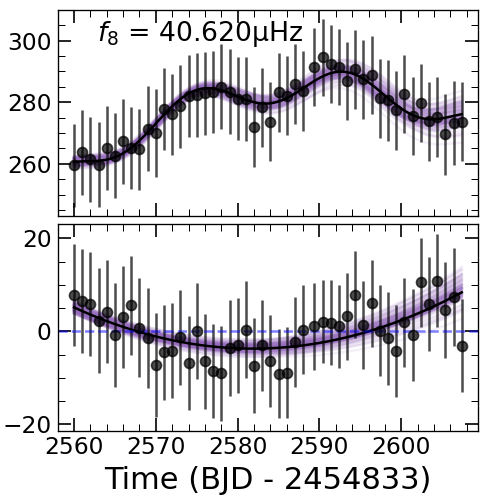}
    \caption{AMs and FMs of the orbital frequency and its harmonic overtone.}
    \label{fig:orb_afm}
\end{figure}

The LSP of the photometry clearly shows low-frequency signals that are attributed to the orbital brightness variations. Their frequencies are derived with $\sim 20.3107\pm0.0015~\mu$Hz, or $0.569851\pm0.000042$\,d, for the primary peak (see Fig.\,\ref{fig:rv-phase}, b).{ We note there is some phase difference between the photometry and RV curve, mainly introduced by the fact that these two curves were folded into phase by their own ephemeris and optimal periods with very slight differences. If we stick folding curves with the same period, the scatter will be a bit larger in one curve, which deserves some particular attention in the future.} When we exploit the dynamics of those two orbital signals, we clearly find that both their frequency and amplitude show quasi-periodic modulations (Fig.\,\ref{fig:orb_afm}). The modulating patterns are not exactly the same as the AMs and FMs observed for pulsation modes if we compare the fitting results in Table\,\ref{Tab:AFM_D}.{ More specifically, it is only the orbital signal $f_3$ exhibits sinusoidal patterns in both the AM and FM}. This can be explained if the companion is indeed an active star presenting magnetic variability (e.g., spot modulation). Recently, \citet{2020ApJ...905...67P} and \citet{2022MNRAS.511.2285W} suggest that spot activity in late-type M dwarf can introduce amplitude and frequency modulations to the signals related to the orbit in the Fourier space. Therefore, {PG~0101+039} could possibly contain an active companion of an M-dwarf star, different from the claiming of a helium white dwarf by \citet{2023A&A...673A..90S}. Moreover, from the summarized information of the sdB pulsator in binary system \citep{2022MNRAS.511.2201S}, there is no sdB+WD system that contains a sdB companion with $P_\mathrm{rot}<10$~days; for instance, EPIC~201206621, with $P_\mathrm{orb}\sim 0.54$\,d and $P_\mathrm{rot} > 45$\,d \citep{2016MNRAS.458.1417R} and EPIC~211696659, with $P_\mathrm{orb}\sim 3.2$\,d and $P_\mathrm{rot} > 28$\,d \citep{2018MNRAS.474.5186R}, whereas a sdB+dM binary usually has a relatively faster rotating sdB pulsator with $P<10$~days {such as EPIC~246023959 with $P_\mathrm{orb}\sim 0.31$\,d and $P_\mathrm{rot}\sim 4.6$\,d and EPIC~246387816 with $P_\mathrm{orb}\sim 0.80$\,d and $P_\mathrm{rot}\sim 9.4$\,d \citep{2019MNRAS.489.1556B}}. To conclude, the type of the faint companion of {PG~0101+039} is still not well determined. Further work to identify the type of companion based on combining photometry and spectroscopy is strongly encouraged. Finally, we note that this unsynchronized system can be used as an independent method to determine the age of the sdB binaries. Because the ratio of the differential rotation rate between the core and envelope can be calculated through the dynamical process of tidal synchronization\citep{1979PhDT........61N,1989ApJ...342.1079G}. This can be further studied after obtaining more precise parameters for the primary and the secondary components.

\subsection{Nonlinear modulation}\label{sec:BP}

\begin{figure}
    \centering
    \includegraphics[width=0.49\textwidth]{ 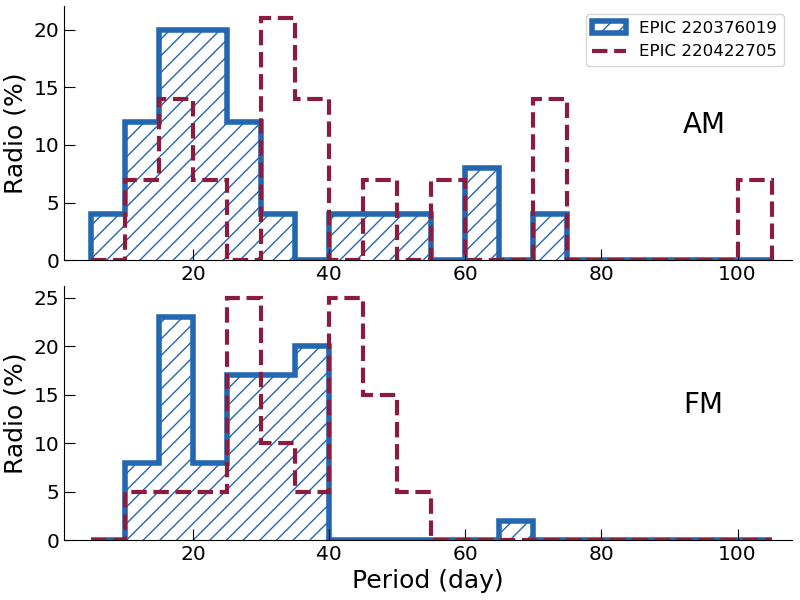}
    \caption{Period distributions of AMs and FMs that are characterized in EPIC~220396019 and EPIC~220422705.}
    \label{fig:P_AFM}
\end{figure}

In {PG~0101+039}, we have clearly detected amplitude and frequency variations in 44 frequencies, most of which are characterized by periodic modulation patterns, for instance, by the fitting of a $G_4$ function. Instead of the AMs and FMs in EPIC~220422705 \citep{2022ApJ...933..211M}, where the close frequencies can introduce the beating effect, the modulations in {PG~0101+039} are mostly intrinsic as the rotational multiplets are well resolved. As first documented in \citet{2016A&A...594A..46Z}, periodic AMs and FMs in sdB pulsator are related to nonlinear resonant mode coupling. In their following results, they found that the pulsation modes in sdB stars are commonly observed to exhibit AM and FM \citep{2018ApJ...853...98Z,2021ApJ...921...37Z}. However, the AMs and FMs measured from {\sl Kepler} photometry exhibit a time resolution of a few tens of days and the 
modulating timescales are characterized from months to years. With {\sl K}2 photometry, the time resolution of AMs and FMs is reduced to 1\,d, which leads to discoveries of short timescale variations of $\sim30$\,d in EPIC~220422705 and {PG~0101+039}. Compared to the former, oscillation modes in {PG~0101+039} generally have a shorter modulating timescale. Figure\,\ref{fig:P_AFM} shows the periods of AMs and FMs in those two sdB pulsators. For instance, the AMs in {PG~0101+039} are found around a period of 20~d, but those of EPIC~220422705 are distributed in a wider range from 20~d to 70~d.

According to the nonlinear regime \citep[see, e.g.,][]{1998BaltA...7...21G}, the modulating timescale could be roughly estimated to the inverse of the frequency mismatch between three resonant modes as 
\begin{equation} \label{emod}
    P_\mathrm{m} \sim 1/\delta f,
\end{equation}
where the $\delta f = i \cdot f_1 + j\cdot f_2 - k\cdot f_3$. We typically consider $[i,j,k]=[1,1,1]$ and $[i,j,k]=[1,1,2]$ for direct resonance and rotational triplet resonance, respectively. In the triplet case, $\delta f \propto \Omega^2 /f$ indicates that a faster rotation introduces a larger frequency mismatch \citep{1992ApJ...394..670D}. It would be natural to foresee a relatively rapid AM and FM in the faster rotating star compared to the slower one if we completely base on Eq.\,\ref{emod}. As we mentioned before, the modulating periods derived for {PG~0101+039} are somewhat shorter than those of EPIC~220422705. {The rotation periods of PG~0101+039 and EPIC~220422705 are $\sim$ 9\,d and 35\,d, respectively, whereas the latter has a typical rotation rate of sdB pulsators, ranging from a few days to several months, for instance, $P_\mathrm{rot}$\,(KIC~5807616) $\sim$ 39\,d \citep{2011Natur.480..496C} and $P_\mathrm{rot}$\,(EPIC~212707862) $\sim$ 80\,d \citep{2016AcA....66..455B}.}
The theoretical prediction can be well consistent with the observation if one imagines that most modes are in triplet resonance. Therefore the ratios of period derived for AMs and FMs in those two stars are clearly distinct. However, we address the point that it is very hard to calculate the modulating timescales precisely for the observed AMs and FMs from the theoretical aspect at present. This is because we have to first construct linear seismic models to obtain linear eigenvalues for the resonant modes, which are the base vectors for further calculation of the complicated nonlinear coefficients that are involved in nonlinear amplitude equations \citep{1982AcA....32..147D}. 

Another important result, related to nonlinear resonance, is that we have observed many pulsation modes exhibiting high (anti-) correlation between AMs and FMs. This finding is also recently disclosed in other sdB pulsators, for instance, KIC~3527751 \citep{2018ApJ...853...98Z} and EPIC~220422705 \citep{2022ApJ...933..211M}. For a three-mode interaction, the nonlinear amplitude equations (AEs) can be simplified as follows: 
\begin{equation}
\mathbf{\dot{A}}_i = \mathbf{Q}(\mathbf{A}_i, \mathbf{A}_j, \mathbf{A}_k),
\end{equation}
where $i,j,k = 1, 2,~\mathrm{or}~3,$ with $i\neq j \neq k$, and the $\mathbf{Q}$ is a function involving nonlinear complicated coupling coefficients. We can see details of nonlinear AEs in, for instance, \citet{1985AcA....35..229M}. The complex amplitude $\mathbf{A}_i$ can be separated into the real part, $A_i$, and the imaginary part, $\phi_i$. The time deviation, or temporal variation, of the real part forecasts AM, whereas the imaginary part is for FM. Numerical explorations of nonlinear AEs explicitly show several cases of anti-correlation between AMs and FMs \citep{1985AcA....35..229M}. As the nonlinear AEs are governed by their amplitudes, it is natural to foresee that 
FMs follow the behavior of AMs. Our findings, therefore, are further consistent with the prediction of nonlinear AEs, which can also be used for the upcoming nonlinear calculations.{ In addition, the observation of non-modulating modes could be non-resonant modes with very slight nonlinear couplings or nonlinear locking modes with too strong couplings. Those modes are suitable for exoplanet detection or monitoring secular evolutionary period changing rates via pulsation-timing technique.}

In {PG~0101+039}, we reported an additional feature that is different from the modulating patterns in previous studies. Within the same multiplet, one of them may exhibit a much shorter timescale compared to others, for instance, the AM of the frequency near $574~\mu$Hz (Figure\,\ref{fig:Mul_570}). To verify the intrinsic of this shallow and rapid AM, we changed the window of the sLSP in various widths and we never saw any significant changes in AM. This suggests that the observed modulations might be more complicated than the simple explorations of nonlinear resonance where only the weak interaction of one kind of resonant mode was considered to avoid complicated calculations of nonlinear AEs \citep[see, e.g.,][]{1994A&A...291..481G,1995A&A...296..405B}. As previously reported in \citet{2016A&A...594A..46Z}, it has been suggested that the observed AMs and FMs could play a role in different kinds of resonance simultaneously, as a consequence that solely the rotation resonance might not be enough to explain their specific AMs and FMs. Therefore, these current discoveries strongly indicate that the nonlinear calculation of AEs with different types of resonance has to be considered in the future, in order to mimic the complicated AMs and FMs from observations. However, in the multiplet component near $574~\mu$Hz, we did not resolve any linear combinations that could be linked to direct resonance at present. There might be several combining modes in potential, but their amplitude is far below the current noise.

Finally, we notice that amplitude and frequency modulating in a short timescale is important to the development of nonlinear stellar theory in the future. Compared to the {\sl Kepler} discoveries, it will extend the investigating sample of nonlinear resonance to a much larger volume for the first time. As complementary, various modulating behaviors and timescales are useful constraints to various nonlinear quantities involved in AEs. 

% TESS will the next
 % \input{Sections/Conclusion}

 \section{Conclusion}\label{sec:con}

PG~0101+039 has been continuously observed by {\sl K}2 in Campaign~8 over a period of $\sim 79$~days. Its photometry was extracted from the TPF file with a series of pixel sizes. The optimal aperture is estimated to be 12 for following Fourier transformation by evaluating the S/N of the primary frequency (Fig.\,\ref{fig:light-curve}). Both the light curves and the Lomb-Scargle periodogram reveal that PG~0101+039 is a $g$-mode-dominated hybrid pulsating sdB star. By setting the threshold of $\mathrm{S/N}=5.2$, we detected 137 independent frequencies (including ten frequencies whose S/N~$< 5.2$ ), two orbital frequencies and 51 linear combinations. The former includes 20 rotational multiplets with 70 identified components (Fig.\,\ref{fig:RotaMult}). Those frequency splittings give a rotational period of $8.81\pm0.06$\,d and $8.60\pm0.16$\,d for the internal part and outer layer, as {\sl g}-modes penetrate deeper than {\sl p}-modes, implying a { marginally} radial differential rotation { with a probability of $\sim 60\%$}. This explanation could be supported that tidal force can initially accelerate the envelope by introducing internal gravity wave \citep{1989ApJ...342.1079G}. We exploit the binary information combined with spectroscopy and photometry, coming up with a solution similar to previous results \citep{1999MNRAS.304..535M,2008A&A...477L..13G}. It suggests that this binary system is still on the way to synchronization with an orbital period of $0.57$~d and a rotational period of $\sim8.8$~d. However, we stress that the conclusion of differential rotation is based on relatively weak evidence if one fully considers the uncertainties in the period of rotation. We then derive the period spacing of $\sim$252\,s and 144\,s for dipole and quadrupole modes, respectively. Subsequently, we identified 28 frequencies in the dipole mode sequence, 28 quadrupole modes, and another 3 in both sequences.

Before we proceeded to characterize amplitude and frequency modulations, we performed a series of testing to prove that modulating pattern is independent of the size of the stamp, thanks to the relative isolate position of PG~0101+039. At this stage, we can exploit the amplitude and frequency modulations for 44 significant pulsations. The majority of those pulsations exhibit clear modulating behavior in various patterns, which can then be characterized by five types of simple functions. The fitting uncertainties were derived with the \texttt{MCMC} technique. We find that the majority of these modulating patterns can be presented with a periodic fitting and these modulating periods are on a timescale of months, or precisely, in the range between $\sim10-70$~days. There are four frequencies modulating in a much more rapid period, slightly less than 10 days. In general, we observe a relatively short timescale of AMs and FMs in PG~0101+039. Interestingly, many frequencies show a clear high (anti-) correlation between their amplitude and frequency. Moreover, we find that two low-frequency signals, related to the orbit, present amplitude and frequency variations as well.

To interpret the discovered AMs and FMs of oscillation modes in PG~0101+039, a natural consequence could be produced by nonlinear resonant couplings which predicts various variations in amplitude and frequency \citep[see, e.g.,][]{1998BaltA...7...21G}. In some particularly resonant conditions, the nonlinear interacting modes undergo periodic variations \citep{1995A&A...296..405B}, as we indeed observe many modes that exhibit periodic modulating patterns; these, in particular, serve as evidence of several multiple components. As the modulating period, $P_\mathrm{m}$, is proportional to the frequency mismatch or higher order effect from rotation, in general, $P_\mathrm{m}$ has somewhat smaller values in PG~0101+039 than EPIC~220422705 as the former has a relatively faster rotation comparing to the latter. 
The (anti-) correlation between AMs and FMs further supports our explanation since the patterns of FMs are governed by that of AMs when one tries to solve the nonlinear amplitude equations in a numerical way \citep{1985AcA....35..229M}. 
However, at the current stage, we are still waiting for the seismic model to calculate the linear eigenvalues for the nonlinear coupling modes, upon which the nonlinear coupling coefficients are built. Moreover, the AMs and FMs exhibit somewhat more complicated patterns than the prediction from nonlinear AEs, in which the assumption only considers three interacting modes within one kind of resonance. These modulating patterns probably offer observational constraints to future calculations of nonlinear amplitude equations. This shorter timescale modulation should be explored in other sdB stars that had been observed by {\sl K}2 and TESS. {We finally note that this finding could jeopardize the exoplanet detection via the time-pulsation method \citep{2007Natur.449..189S,2018A&A...611A..85S}. Because the phase modulation can hardly be well derived if we do not have information a priori on the frequency modulations.}

\begin{acknowledgements}

We acknowledge the support from the National Natural Science Foundation of China (NSFC) through grants 11833002, 12273002, 12090040, 12090042, 11903005 and 12203010. 
S.C. is supported by the Agence Nationale de la Recherche (ANR, France) under grant ANR-17-CE31-0018, funding the INSIDE project, and financial support from the Centre National d'Études Spatiales (CNES, France). The authors gratefully acknowledge the {\em Kepler} team and all who have contributed to making this mission possible. Funding for the {\em Kepler} mission is provided by NASA's Science Mission Directorate. The LAMOST Telescope is a National Major Scientific Project built by the Chinese Academy of Sciences. Funding for the project has been provided by the National Development and Reform Commission. {JNF, WZ, JW acknowledge the science research grants from the China Manned Space Project.}

\end{acknowledgements}

\bibliographystyle{aa} 
\bibliography{reference}

\begin{appendix}

\section{The optimal stamp for the light curve}\label{appecd:lc_stamp}

\begin{figure}
\centering 
\includegraphics[width=0.5\textwidth]{ 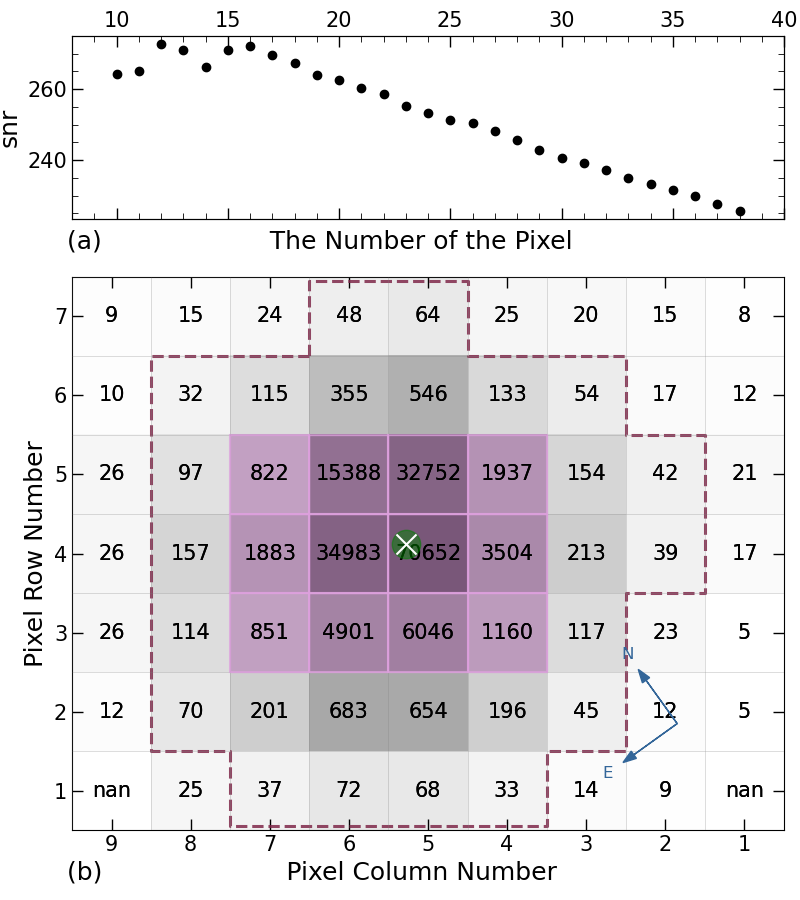}
\caption{ Aperture selection of TPFs.
Fig\,(a) shows how the SNR of the frequency with the highest amplitude varies from the number of pixels.
(b) The TPFs with the median value of {PG~0101+039}. The size of the green dot represents the relative brightness of the star, and the white '$\times$' is the location of our target star. {The plum boxes are the pixels used for extraction.}
}
\label{fig:lc_stemp}
\end{figure}

To find the optimal stamp for {PG0101~039}, we considered a series of stamps with different pixel sizes from the TPFs. 
The stamp was initially covered by ten pixels with the highest values of flux, thanks to {PG~0101+039} being free of flux contamination. Then the stamp was expanded outward to nearby pixels in sequence according to their values of flux down to around 30, a level similar to a few times of background. In our case, the stamp is composed of 38 pixels that completely cover the target, as shown in Fig\,\ref{fig:lc_stemp}\,(b) by the pixels in grey shadow and surrounded by the dashed lines. For each stamp enlargement, we extracted the light curves for {PG~0101+039} and performed some preliminary processes on them, including detrending and outlier clipping. A few outliers were clipped off from the overall flattened light curves by filtering with a local 3.5$\sigma$ threshold. Then the light curves were normalized and performed with Fourier transformation for quality comparison. The signal-to-noise ratio (S/N) of the signal with the highest amplitude ($\sim191.5\,\mu$Hz) was denoted to represent the quality of the light curves. The recorded S/N for each light curve is shown in Fig\,\ref{fig:light-curve}\,(a), with values from 222 to 273. The highest S/Ns are found around pixel sizes of around 12 to 16.
We thus took the optimal stamp with 12 central pixels and the corresponding light curves were used according to our following analysis, a similar strategy to \citet{2021ApJ...922....2D}. 

\FloatBarrier

\FloatBarrier

\section{All frequencies detected in PG~0101+039}

\begin{table*} 
\renewcommand{\arraystretch}{1.2}%设置行高
\caption[]{\label{Tab:freq_all} All frequencies we detected in {PG~0101+039} by order of increasing frequency.}
\small
\begin{center}
\begin{tabular}{lccccccrcccc}
\hline
\hline
{ID} & {Frequency} & {$\sigma$f} & {Period} & {$\sigma$P} & {Amplitude} & {$\sigma$A}  & {S/N} & {$\ell$} & {$m$} & {{Mode}} & {Comments}  \\
{}  & {($\mu$Hz)} & {($\mu$Hz)} & {(s)} & {(s)} 
& {(ppm)} & {(ppm)} & {}& {} & {} & {}& {}\\
\hline
$f_{74}$   &      84.1622  &     0.0149  &      11881.8220  &     2.1086  &    39.11  &    7.21  &    5.4  &   $1^*$   &  $-1^*$  &  $g$  &  ...  \\ 
$f_{55}$   &     109.5365  &     0.0116  &       9129.3736  &     0.9698  &    52.20  &    7.49  &    7.0  &      &    &  $g$  &  ...  \\ 
$f_{62}$   &     111.6446  &     0.0128  &       8956.9917  &     1.0249  &    48.05  &    7.57  &    6.3  &      &    &  $g$  &  ...  \\ 
$f_{54}$   &     112.4960  &     0.0117  &       8889.2042  &     0.9258  &    52.62  &    7.60  &    6.9  &   $2^*$   &  $-2^*$  &  $g$  &  ...  \\ 
$f_{63}$   &     114.3497  &     0.0128  &       8745.1024  &     0.9756  &    47.73  &    7.51  &    6.3  &    $2^*$  &  $-2^*$  &  $g$  &  ...  \\ 
$f_{53}$   &     119.0652  &     0.0116  &       8398.7592  &     0.8153  &    53.25  &    7.59  &    7.0  &   $1^*$   &  $-1^*$  &  $g$  &  ...  \\ 
$f_{49}$   &     121.3440  &     0.0108  &       8241.0347  &     0.7321  &    57.56  &    7.65  &    7.5  &   $1^*$   &  $1^*$  & $g$   &  ...  \\ 
$f_{26}$   &     124.8665  &     0.0070  &       8008.5516  &     0.4500  &    89.49  &    7.74  &   11.6  &      &    &  $g$  &  ...  \\ 
$f_{61}$   &     138.5549  &     0.0134  &       7217.3580  &     0.6996  &    48.55  &    8.05  &    6.0  &   $2^*$   &  $-2^*$  &  $g$  &  ...  \\ 
$f_{44}$   &     144.0292  &     0.0104  &       6943.0373  &     0.5018  &    62.52  &    8.03  &    7.8  &  1  &  0 &  $g$  &  ...  \\ 
$f_{20}$   &     144.7187  &     0.0055  &       6909.9594  &     0.2630  &   117.88  &    8.01  &   14.7  &  1  & 1 &  $g$  &  AFM  \\ 
$f_{30}$   &     155.3015  &     0.0076  &       6439.0879  &     0.3169  &    84.75  &    7.99  &   10.6  &  1  &  0  &  $g$  &  ...  \\ 
$f_{37}$   &     155.9591  &     0.0090  &       6411.9392  &     0.3690  &    72.00  &    7.97  &    9.0  &  1  & 1 &  $g$  &  ...  \\ 
$f_{56}$   &     157.1873  &     0.0127  &       6361.8370  &     0.5152  &    50.79  &    7.98  &    6.4  &      &    &  $g$  &  ...  \\ 
$f_{32}$   &     163.1337  &     0.0082  &       6129.9395  &     0.3063  &    79.04  &    7.95  &    9.9  &  $2^*$    &  $-1^*$  &  $g$  &  ...  \\ 
$f_{7}$   &     167.6954  &     0.0021  &       5963.1940  &     0.0739  &   302.10  &    7.74  &   39.0  &  1  &  -1  &  $g$  &  AFM  \\ 
$f_{27}$   &     168.4449  &     0.0070  &       5936.6595  &     0.2469  &    89.20  &    7.71  &   11.6  &  1 & 0   & $g$   &  ...  \\ 
$f_{9}$   &     169.0794  &     0.0030  &       5914.3814  &     0.1045  &   209.27  &    7.71  &   27.1  &   1 & 1 &  $g$  &  AFM  \\ 
$f_{23}$   &     176.0551  &     0.0063  &       5680.0400  &     0.2049  &    99.85  &    7.82  &   12.8  &   $1^*|2^*$   &  $0^*|-2^*$  &  $g$  &  AFM  \\ 
$f_{34}$   &     180.3555  &     0.0082  &       5544.6053  &     0.2524  &    75.95  &    7.69  &    9.9  &   $2^*$   &  $-2^*$  & $g$  &  ...  \\ 
$f_{35}$   &     182.5338  &     0.0082  &       5478.4369  &     0.2454  &    75.49  &    7.62  &    9.9  &  1  &   -1 &  $g$  &  ...  \\ 
$f_{11}$   &     183.1629  &     0.0031  &       5459.6210  &     0.0916  &   200.91  &    7.61  &   26.4  &   1 &   0 &  $g$  &  AFM  \\ 
$f_{58}$   &     183.8808  &     0.0122  &       5438.3060  &     0.3616  &    50.05  &    7.55  &    6.6  &   1 & 1 &  $g$  &  ...  \\ 
$f_{42}$   &     185.6676  &     0.0092  &       5385.9685  &     0.2664  &    66.64  &    7.55  &    8.8  &    $2^*$  &   $-2^*$ &  $g$  &  ...  \\ 
$f_{1}$   &     191.4365  &     0.0004  &       5223.6643  &     0.0108  &  1532.74  &    7.52  &  203.9  &    $1^*|2^*$  &  $0^*|2^*$  &  $g$  &  AFM  \\ 
$f_{66}$   &     200.5844  &     0.0130  &       4985.4327  &     0.3234  &    45.26  &    7.26  &    6.2  &   2   &  -1  &  $g$  &  ...  \\ 
$f_{13}$   &     201.6434  &     0.0032  &       4959.2499  &     0.0785  &   181.95  &    7.16  &   25.4  &   2   &  0  &  $g$  &  AFM  \\ 
$f_{40}$   &     202.5375  &     0.0086  &       4937.3562  &     0.2105  &    67.37  &    7.18  &    9.4  &   2   &  1?  &  $g$  &  ...  \\ 
$f_{2}$   &     211.0751  &     0.0012  &       4737.6492  &     0.0264  &   466.68  &    6.77  &   69.0  &    $2^*$  &  $0^*$  &  $g$  &  AFM  \\ 
$f_{19}$   &     212.3919  &     0.0041  &       4708.2776  &     0.0918  &   130.76  &    6.68  &   19.6  &   $1^*$   &  $0^*$  &  $g$  &  AFM  \\ 
$f_{18}$   &     216.2481  &     0.0036  &       4624.3181  &     0.0761  &   146.22  &    6.42  &   22.8  &      &    &  $g$  &  AFM  \\ 
$f_{70}$   &     218.8927  &     0.0124  &       4568.4489  &     0.2582  &    41.96  &    6.40  &    6.5  &    $2^*$  &  $-1^*$  &  $g$  &  ...  \\ 
$f_{81}$   &     225.4701  &     0.0139  &       4435.1780  &     0.2736  &    34.96  &    6.00  &    5.8  &    $2^*$  &  $-1^*$  &  $g$  &  ...  \\ 
$f_{43}$   &     231.0496  &     0.0074  &       4328.0757  &     0.1381  &    64.10  &    5.83  &   11.0  &   2   &   -1 &  $g$  &  ...  \\ 
$f_{46}$   &     232.1609  &     0.0076  &       4307.3583  &     0.1407  &    61.37  &    5.74  &   10.7  &   $2^*$   &  $0^*$  &  $g$  &  AFM  \\ 
$f_{64}$   &     234.3496  &     0.0096  &       4267.1295  &     0.1757  &    47.33  &    5.63  &    8.4  &  2    &  2  &  $g$  & ... \\ 
$f_{21}$   &     238.2989  &     0.0040  &       4196.4107  &     0.0698  &   110.22  &    5.39  &   20.4  &   $1^*$   &  $0^*$  &  $g$  &  AFM  \\ 
$f_{45}$   &     239.5982  &     0.0070  &       4173.6544  &     0.1211  &    62.25  &    5.34  &   11.7  &   $2^*$   &  $1^*$  &  $g$  &  ...  \\ 
$f_{51}$   &     250.3381  &     0.0072  &       3994.5983  &     0.1150  &    55.15  &    4.90  &   11.2  &      &    &  $g$  &  ...  \\ 
$f_{4}$   &     254.6364  &     0.0009  &       3927.1685  &     0.0146  &   404.06  &    4.71  &   85.8  &  1  &  0  &  $g$  &  S  \\ 
$f_{28}$   &     255.2867  &     0.0043  &       3917.1639  &     0.0665  &    88.47  &    4.73  &   18.7  &  1 & 1 &  $g$  &  AFM  \\ 
$f_{10}$   &     258.4635  &     0.0018  &       3869.0180  &     0.0273  &   207.41  &    4.67  &   44.4  &      &    &  $g$  &  AFM  \\ 
$f_{97}$   &     260.7369  &     0.0134  &       3835.2831  &     0.1977  &    27.72  &    4.60  &    6.0  &      &    &  $g$  &  ...  \\ 
$f_{114}$   &     281.9538  &     0.0150  &       3546.6802  &     0.1887  &    22.08  &    4.09  &    5.4 &   $2^*$   &  $-2^*$  &  $g$  &  ...  \\ 
$f_{22}$   &     291.0652  &     0.0031  &       3435.6564  &     0.0362  &   101.26  &    3.84  &   26.4  &  1  &  -1  &   $g$ &  AFM  \\ 
$f_{17}$   &     291.7374  &     0.0021  &       3427.7404  &     0.0244  &   149.76  &    3.83  &   39.1  &  1  &  0  &  $g$  &  AFM  \\ 
$f_{47}$   &     293.5365  &     0.0052  &       3406.7318  &     0.0598  &    59.65  &    3.79  &   15.7  &   $2^*$   &  $-2^*$  &  $g$  &  ...  \\ 
$f_{16}$   &     305.5117  &     0.0019  &       3273.1966  &     0.0202  &   152.56  &    3.55  &   43.0  &   $2^*$   &  $-2^*$  &  $g$  &  AFM  \\ 
$f_{12}$   &     314.0817  &     0.0015  &       3183.8846  &     0.0150  &   187.03  &    3.42  &   54.7  &   $1^*$  &  $0^*$  &  $g$  &  AFM  \\ 
$f_{14}$   &     315.3790  &     0.0016  &       3170.7879  &     0.0159  &   174.72  &    3.41  &   51.3  &      &    &  $g$  &  AFM  \\ 
$f_{31}$   &     316.7995  &     0.0034  &       3156.5707  &     0.0341  &    80.68  &    3.41  &   23.7  &     &    &  $g$  &  AFM  \\ 
$f^*_{174}$   &     329.9880  &     0.0187  &       3030.4133  &     0.1716  &    14.22  &    3.28  &    4.3  &   2   &  -2  &   $g$ & ...  \\ 
 $f^*_{145}$   &     331.0714  &     0.0158  &       3020.4962  &     0.1442  &    16.80  &    3.28  &    5.1  &   2   &  -1  &  $g$  & ...  \\ 
$f_{84}$   &     332.1861  &     0.0079  &       3010.3603  &     0.0711  &    33.87  &    3.28  &   10.3  &   $2^*$   &  $0^*$  & $g$  & ...  \\ 
\hline
\end{tabular}
\end{center}

% \tablefoot{AM/FM/AFM indicates that the frequency has modulation of amplitude~(AM)| frequency~(FM) or both~(AFM). '*' means the S/N of the mode is below the threshold limit.}
\end{table*}

\addtocounter{table}{-1} 
\begin{table*} \caption[]{continued.}
\renewcommand{\arraystretch}{1.2}%设置行高
% \small
\begin{center}
\begin{tabular}{lccccccrcccc}
\hline
\hline
{ID} & {Frequency} & {$\sigma$f} & {Period} & {$\sigma$P} & {Amplitude} & {$\sigma$A}  & {S/N} & {$\ell$} & {$m$} & {{Mode}} & {Comments}  \\
{}  & {($\mu$Hz)} & {($\mu$Hz)} & {(s)} & {(s)} 
& {(ppm)} & {(ppm)} & {}& {} & {} & {}& {}\\             
\hline
$f_{94}$   &     333.2654  &     0.0091  &       3000.6119  &     0.0822  &    29.11  &    3.28  &    8.9  &    2  & 1 &  $g$  & ...  \\ 
$f_{103}$   &     334.3583  &     0.0107  &       2990.8035  &     0.0960  &    24.70  &    3.27  &    7.5  &   2   &  2  &  $g$  & ... \\ 
$f_{69}$   &     343.3255  &     0.0061  &       2912.6879  &     0.0515  &    42.30  &    3.17  &   13.3  &   1 &  -1  &   $g$ &  ...  \\ 
$f_{15}$   &     343.9567  &     0.0015  &       2907.3428  &     0.0128  &   169.36  &    3.15  &   53.7  &   1 &  0  &  $g$  &  AFM  \\ 
$f_{78}$   &     350.7326  &     0.0066  &       2851.1747  &     0.0534  &    37.44  &    3.03  &   12.3  &   $2^*$   &  $-2^*$  &  $g$  &   \\ 
$f_{06}$   &     377.4207  &     0.0008  &       2649.5631  &     0.0053  &   302.17  &    2.84  &  106.4  &  1  &  -1  &  $g$  &  AFM  \\ 
$f_{25}$   &     378.0847  &     0.0026  &       2644.9102  &     0.0179  &    90.28  &    2.84  &   31.8  &  1  &  0  &  $g$  &  AFM  \\ 
$f_{05}$   &     378.7191  &     0.0008  &       2640.4797  &     0.0053  &   305.51  &    2.84  &  107.6  &  1  & 1 &  $g$  &  AFM \\ 
$f_{39}$   &     390.5330  &     0.0034  &       2560.6031  &     0.0220  &    67.72  &    2.80  &   24.2  &   $2^*$   &  $-2^*$  & $g$  &  FM  \\ 
$f_{125}$   &     412.8598  &     0.0107  &       2422.1296  &     0.0625  &    20.60  &    2.71  &    7.6  &      &    &  $g$  & ... \\ 
$f_{41}$   &     414.4970  &     0.0033  &       2412.5626  &     0.0192  &    66.65  &    2.71  &   24.6  &   $1^*$   &  $0^*$  &  $g$  &  AFM  \\ 
$f_{111}$   &     450.1190  &     0.0091  &       2221.6350  &     0.0448  &    22.56  &    2.52  &    8.9  & 1   &  -1  &  $g$  & ... \\ 
$f_{24}$   &     450.8067  &     0.0021  &       2218.2457  &     0.0105  &    96.53  &    2.53  &   38.1  &  1  &   0 &  $g$  &  AFM  \\ 
$f_{105}$   &     544.1181  &     0.0085  &       1837.8364  &     0.0287  &    23.61  &    2.47  &    9.5 &      &     &  $g$  &  ...  \\ 
$f_{142}$   &     549.6182  &     0.0113  &       1819.4448  &     0.0373  &    17.63  &    2.45  &    7.2  &      &    &  $g$  & ... \\ 
$f_{33}$   &     570.2222  &     0.0025  &       1753.7022  &     0.0077  &    78.16  &    2.41  &   32.4  &   4   &  -3  & $g$   &  AFM  \\ 
$f_{60}$   &     571.5085  &     0.0040  &       1749.7552  &     0.0122  &    49.33  &    2.43  &   20.3  &   4   &  -2  &  $g$  &  AFM  \\ 
$f_{102}$   &     572.7683  &     0.0079  &       1745.9066  &     0.0240  &    25.13  &    2.44  &   10.3  &   4   &  -1  &  $g$  & ...  \\ 
$f_{36}$   &     574.0246  &     0.0026  &       1742.0856  &     0.0079  &    75.38  &    2.43  &   31.0  &   4   &  0  &  $g$  &  AFM  \\ 
 $f^*_{185}$   &     575.2998  &     0.0175  &       1738.2241  &     0.0530  &    11.27  &    2.44  &    4.6  &   4   & 1 &  $g$  & ...  \\ 
$f_{99}$   &     576.5143  &     0.0076  &       1734.5624  &     0.0227  &    26.17  &    2.44  &   10.7  &   4   &  2  &  $g$  &  ...  \\ 
 $f^*_{190}$   &     577.7372  &     0.0220  &       1730.8908  &     0.0659  &     8.98  &    2.44  &    3.7  &   4   &  3  &  $g$  & ...  \\ 
$f_{146}$   &     601.5151  &     0.0119  &       1662.4687  &     0.0329  &    16.74  &    2.46  &    6.8  &      &    &  $g$  &  ...  \\ 
$f_{65}$   &     607.7807  &     0.0044  &       1645.3304  &     0.0120  &    45.86  &    2.51  &   18.3  &      &    &  $g$  &  AFM  \\ 
$f_{73}$   &     617.1960  &     0.0052  &       1620.2307  &     0.0136  &    39.34  &    2.51  &   15.7  &      &    &  $g$  &  AFM  \\ 
$f_{104}$   &     635.5326  &     0.0084  &       1573.4833  &     0.0208  &    24.38  &    2.53  &    9.6  &      &    &  $g$  &  ...  \\ 
$f_{79}$   &     651.3998  &     0.0056  &       1535.1555  &     0.0132  &    36.24  &    2.51  &   14.4  &   2   &  -1  &  $g$  & ... \\ 
$f_{176}$   &     652.5496  &     0.0148  &       1532.4506  &     0.0348  &    13.83  &    2.53  &    5.5  &  2   &   0 &  $g$  & ... \\ 
$f_{50}$   &     653.5355  &     0.0036  &       1530.1388  &     0.0083  &    57.54  &    2.53  &   22.8  &   2   &  1 &  $g$  &  AM  \\ 
$f_{87}$   &     702.6498  &     0.0063  &       1423.1840  &     0.0128  &    32.14  &    2.51  &   12.8  &   6   &    &  $g$  &  ...  \\ 
$f_{88}$   &     705.2075  &     0.0066  &       1418.0224  &     0.0133  &    30.93  &    2.53  &   12.2  &   6   &    &  $g$  &  ...  \\ 
$f_{38}$   &     707.7781  &     0.0030  &       1412.8722  &     0.0060  &    69.22  &    2.55  &   27.1  &    6|8  &    &  $g$  &  AFM  \\ 
 $f^*_{188}$   &     709.0981  &     0.0194  &       1410.2421  &     0.0385  &    10.67  &    2.55  &    4.2  &   8   &    &  $g$  & ...  \\ 
$f_{67}$   &     710.3852  &     0.0046  &       1407.6870  &     0.0091  &    45.19  &    2.55  &   17.7  &   8   &    &  $g$  &  ...  \\ 
$f_{80}$   &     711.7116  &     0.0059  &       1405.0636  &     0.0117  &    34.99  &    2.56  &   13.7  &   8   &    &  $g$  &  ...  \\ 
$f_{68}$   &     732.8429  &     0.0047  &       1364.5489  &     0.0088  &    44.95  &    2.61  &   17.2  &      &    &  $g$  &  AFM  \\ 
$f_{129}$   &     739.7114  &     0.0109  &       1351.8785  &     0.0199  &    19.79  &    2.65  &    7.5  &      &    &  $g$  &  ...  \\ 
$f_{120}$   &     849.7909  &     0.0099  &       1176.7600  &     0.0137  &    20.93  &    2.55  &    8.2  &      &    &  $g$  &  ...  \\ 
$f_{93}$   &     871.4793  &     0.0073  &       1147.4742  &     0.0096  &    29.65  &    2.68  &   11.1  &      &    &  $g$  &  ...  \\ 
$f_{108}$   &     871.9024  &     0.0093  &       1146.9173  &     0.0123  &    23.13  &    2.66  &    8.7  &      &    &  $g$  &  ...  \\ 
$f_{71}$   &     872.8334  &     0.0053  &       1145.6940  &     0.0070  &    40.44  &    2.66  &   15.2  &      &    &  $g$  &  ...  \\ 
$f_{131}$   &     884.2697  &     0.0109  &       1130.8767  &     0.0139  &    19.76  &    2.65  &    7.5  &   $\ge 8?$   &    &  $g$  & ... \\ 
$f_{115}$   &     886.9046  &     0.0097  &       1127.5170  &     0.0123  &    22.01  &    2.63  &    8.4  &   $\ge 8?$   &    &  $g$  & ...  \\ 
$f_{96}$   &     889.5773  &     0.0075  &       1124.1294  &     0.0095  &    28.02  &    2.61  &   10.7  &    $\ge 8?$  &    &  $g$  &  ...  \\ 
$f_{86}$   &     890.9649  &     0.0065  &       1122.3786  &     0.0082  &    32.78  &    2.63  &   12.5  &   $\ge 8?$   &    &  $g$  & ...  \\ 
$f_{127}$   &     929.4440  &     0.0103  &       1075.9121  &     0.0119  &    20.49  &    2.60  &    7.9  &   $\ge 8?$   &    &  $g$  &  ...  \\ 
 $f_{179}^*$   &     932.0668  &     0.0177  &       1072.8845  &     0.0204  &    11.92  &    2.61  &    4.6  &   $\ge 8?$   &    &  $g$  & ...  \\ 
$f_{170}$   &     933.3882  &     0.0146  &       1071.3656  &     0.0168  &    14.49  &    2.61  &    5.5  &    $\ge 8?$  &    &  $g$  & ...  \\ 
$f_{163}$   &     934.7951  &     0.0142  &       1069.7532  &     0.0162  &    15.08  &    2.64  &    5.7  &   $\ge 8?$   &    &  $g$  & ...  \\ 
 $f^*_{178}$   &     936.1284  &     0.0180  &       1068.2296  &     0.0205  &    11.96  &    2.66  &    4.5  &   $\ge 8?$   &    &  $g$  & ...  \\ 
\hline
\end{tabular}
\end{center}
% \tablefoot{
% AM/FM/AFM indicates that the frequency has modulation of amplitude~(AM), frequency~(FM) or both~(AFM). '*' means the S/N of the mode is below the threshold limit.}
\end{table*}

\addtocounter{table}{-1} 
\begin{table*} \caption[]{continued.}
\renewcommand{\arraystretch}{1.2}%设置行高
% \small
\begin{center}
\begin{tabular}{lccccccrcccc}
\hline
\hline
{ID} & {Frequency} & {$\sigma$f} & {Period} & {$\sigma$P} & {Amplitude} & {$\sigma$A}  & {S/N} & {$\ell$} & {$m$} & {{Mode}} & {Comments}  \\
{}  & {($\mu$Hz)} & {($\mu$Hz)} & {(s)} & {(s)} 
& {(ppm)} & {(ppm)} & {}& {} & {} & {}& {}\\             
\hline
$f_{92}$   &     953.6107  &     0.0072  &       1048.6460  &     0.0079  &    29.88  &    2.64  &   11.3  &      &    &  $g$  &  ...  \\ 
$f_{72}$   &     955.0351  &     0.0054  &       1047.0820  &     0.0059  &    39.84  &    2.64  &   15.1  &      &    &  $g$  &  ...  \\ 
$f_{135}$   &    1060.5123  &     0.0107  &        942.9405  &     0.0095  &    18.60  &    2.46  &    7.6  &   8   &    &  $g$  & ...  \\ 
$f_{133}$   &    1063.1768  &     0.0103  &        940.5773  &     0.0091  &    19.34  &    2.45  &    7.9  &    8  &    &  $g$  & ... \\ 
 $f_{189}^*$   &    1064.4641  &     0.0215  &        939.4398  &     0.0190  &     9.26  &    2.46  &    3.8  &   8   &    &  $g$  & ...  \\ 
$f_{106}$   &    1065.7602  &     0.0085  &        938.2973  &     0.0075  &    23.46  &    2.45  &    9.6  &   8   &    &  $g$  & ...  \\ 
 $f^*_{180}$   &    1068.3546  &     0.0167  &        936.0188  &     0.0146  &    11.89  &    2.44  &    4.9  &    8  &    & $g$   & ...  \\ 
 $f_{187}^*$   &    1365.0343  &     0.0192  &        732.5823  &     0.0103  &    10.78  &    2.56  &    4.2  &   4   &    &  mixed &  ...  \\ 
$f_{82}$   &    1367.6826  &     0.0060  &        731.1638  &     0.0032  &    34.83  &    2.56  &   13.6  &    4  &    &  mixed &  ...  \\ 
$f_{107}$   &    1370.3923  &     0.0090  &        729.7180  &     0.0048  &    23.21  &    2.57  &    9.0  &   4   &    &  mixed &  ...  \\ 
$f_{101}$   &    1373.1646  &     0.0083  &        728.2448  &     0.0044  &    25.15  &    2.59  &    9.7  &  4    &    &  mixed &  ...  \\ 
$f_{91}$   &    1478.3042  &     0.0077  &        676.4508  &     0.0035  &    30.00  &    2.84  &   10.6  &      &    &  mixed &  ...  \\ 
$f_{117}$   &    1568.3730  &     0.0104  &        637.6034  &     0.0042  &    21.58  &    2.76  &    7.8  &      &    &  mixed &  ...  \\ 
$f_{175}$   &    1626.2233  &     0.0152  &        614.9217  &     0.0057  &    14.07  &    2.63  &    5.3  &      &    &  mixed &  ...  \\ 
$f_{134}$   &    1751.3999  &     0.0110  &        570.9718  &     0.0036  &    18.95  &    2.57  &    7.4  &      &    &  mixed &  ...  \\ 
$f_{110}$   &    1824.3909  &     0.0089  &        548.1281  &     0.0027  &    22.65  &    2.50  &    9.1  &      &    &   mixed &  ...  \\ 
$f_{98}$   &    1888.4310  &     0.0072  &        529.5401  &     0.0020  &    27.28  &    2.42  &   11.3  &      &    &  mixed  &  ...  \\ 
$f_{85}$   &    3269.8990  &     0.0056  &        305.8198  &     0.0005  &    32.85  &    2.29  &   14.3  &      &    &  $p$  &  AFM  \\ 
$f_{57}$   &    3306.9347  &     0.0036  &        302.3948  &     0.0003  &    50.25  &    2.25  &   22.3  &  4    &    &  $p$  &  AFM  \\ 
$f_{128}$   &    3310.7620  &     0.0090  &        302.0453  &     0.0008  &    20.33  &    2.26  &    9.0  &   4   &    & $p$   & ...  \\ 
$f_{177}$   &    3312.2120  &     0.0147  &        301.9130  &     0.0013  &    12.34  &    2.23  &    5.5  &   4   &    & $p$   & ...  \\ 
$f_{167}$   &    3313.5855  &     0.0123  &        301.7879  &     0.0011  &    14.68  &    2.22  &    6.6  &   4   &    &  $p$  &  ...  \\ 
$f_{90}$   &    4318.7076  &     0.0067  &        231.5508  &     0.0004  &    30.30  &    2.52  &   12.0  &      &    &  $p$  & AFM  \\ 
$f_{89}$   &    4320.7351  &     0.0068  &        231.4421  &     0.0004  &    30.48  &    2.56  &   11.9  &      &    &  $p$  & AFM  \\ 
$f_{171}$   &    4326.7505  &     0.0151  &        231.1203  &     0.0008  &    14.44  &    2.70  &    5.4  &      &    &  $p$  &  ...  \\ 
$f_{122}$   &    4337.3898  &     0.0105  &        230.5534  &     0.0006  &    20.80  &    2.69  &    7.7  &      &    &  $p$  &  ...  \\ 
$f_{113}$   &    4340.1080  &     0.0099  &        230.4090  &     0.0005  &    22.18  &    2.70  &    8.2  &      &    &  $p$  &  ...  \\ 
$f_{151}$   &    4362.3907  &     0.0134  &        229.2321  &     0.0007  &    15.86  &    2.63  &    6.0  &      &    &  $p$  &  ...  \\ 
$f_{123}$   &    4371.2133  &     0.0095  &        228.7694  &     0.0005  &    20.78  &    2.44  &    8.5  &      &    &  $p$  &  ...  \\ 
$f_{184}$   &    6042.4267  &     0.0150  &        165.4964  &     0.0004  &    11.27  &    2.09  &    5.4  &      &    &  $p$  &  ...  \\ 
$f_{182}$   &    7505.8215  &     0.0152  &        133.2299  &     0.0003  &    11.30  &    2.12  &    5.3  &      &    &  $p$  &  ...  \\ 
% $f_{178}$   &    7930.6988  &     0.0120  &        126.0923  &     0.0002  &    13.90  &    2.06  &    6.7  &      &    & $p$   &  ...  \\ 
$f_{183}$   &    8025.0926  &     0.0152  &        124.6092  &     0.0002  &    11.29  &    2.12  &    5.3  &      &    &  $p$  &  ...  \\ 
$f_{186}$   &    8308.3542  &     0.0150  &        120.3608  &     0.0002  &    10.85  &    2.00  &    5.4  &      &    &  $p$  &  ...  \\ 
\hline
\multicolumn{12}{c}{Orbitial information}  \\
\hline 
$f_{3}$   &      20.3085  &     0.0014  &      49240.4255  &     3.4977  &   438.78   &    7.81  &   56.2  &      &    &    &  AFM  \\ 
$f_{8}$   &      40.6194  &     0.0022  &      24618.7763  &     1.3508  &   272.22  &    7.49  &   36.4  &      &    &    & AFM  \\ 
\hline 
\multicolumn{12}{c}{Combination frequencies}  \\
\hline 
$f_{52}$   &      98.6834  &     0.0109  &      10133.4125  &     1.1190  &    53.64  &    7.21  &    7.4  &      &    &    & $f_{4}-f_{37}$  \\ 
$f_{29}$   &     130.8713  &     0.0074  &       7641.0956  &     0.4302  &    86.05  &    7.82  &   11.0  &      &    &    &  $f_{90}/33$  \\ 
$f_{59}$   &     251.7220  &     0.0079  &       3972.6357  &     0.1240  &    49.37  &    4.79  &   10.3  &      &    &    & $f_{8}+f_{2}$  \\ 
$f_{95}$   &     257.2430  &     0.0134  &       3887.3742  &     0.2023  &    28.55  &    4.71  &    6.1  &      &    &    & $f_{36}-f_{31}$  \\ 
$f_{100}$   &     289.0481  &     0.0122  &       3459.6319  &     0.1458  &    25.68  &    3.86  &    6.7  &      &    &    & $f_{49}+f_{7}$  \\ 
$f_{48}$   &     346.2998  &     0.0042  &       2887.6715  &     0.0354  &    59.60  &    3.12  &   19.1  &      &    &    & $f_{32}+f_{11}$  \\ 
$f_{76}$   &     347.4153  &     0.0067  &       2878.3998  &     0.0557  &    37.55  &    3.11  &   12.1  &      &    &    & $f_{37}+f_{1}$  \\ 
$f_{124}$   &     368.0170  &     0.0113  &       2717.2658  &     0.0834  &    20.77  &    2.89  &    7.2  &      &    &    & $f_{55}+f_{10}$  \\ 
$f_{75}$   &     400.6151  &     0.0059  &       2496.1617  &     0.0370  &    37.83  &    2.77  &   13.7  &      &    &    & $f_{27}+f_{46}$  \\ 
$f_{77}$   &     414.9698  &     0.0058  &       2409.8139  &     0.0340  &    37.47  &    2.70  &   13.9  &      &    &    & $f_{58}+f_{43}$  \\ 
$f_{109}$   &     550.7882  &     0.0087  &       1815.5798  &     0.0287  &    22.73  &    2.44  &    9.3  &     &     &    & $f_{95}+f_{47}$  \\ 
$f_{112}$   &     597.7314  &     0.0090  &       1672.9921  &     0.0252  &    22.26  &    2.47  &    9.0  &      &    &    & $f_{51}+f_{76}$  \\ 
\hline
\end{tabular}
\end{center}
% \tablefoot{
% AM/FM/AFM indicates that the frequency has modulation of amplitude~(AM), frequency~(FM) or both~(AFM). '*' means the S/N of the mode is below the threshold limit.
% }
\end{table*}

\addtocounter{table}{-1} 
\begin{table*} \caption[]{continued.}
\renewcommand{\arraystretch}{1.2}%设置行高
% \small
\begin{center}
\begin{tabular}{lccccccrcccc}
\hline
\hline
{ID} & {Frequency} & {$\sigma$f} & {Period} & {$\sigma$P} & {Amplitude} & {$\sigma$A}  & {S/N} & {$\ell$} & {$m$} & {{Mode}} & {Comments}  \\
{}  & {($\mu$Hz)} & {($\mu$Hz)} & {(s)} & {(s)} 
& {(ppm)} & {(ppm)} & {}& {} & {} & {}& {}\\             
\hline
$f_{119}$   &     639.2478  &     0.0098  &       1564.3386  &     0.0240  &    21.09  &    2.56  &    8.2  &      &    &    & $f_{86}-f_{59}$  \\ 
$f_{150}$   &     645.5340  &     0.0130  &       1549.1051  &     0.0312  &    16.07  &    2.58  &    6.2  &      &    &    & $f_{136}-f_{77}$  \\ 
$f_{137}$   &     649.6020  &     0.0112  &       1539.4041  &     0.0266  &    18.29  &    2.53  &    7.2  &      &    &    & $f_{118}-f_{4}$  \\ 
$f_{148}$   &     678.8064  &     0.0124  &       1473.1740  &     0.0269  &    16.32  &    2.49  &    6.5  &      &    &    & $f_{117}-f_{96}$  \\ 
$f_{136}$   &     680.4659  &     0.0111  &       1469.5814  &     0.0239  &    18.35  &    2.50  &    7.3  &      &    &    & $f_{108}-f_{1}$  \\ 
$f_{144}$   &     681.4268  &     0.0119  &       1467.5091  &     0.0255  &    17.01  &    2.49  &    6.8  &      &    &    & $f_{71}-f_{1}$  \\ 
$f_{132}$   &     718.5718  &     0.0108  &       1391.6494  &     0.0208  &    19.61  &    2.60  &    7.5  &      &    &    & $f_{117}-f_{121}$  \\ 
$f_{164}$   &     721.2262  &     0.0141  &       1386.5276  &     0.0271  &    14.88  &    2.59  &    5.8  &      &    &    & $f_{20}+f_{99}$  \\ 
$f_{154}$   &     734.2327  &     0.0136  &       1361.9661  &     0.0252  &    15.67  &    2.63  &    6.0  &      &    &    & $f_{52}+f_{104}$  \\ 
$f_{153}$   &     740.1387  &     0.0137  &       1351.0981  &     0.0250  &    15.68  &    2.65  &    5.9  &      &    &    & $f_{61}+f_{147}$  \\ 
$f_{140}$   &     745.0942  &     0.0119  &       1342.1122  &     0.0214  &    18.08  &    2.65  &    6.8  &      &    &    & $f_{55}+f_{104}$  \\ 
$f_{155}$   &     750.5260  &     0.0135  &       1332.3989  &     0.0240  &    15.67  &    2.61  &    6.0  &      &    &    & $f_{108}-f_{49}$  \\ 
$f_{149}$   &     753.2674  &     0.0130  &       1327.5499  &     0.0230  &    16.14  &    2.60  &    6.2  &      &    &    & $f_{40}+f_{109}$  \\ 
$f_{157}$   &     759.7478  &     0.0137  &       1316.2263  &     0.0238  &    15.53  &    2.63  &    5.9  &      &    &    & $f_{91}-f_{133}$  \\ 
$f_{168}$   &     792.9694  &     0.0144  &       1261.0827  &     0.0229  &    14.66  &    2.60  &    5.6  &      &    &    & $f_{54}+f_{137}$  \\ 
$f_{169}$   &     830.0562  &     0.0145  &       1204.7377  &     0.0211  &    14.58  &    2.61  &    5.6  &      &    &    & $f_{26}+f_{88}$  \\ 
$f_{83}$   &     831.4640  &     0.0061  &       1202.6979  &     0.0089  &    34.81  &    2.63  &   13.2  &      &    &    & $6f_{61}$  \\ 
$f_{126}$   &     851.2135  &     0.0100  &       1174.7934  &     0.0139  &    20.57  &    2.55  &    8.1  &      &    &    & $f_{93}-f_{3}$  \\ 
$f_{118}$   &     904.2540  &     0.0101  &       1105.8840  &     0.0123  &    21.18  &    2.63  &    8.1  &      &    &    & $f_{91}-f_{36}$  \\ 
$f_{172}$   &     992.8853  &     0.0149  &       1007.1657  &     0.0151  &    14.36  &    2.64  &    5.4  &      &    &    & $f_{12}+f_{149}$  \\ 
$f_{147}$   &    1032.9838  &     0.0126  &        968.0694  &     0.0118  &    16.58  &    2.59  &    6.4  &      &    &    & $f_{139}-f_{105}$  \\ 
$f_{143}$   &    1080.2154  &     0.0113  &        925.7413  &     0.0097  &    17.57  &    2.45  &    7.2  &      &    &    & $f_{76}+f_{68}$  \\ 
$f_{160}$   &    1083.4305  &     0.0132  &        922.9941  &     0.0113  &    15.13  &    2.47  &    6.1  &      &    &    & $f_{69}+f_{154}$  \\ 
$f_{141}$   &    1109.5521  &     0.0120  &        901.2646  &     0.0098  &    17.77  &    2.64  &    6.7  &      &    &    & $f_{37}+f_{92}$  \\ 
$f_{158}$   &    1172.7899  &     0.0143  &        852.6676  &     0.0104  &    15.33  &    2.71  &    5.7  &      &    &    & $f_{91}-f_{16}$  \\ 
$f_{161}$   &    1178.0210  &     0.0145  &        848.8813  &     0.0104  &    15.13  &    2.70  &    5.6  &      &    &    & $29f_{8}$  \\ 
$f_{152}$   &    1377.2636  &     0.0133  &        726.0774  &     0.0070  &    15.70  &    2.58  &    6.1  &      &    &    & $f_{12}+f_{134}$  \\ 
$f_{166}$   &    1429.0765  &     0.0142  &        699.7526  &     0.0070  &    14.83  &    2.60  &    5.7  &      &    &    & $f_{38}+f_{166}$  \\ 
$f_{162}$   &    1499.6580  &     0.0149  &        666.8187  &     0.0066  &    15.11  &    2.78  &    5.4  &      &    &    & $f_{33}+f_{128}$  \\ 
$f_{130}$   &    1555.0277  &     0.0113  &        643.0754  &     0.0047  &    19.77  &    2.75  &    7.2  &      &    &    & $f_{88}+f_{121}$  \\ 
$f_{138}$   &    1577.1092  &     0.0123  &        634.0715  &     0.0050  &    18.26  &    2.78  &    6.6  &      &    &    & $f_{88}+f_{108}$  \\ 
$f_{139}$   &    1584.0543  &     0.0125  &        631.2915  &     0.0050  &    18.22  &    2.81  &    6.5  &      &    &    & $78f_{3}$  \\ 
$f_{165}$   &    1601.1127  &     0.0150  &        624.5656  &     0.0059  &    14.83  &    2.75  &    5.4  &      &    &    & $14f_{63}$  \\ 
$f_{159}$   &    1623.8112  &     0.0142  &        615.8351  &     0.0054  &    15.20  &    2.66  &    5.7  &      &    &    & $f_{155}+f_{96}$  \\ 
$f_{156}$   &    1687.8814  &     0.0130  &        592.4587  &     0.0046  &    15.60  &    2.50  &    6.2  &      &    &    & $f_{68}+f_{72}$  \\ 
$f_{173}$   &    1694.5446  &     0.0142  &        590.1290  &     0.0050  &    14.30  &    2.51  &    5.7  &      &    &    & $f_{158}+f_{164}$  \\ 
$f_{116}$   &    1749.1470  &     0.0096  &        571.7073  &     0.0031  &    21.80  &    2.57  &    8.5  &      &    &    & $f_{5}+f_{107}$  \\ 
$f_{121}$   &    3282.5175  &     0.0089  &        304.6442  &     0.0008  &    20.85  &    2.29  &    9.1  &      &    &    & $39f_{74}$  \\ 
% $f_{165}$   &    4363.3811  &     0.0142  &        229.1801  &     0.0007  &    14.90  &    2.61  &    5.7  &      &    &    & $f_{119}-f_{27}$  \\ 
% $f_{119}$   &    4531.8317  &     0.0077  &        220.6613  &     0.0004  &    21.13  &    2.01  &   10.5  &      &    &    & $f_{2}+f_{89}$  \\ 
$f_{181}$   &    7553.0399  &     0.0145  &        132.3970  &     0.0003  &    11.56  &    2.07  &    5.6  &      &    &    & $22f_{69}$  \\ 
\hline 
\end{tabular}
\end{center}
\tablefoot{
Column 1 shows the  ID in order of decreasing amplitude; columns 2 and 3 give the frequencies in $\mu$Hz and errors; columns 4 and 5 give the periods in seconds and errors; columns 6 and 7 show the amplitudes in ppm (parts per million) and the errors; column 8 gives the (S/N) level; columns 9 and 10 show the quantum number identified by the asymptotic regime and rotation (see Sects.~\ref{sec:RoMul} and \ref{sec:Period_Spacing});{ column 11 shows the preliminary classification of modes (see Sect.~\ref{sec:Fre_extra_class})} and column 12 comments on whether amplitude or frequency modulations were measured or not.
AM/FM/AFM indicates that the frequency has modulations of amplitude~(AM), frequency~(FM) or both~(AFM) and 'S' stands for stable. '*' in $\ell$ and $n$ means they are identified by period spacing or the \'Echelle diagram and in ID means the S/N of the frequency is below the significant value of 5.2, but the frequency locates the position of a rotational component. }
\end{table*}

\FloatBarrier

\section{Period spacing}\label{sec:Period_Spacing}
\begin{figure*}[t]
\centering
\includegraphics[width=\textwidth]{ 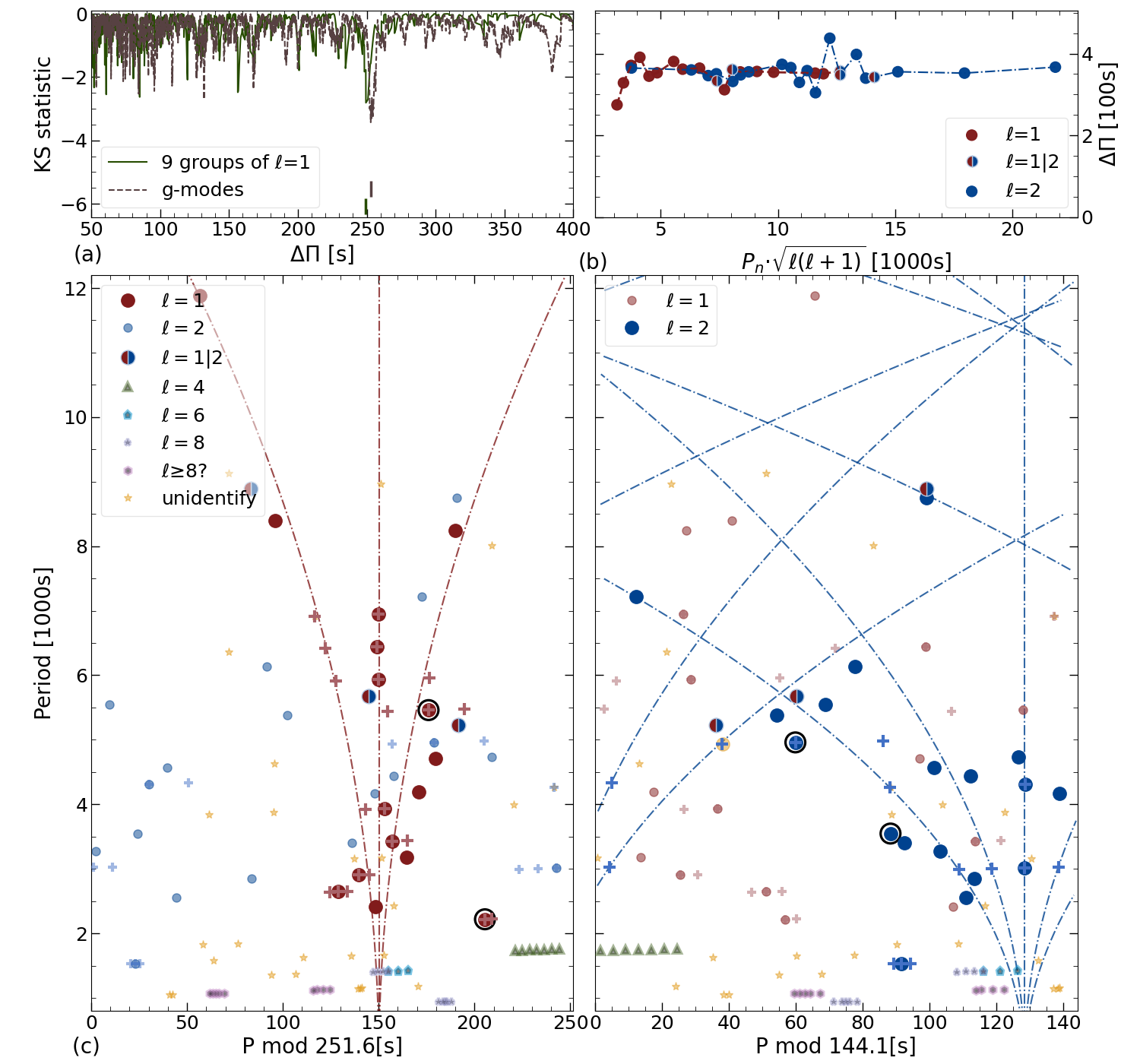}
\caption{Period spacing and mode identification for independent {\sl g}-modes. (a) Kolmogorov-Smirnov (KS) tests on two period sets with $\ell = 1$. Two vertical segments locate the minimum values for preliminary spacing.
(b) Period difference between consecutive modes of the $\ell = 1$ and $\ell = 2$ sequences, after converting to reduced periods.
(c) \'Echelle diagram for dipole (left) and quadruple (right) modes. The vertical curves represent the corresponding period for rotational components, with 3 and 5 for triplet and quintuplet, respectively.
The red and blue + symbols represent the splitting frequency components for $\ell=1$ and $\ell=2$, respectively.
% The mode marked with white '$\times$' may be a trapped mode.
% gets a larger period deviation 
And the yellow cycle indicates the mode in $\ell = 2$ with a larger deviation in frequency splitting values.
The modes surrounded by the solid circles are possibly identified to be trapped modes.}
\label{fig:PS}
\end{figure*}

Another technique that helps mode identification relies on the seismic property of periods of {\sl g}-modes or frequency of {\sl p}-modes. Here, we only analyzed the property of {\sl g}-modes as high-frequency {\sl p}-modes are only a few in {PG~0101+039}. In the asymptotic limit ($n \gg \ell$), consecutive overtones of {\sl g}-modes are nearly equally spaced in period \cite[see, e.g., ][]{2010aste.book.....A}.
The pulsation period of a given mode with degree $\ell$ and radial order $n$ can be expressed as $\Pi_{l,n} \approx  n \cdot \Pi_{0} /\sqrt{(l(l+1))} + \epsilon$ 
where $\Pi_0$ is the fundamental radial period and $\epsilon$ is an offset \citep{1979nos..book.....U}.
The period difference (commonly called period spacing) of two consecutive radial overtones with a given modal degree can be estimated as follows: 
\begin{equation}\label{eq:Ps2}
        \Delta \Pi_{l} \equiv \Pi_{l,n+1} - \Pi_{l,n} \approx \frac{\Pi_{0}}{\sqrt{(l(l+1))}}, 
\end{equation}
It is convenient to see that the period spacing of $\ell$ = 2 sequence is relative to $\ell$ = 1 sequence with $\Delta \Pi_{\ell=2} \approx \Delta\, \Pi_{\ell=1}/\sqrt{3}$.

Similar to \citet{2022ApJ...933..211M}, we performed the Kolmogorov-Smirnov (KS) test to identify the average spacing of {\sl g}-modes in {PG~0101+039}. This popular test returns spacing correlations as highly negative values for the most common spacing in a data set \citep{1988IAUS..123..329K}. As identified to be rotational dipole modes, the central components of nine $\ell=1$ multiplets were fed for KS test.
Then we applied the second KS test for all the independent frequencies lower than 1200~$\mu$Hz to be probable {\sl g}-modes. Figure\,\ref{fig:PS}(a) shows the results of the KS tests that the minimum values locate around 252.9~s and 249.4~s for the first and second tests, respectively.
Both values are consistent with that of dipole modes in many sdB pulsators \citep[see, e.g.,][]{ 2011MNRAS.414.2885R,2020MNRAS.495.2844S}.

We identified 21 frequencies to be dipole modes, including 9 central ($m=0$) components, based on the rotation splitting (see Fig.\ref{fig:RotaMult}, a). Similarly, 14 frequencies are identified to be quadruple modes, including four central ($m=0$) ones (see Fig.\ref{fig:RotaMult}(b)). In addition, 35 frequencies can be associated with high degree modes forming seven incomplete multiplets (see Fig.\ref{fig:RotaMult}(c-h)). As the high degree modes are very few, we do not provide any identified results to their radial orders, $n,$ and the azimuthal numbers, $m$, except the frequencies near 574~$\mu$Hz. Besides the above modes, we adopted a preliminary period spacing of 251\,s and 145\,s to identify the $\ell=1$ and $\ell=2$ modes, respectively, based on the KS test. If a mode is satisfied within the value of less than 35\,s and 25\,s to the radial order of the $\ell=1$ and $\ell=2$ sequences, respectively, we will classify it to that sequence. We then applied a linear fitting to the identified modes and derived the values of $\Delta \Pi_{\ell = 1} = 249.2 \pm 1.5$~s and $\Delta \Pi_{\ell = 2} = 146.2 \pm 1.7$~s, respectively.

To test the fitting results of the asymptotic period spacing, we then constructed the \'echelle diagrams for the identified modes. In order to give optimal diagrams, 
we change the period spacing with a step of 0.1~s relative to $\Pi_{\ell = 1} = 249.2 \pm 3 \times 1.5$~s and $\Delta \Pi_{\ell = 2} = 146.2 \pm 3 \times 1.7$~s, respectively. The optimal \'echelle diagrams are chosen by maximum the rotational components close to their calculated values. We finally obtained a period spacing of 251.6\,s and 144.1\,s for the $\ell=1$ and $\ell=2$ sequences, respectively. Figure\,\ref{fig:PS}\,(c) shows the \'echelle diagrams of the two sequences, where the most dipole modes locate very close to the calculated period. We observed five multiplets whose components are consistent with the rotational splittings. 
We assigned the $m$ value with the identified modes by their location to the closest rotational splitting. We identified 28 $\ell=1$ and 28 $\ell=2$ modes, while 3 may be related to both sequences. All identified modes are listed in columns~9-11 of Table\,\ref{Tab:freq} with $\ell$, $m,$ and relative $n$ values. We note that the real radial order $n$ can only be derived through seismic modeling. Figure \,\ref{fig:PS}\,(b) presents the deviation of period spacing as a function of the reduced pulsation period spacing. We observe a jitter pattern in the $\ell =1$ sequence, which is very common in many sdB stars \citep[see, e.g.,][]{2014A&A...569A..15O,2020MNRAS.495.2844S}. { We note that the \'echelle diagram does not show any clear pattern of ridge in period spacing. From a theoretical aspect, a young sdB does not develop the regularity in period spacing \citep{2014ASPC..481..179C}. PG~0101+039 locates the lower part of { g}-mode puslating sdB stars on the $\log g-T_\mathrm{eff}$ diagram, which 
indeed indicates that PG~0101+039 is a relatively young sdB star, { roughly estimated with an age about 50~Myr after the zero-age extreme horizontal branch} \citep{2002ApJS..140..469C,2012A&A...539A..12F}}. A more precise age can be foreseen in future modeling.

\section{AM and FM comparison from various stamped pixels}\label{sec:AFM_T}

\begin{figure*}[t]
\centering
\includegraphics[width=\textwidth]{ 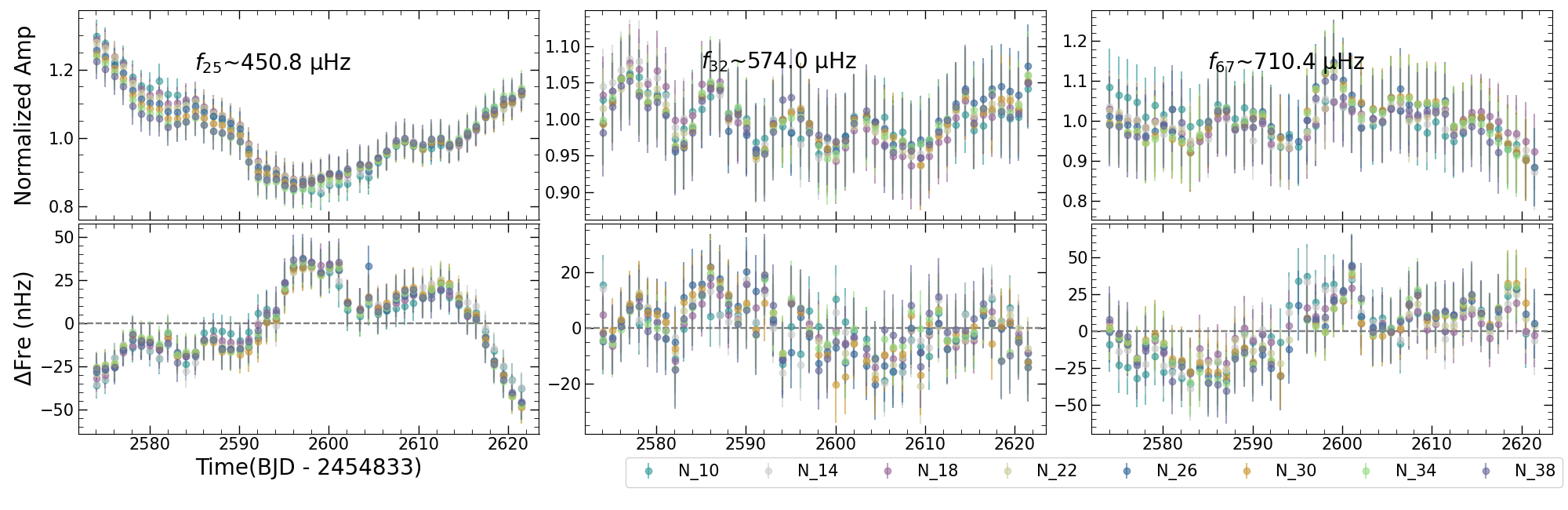}
\caption{Examples of three frequencies ($f\sim450.8\,\mu$Hz, $f\sim574.0\,\mu$Hz and $f\sim710.4\,\mu$Hz) as a comparison of the amplitude and frequency modulation in {PG~0101+039} from stamp-decisive photometry. The sizes of eight types of pixels are labeled in the outside legend.
Amplitudes are normalized to units and frequencies are shifted to their averages as represented by the dashed horizontal lines.}
\label{fig:AFM_test}
\end{figure*}

As we mention in Section~\ref{sec:APT}, the optimal photometry was extracted with a 12-pixel stamp to provide a good number of low-amplitude signals. However, whether the change of stamped size poses a challenge to destroy the intrinsic modulation patterns or not is critical to nonlinear calculation in the future. As documented in \citet{2021ApJ...921...37Z}, they measured somewhat different patterns of AM between the SAP and PDC-SAP photometry, in particular, a star suffers heavy light pollution from nearby stars. We thus evaluate a similar comparison of different AMs and FMs from the photometry extracted with different sizes of stamped pixels. This step is important for further analysis of the dynamics of oscillation modes in various pulsators from the {\sl K}2 photometry. During the process of searching for the optimal photometry, we recorded the light curves constructed by various stamps. In our comparison of AM and FM, we again adopted eight of those recording light curves to prewhiten frequencies piecewise in order to save computation time since AM and FM measurements consume the most computational resource. These eight light curves display stamps covering a fraction from 15\% to 60\% of the entire flux, with pixel sizes ranging from 10 to 38 with a step of four pixels. 
%这次我们开始每隔开四个像素取一次amfm的计算，然后我们得到了他们的结果。amf过程跟zong2018年提出的方法是一致的。

The AM and FM were measured from these piecewise light curves that are sliding along the eight recorded light curves with an identical time step of 1~d and window width of 30~d through a similar method to that of \citet{2022ApJ...933..211M} and initialed by \citet{2018ApJ...853...98Z}. Compared to EPIC~220422705, rotational components can be well resolved in {PG~0101+039} since the 30-d length covers more than three times the length of of its rotation period. We note that close peaks within the frequency resolution ($\sim$ 0.4 $\mu$Hz) are detected, we keep the weighted average peak as the measured value for that frequency. There are 35 measurements of AM and FM from the $8\times48$ constructed light curves. For direct comparison, amplitudes were normalized to their averages and frequencies were shifted to their averages. 
Figure \,\ref{fig:AFM_test} shows examples of three frequencies as representative results for our comparison. Those three frequencies present various kinds of AM and FM from relatively large scale to small. We clearly see that the patterns of AMs and FMs are consistent within the measuring uncertainties for all eight types of photometry in those three frequencies. For instance, all AM and FM of $f\sim450.8\,\mu$Hz follow nearly the same trend both in relatively short and long terms. The entire comparison, although not fully provided, gives a similar result to those of the representatives. We, therefore, conclude that both AM and FM patterns do not change when aperture size varies during the extraction of photometry. We note that this conclusion is only valid for targets that are free from contamination by nearby stars. With this test, we can proceed with our characterization of AM and FM in {PG~0101+039}, without any worries related to the photometric factor.

\section{AFMs in PG~0101+039}

\subsection{AFMs in multiplets}\label{sec:AFM_ap}

\begin{figure*}
\centering
\includegraphics[width=\textwidth]{ 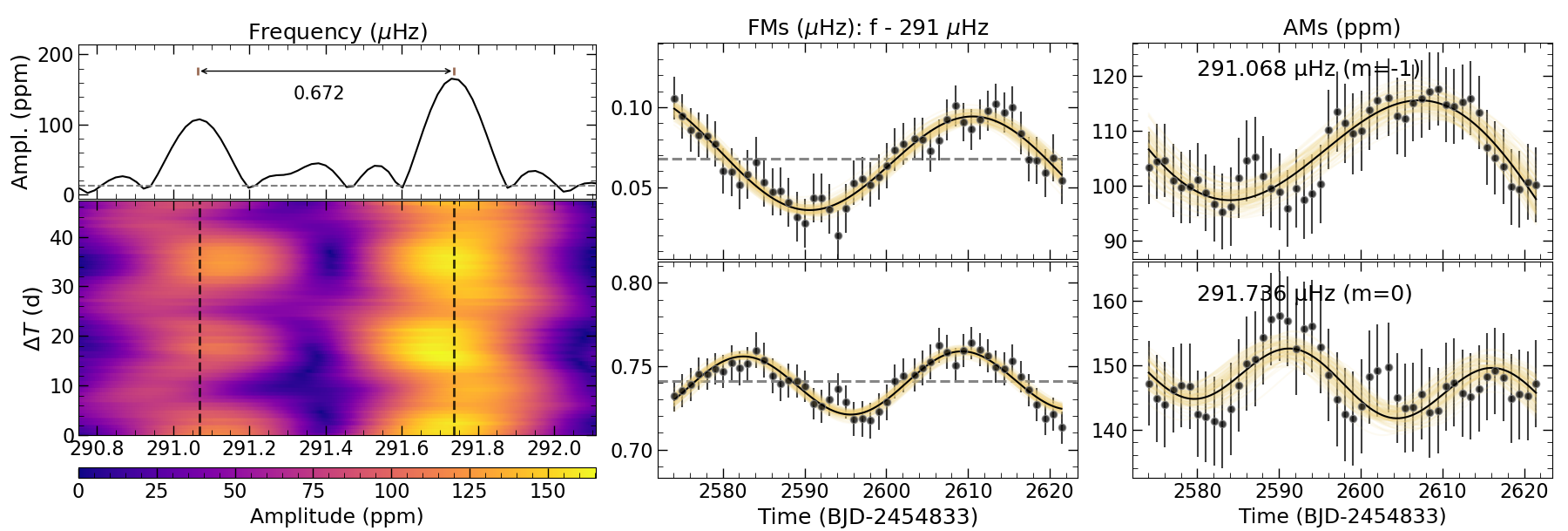}
\caption{Same as Fig.\,\ref{fig:Mul_167}, but for the g-mode doublet near 291\,$\mu$Hz.}
\label{fig:Mul_291}
\end{figure*}

Figure\,\ref{fig:Mul_291} presents the AMs and FMs of the two ($m=-1$ and $m=0$) components forming the $g$-mode doublet near 291\,$\mu$Hz. Both the two components exhibit periodic variations of AMs and FMs on timescales approximately of $30-40$~d. By merely comparing the sinusoidal part of the fitting $G_4$, the two components display similar scales in AM and FM of $20-30$~ppm and $5-10$~nano~Hz, respectively.

\begin{figure*}
\centering
\includegraphics[width=\textwidth]{ 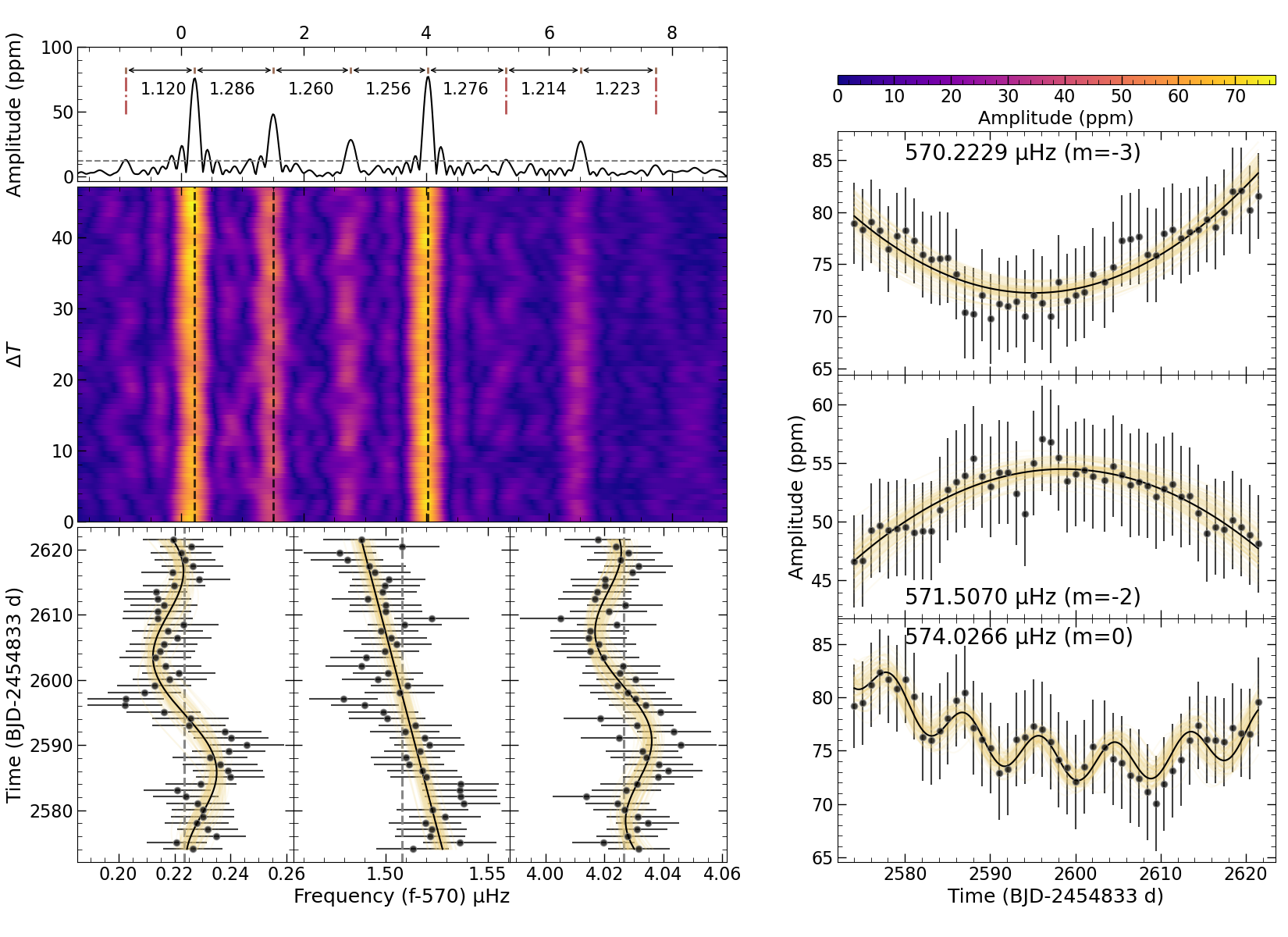}
\caption{Similar to Fig.\,\ref{fig:Mul_377}, but for the $\ell = 4$ g-mode multiplet near 570\,$\mu$Hz. We note that at least four components are missing in this multiplet and the brown vertical dashed lines indicate the expected position for three of them}
\label{fig:Mul_570}
\end{figure*}

\begin{figure*}
\centering
\includegraphics[width=\textwidth]{ 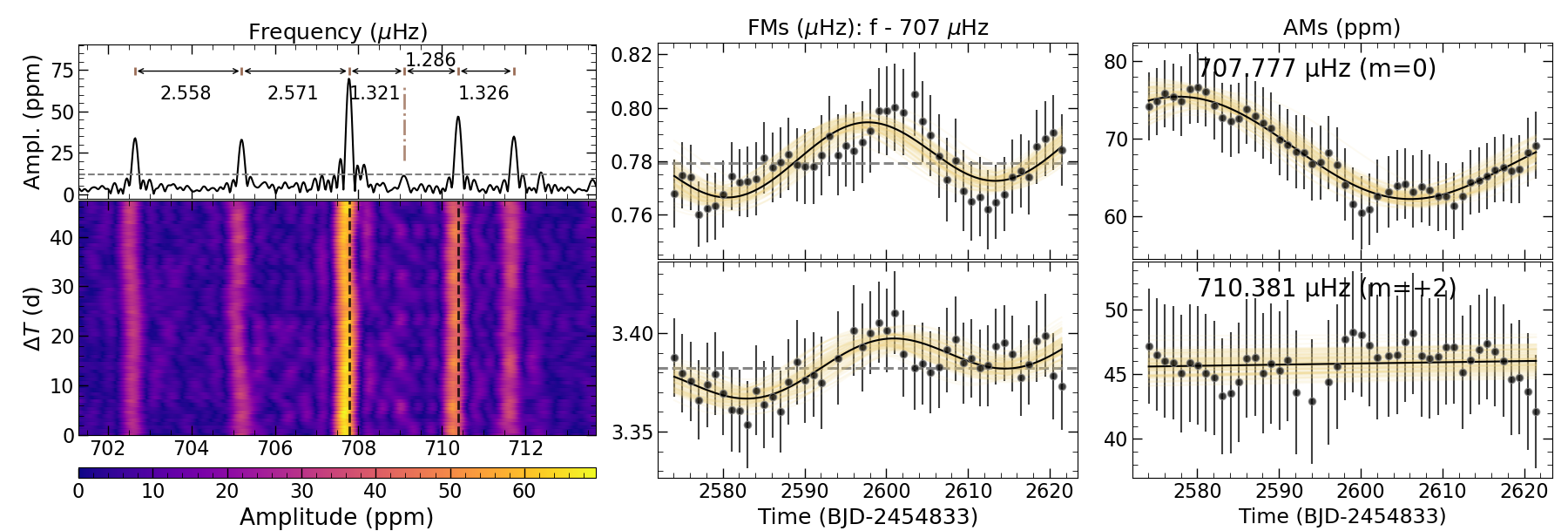}
\caption{Similar to Fig.\,\ref{fig:Mul_167}, but for the $\ell =6$ and $8$ g-mode multiplets near 707\,$\mu$Hz. 
}
\label{fig:Mul_707}
\end{figure*}

Figure\,\ref{fig:Mul_570} displays the AMs and FMs occurring in three components that form the incomplete high degree ($\ell = 4$) multiplet near 570\,$\mu$Hz. 
From the sliding LSP, the five components seem stable in amplitude and frequency, whereas the others cannot be distinguished from the noise. Due to the low amplitude in this multiplet, we only measured AMs and FMs for three components ($m=-3, m=-2, m=0$) above our current criteria as revealed by piecewise measurements. With this precision in amplitude and frequency, the behaviors of AMs and FMs have been observed to be variable throughout the entire observation. However, the varying scales both in amplitude and frequency are comparable to their measuring uncertainties of a few tenths of micro Hz or a few ppm. For instance, the FM of the ($m=-2$) component near $f\sim571.5\,\mu$Hz is a $G_1$ fitting but seems periodic if the uncertainties were much smaller.
The AMs of $m=-3$ and $m=-2$ evolve nearly in an anti-phase with a parabolic behavior and fitted with a $G_2$ function. However, the frequencies of $m=-3$ and $m=0$ seem to evolve in phase with a quasi-periodic behavior on a timescale of approximately 30\,days. We note that the AM of the $m=0$ component shows a periodic pattern on a quite short timescale of $\sim9$\,days. However,  we also note that this periodic AM has a scale, namely, $\sim 2$~ppm, smaller than the amplitude uncertainties. Similarly, Fig.\,\ref{fig:Mul_707} illustrates the AMs and FMs of two components in the $\ell > 3$ multiplet near 707\,$\mu$Hz. Both the two FMs exhibit a periodic pattern with a similar timescale of $\sim30$\,days, and a modulating scale of $10-20$ nano Hz. The AMs of $f_{67}\sim710.4~\mu$Hz seems relatively stable during the observation, whereas the other exhibits quasi-periodic AMs with a scale of $\sim10$\,ppm.


\subsection{The other AFMs}\label{Ap:AFM_L}
\clearpage
\begin{figure*}
\centering
\includegraphics[width=0.2693\textwidth]{ 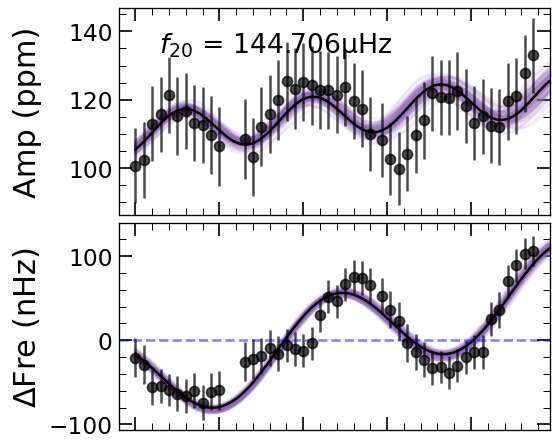}
\includegraphics[width=0.234\textwidth]{ 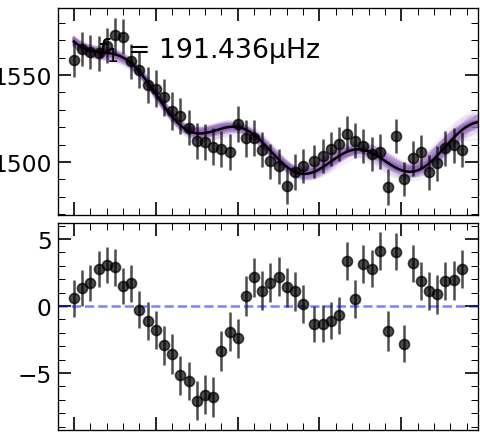}
\includegraphics[width=0.234\textwidth]{ 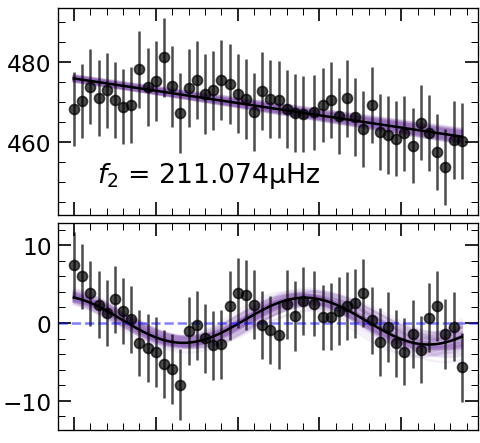}
\includegraphics[width=0.234\textwidth]{ 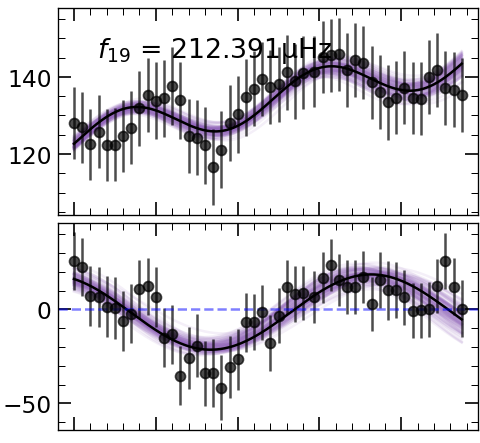}

\includegraphics[width=0.2693\textwidth]{ 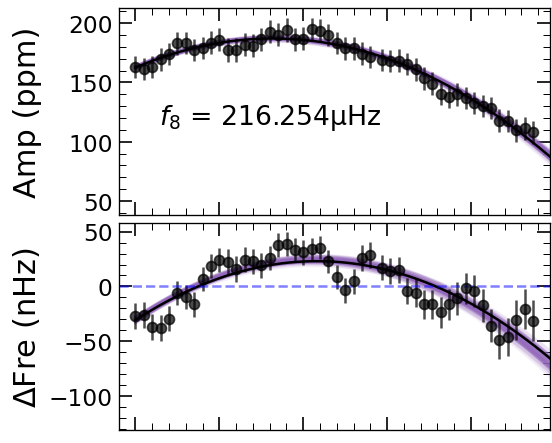}
\includegraphics[width=0.234\textwidth]{ 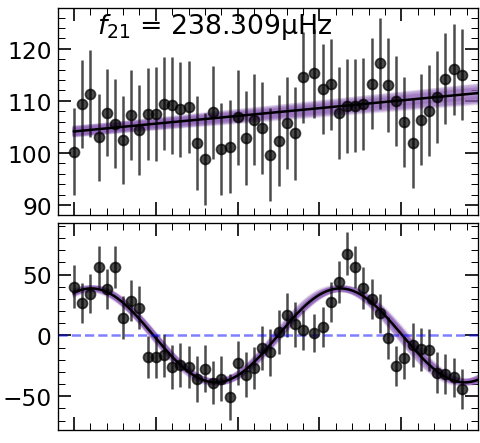}
\includegraphics[width=0.234\textwidth]{ 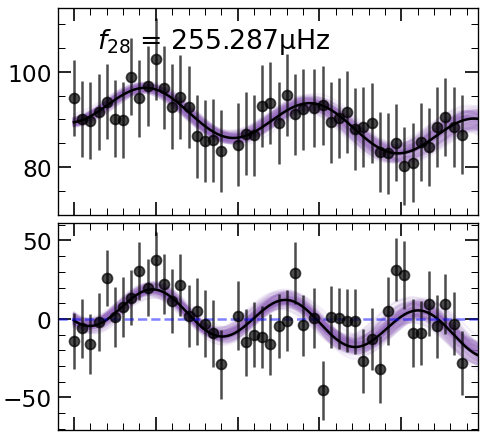}
\includegraphics[width=0.234\textwidth]{ 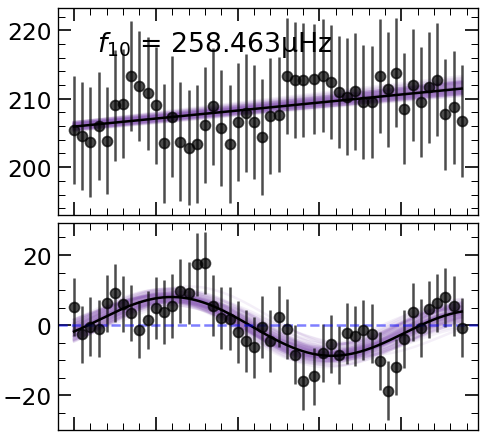}

\includegraphics[width=0.2693\textwidth]{ 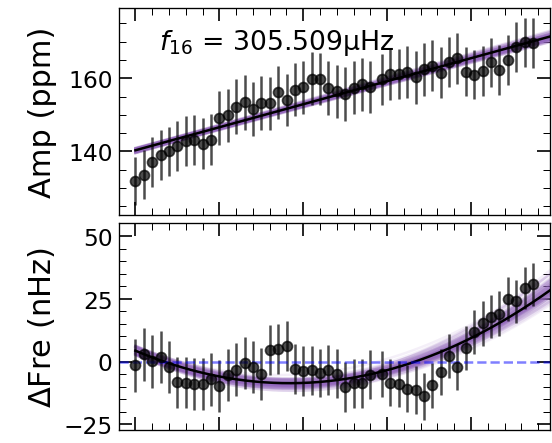}
\includegraphics[width=0.234\textwidth]{ 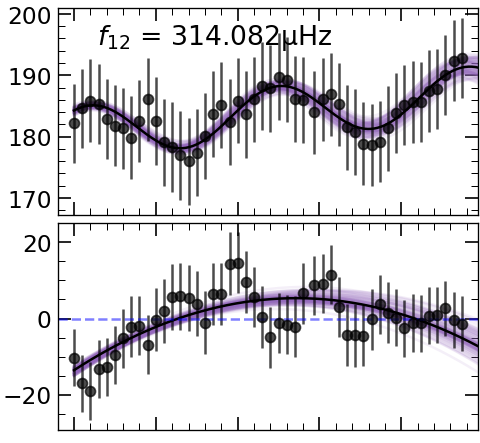}
\includegraphics[width=0.234\textwidth]{ 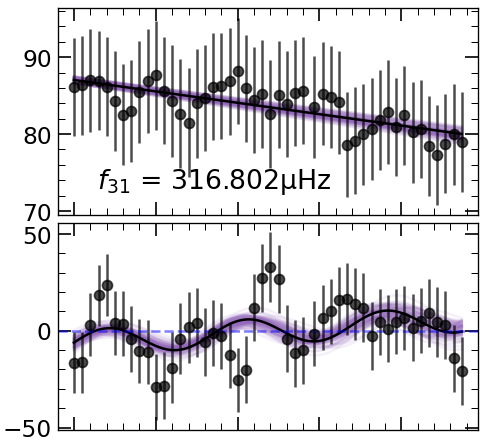}
\includegraphics[width=0.234\textwidth]{ 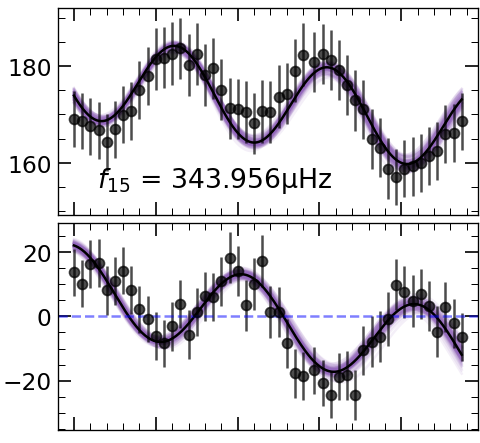}

\includegraphics[width=0.2693\textwidth]{ 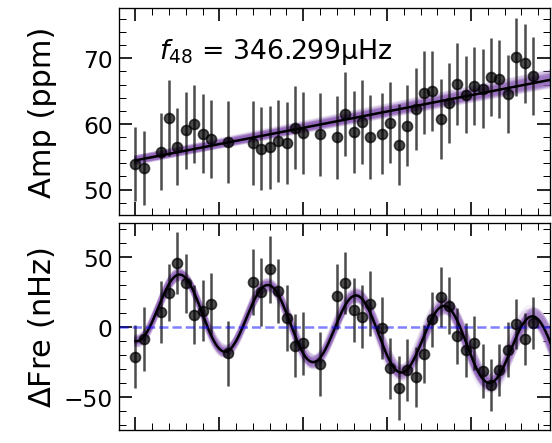}
\includegraphics[width=0.234\textwidth]{ 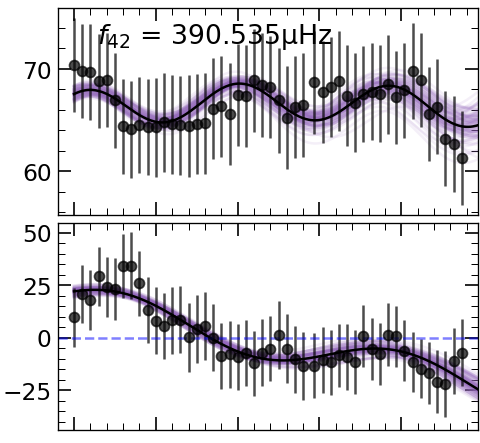}
\includegraphics[width=0.234\textwidth]{ 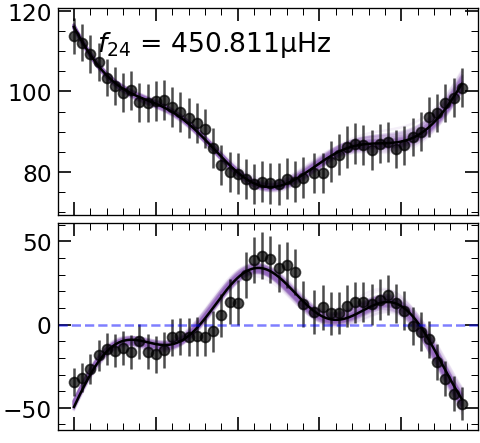}
\includegraphics[width=0.234\textwidth]{ 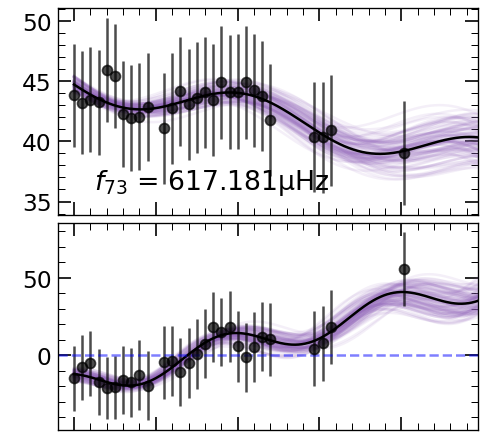}

\includegraphics[width=0.2693\textwidth]{ 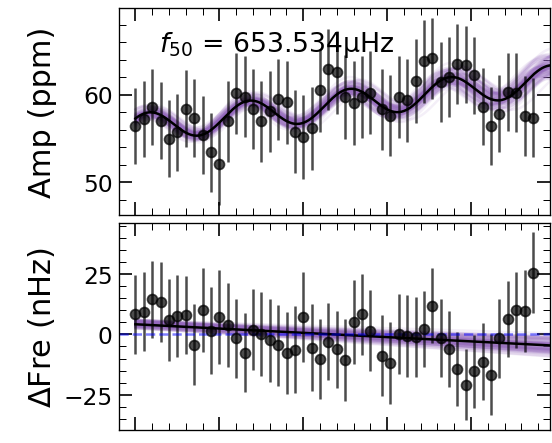}
\includegraphics[width=0.234\textwidth]{ 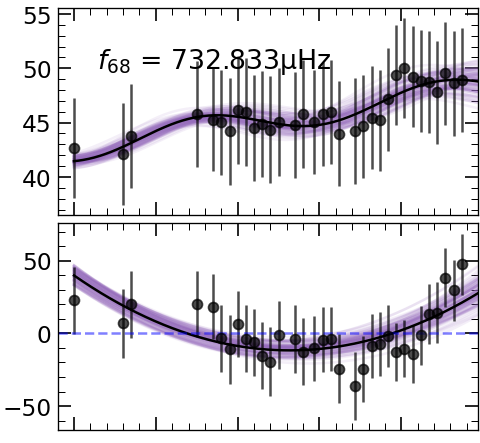}
\includegraphics[width=0.234\textwidth]{ 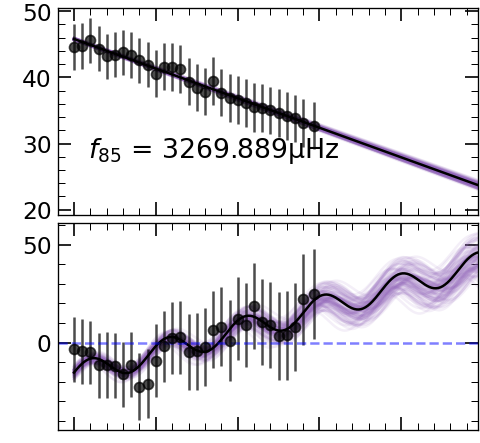}
\includegraphics[width=0.234\textwidth]{ 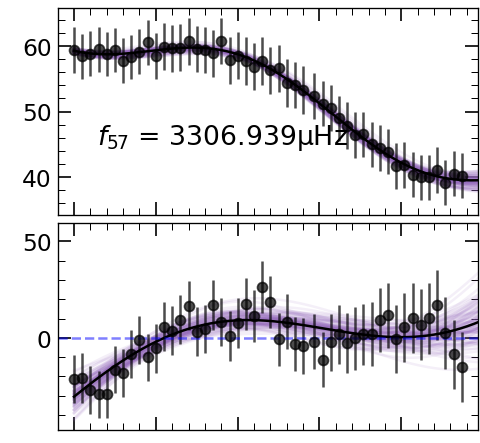}

\includegraphics[width=0.2693\textwidth]{ 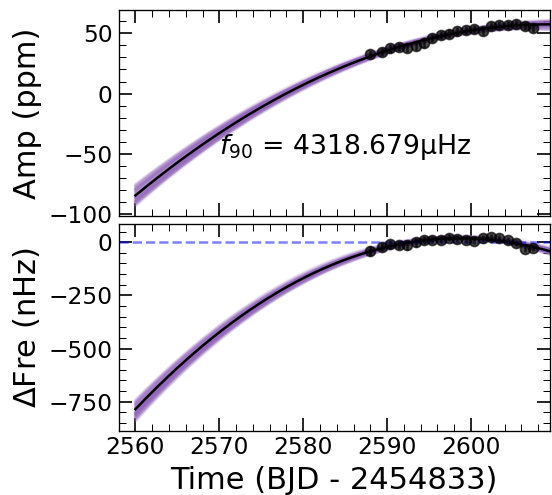}
\includegraphics[width=0.234\textwidth]{ 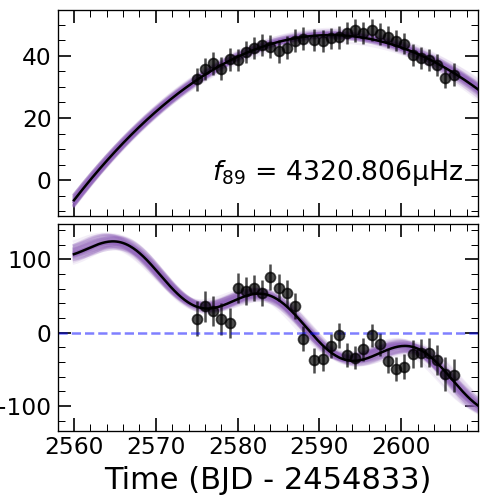}
\includegraphics[width=0.234\textwidth]{ 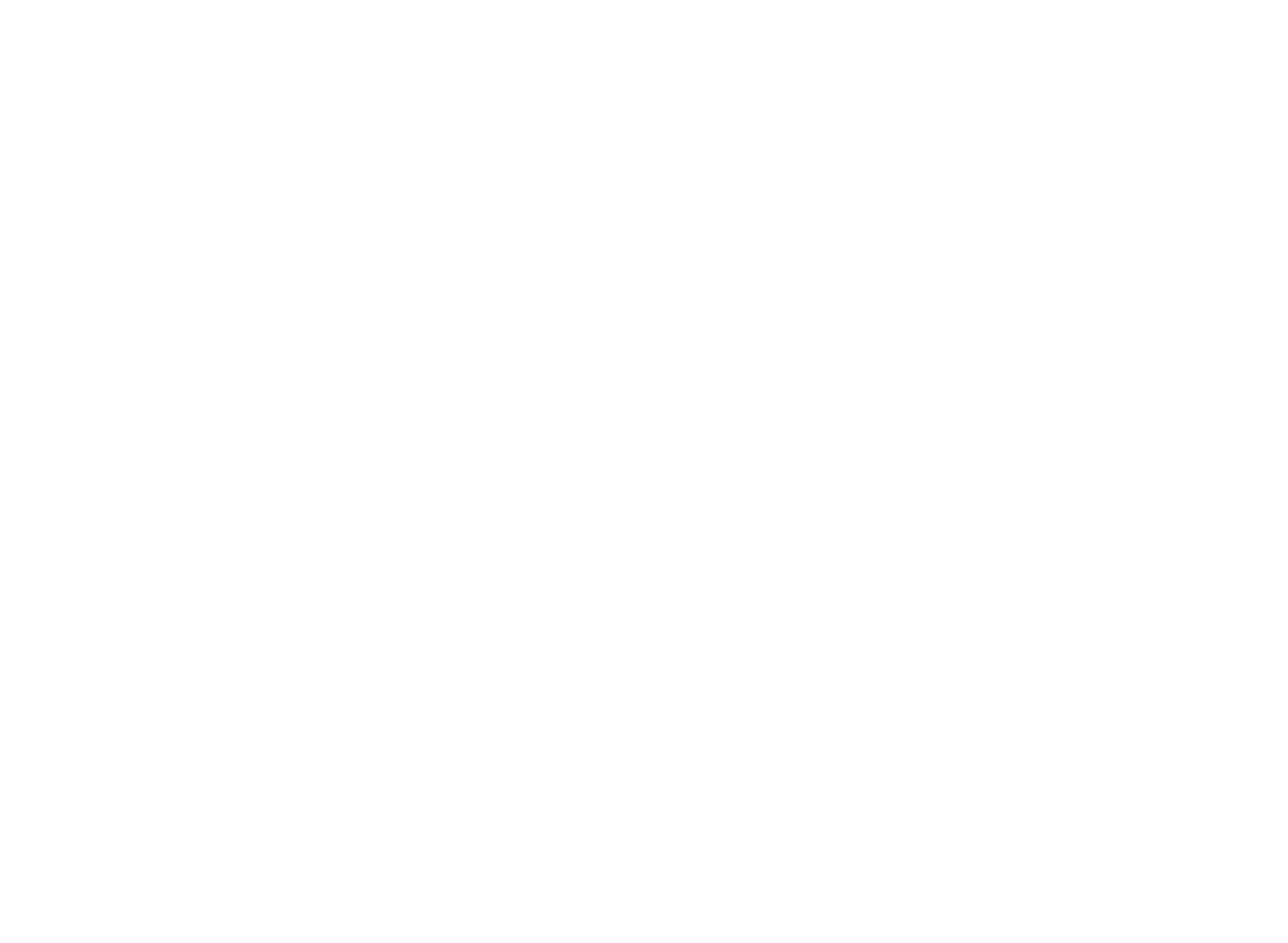}
\includegraphics[width=0.234\textwidth]{ white.pdf}
\caption{Gallery of 35 typical frequencies with AM/FM variations. Each module refers to the AM (top panel) and FM (bottom panel) variations of one single frequency. The frequencies are shifted to the average values as represented by the dashed horizontal lines, which are indicated in the bottom panel. The solid curves in purple and black represent the fitting results from the {\texttt MCMC} method and the optimal fitting, respectively.\label{fig:AF_L}
}
\end{figure*}

\FloatBarrier

\section{Radial velocity information}
\label{AP:Spectrum}
\begin{table}
\caption{Journal of 50 spectra of LAMOST DR9 and their Radial-velocity (RV) measurements presented in this paper.}    
\label{table:spe}     
\centering         
\begin{tabular}{l|cccc}     
\hline
\hline 
Obs. date & BJD & RV & RV error & S/N  \\
\hline     
\multirow{6}{*}{2018-10-28} & 3587.1122 & -25.62 & 2.92 & 15.5 \\
            & 3587.1292 & -2.07 & 2.02 & 13.5 \\
            & 3587.1459 & 21.70 & 3.35 & 14.3 \\
            & 3587.1624 & 24.90 & 3.09 & 15.9 \\
            & 3587.1788 & 48.67 & 4.31 & 13.5 \\
            & 3587.1954 & 57.68 & 2.63 & 15.7 \\
\hline
\multirow{5}{*}{2019-11-03} & 3958.1088 & -37.57 & 0.51 & 81.4 \\
            & 3958.1250 & -22.09 & 0.62 & 85.6 \\
            & 3958.1412 & -7.53 & 0.56 & 75.0 \\
            & 3958.1574 & 9.75 & 0.73 & 70.3 \\
            & 3958.1736 & 27.71 & 0.78 & 69.3 \\
\hline
\multirow{6}{*}{2019-11-08} & 3963.0716 & -81.37 & 0.48 & 73.0 \\
            & 3963.0879 & -90.14 & 0.67 & 64.2 \\
            & 3963.1042 & -97.06 & 0.83 & 63.4 \\
            & 3963.1204 & -98.89 & 0.87 & 65.1 \\
            & 3963.1366 & -97.61 & 0.63 & 66.3 \\
            & 3963.1529 & -91.85 & 0.46 & 66.0 \\
\hline
\multirow{1}{*}{2019-11-11} & 3966.0807 & -32.19 & 3.35 & 16.8 \\
\hline
\multirow{5}{*}{2019-11-13} & 3968.0821 & 46.57 & 1.04 & 47.0 \\
            & 3968.0983 & 28.85 & 0.73 & 53.4 \\
            & 3968.1145 & 6.61 & 1.03 & 50.0 \\
            & 3968.1308 & -12.87 & 0.97 & 50.4 \\
            & 3968.1471 & -33.32 & 0.91 & 44.4 \\
\hline
\multirow{4}{*}{2019-11-18} & 3973.0914 & 109.92 & 0.71 & 64.6 \\
            & 3973.1076 & 111.94 & 0.66 & 66.3 \\
            & 3973.1238 & 109.69 & 0.58 & 61.7 \\
            & 3973.1400 & 105.23 & 0.84 & 58.4 \\
\hline
\multirow{5}{*}{2019-11-19} & 3974.0739 & -20.39 & 0.54 & 79.2 \\
            & 3974.0901 & -3.59 & 0.65 & 78.0 \\
            & 3974.1064 & 15.86 & 0.93 & 77.3 \\
            & 3974.1226 & 32.46 & 0.64 & 77.9 \\
            & 3974.1380 & 47.85 & 0.86 & 71.1 \\
\hline
\multirow{1}{*}{2019-12-02} & 3987.0117 & -86.97 & 0.76 & 79.4 \\
\hline
\multirow{9}{*}{2019-12-04} & 3988.9611 & 53.54 & 0.91 & 68.0 \\
            & 3988.9774 & 69.65 & 0.44 & 67.8 \\
            & 3988.9936 & 82.86 & 0.67 & 73.2 \\
            & 3989.0098 & 93.37 & 0.64 & 73.4 \\
            & 3989.0261 & 102.53 & 0.65 & 70.8 \\
            & 3989.0423 & 108.69 & 0.61 & 67.1 \\
            & 3989.0585 & 110.27 & 0.68 & 65.2 \\
            & 3989.0748 & 112.07 & 0.56 & 60.8 \\
            & 3989.0910 & 107.24 & 1.04 & 54.3 \\
\hline
\multirow{8}{*}{2019-12-12}& 3996.9492 & 61.67 & 1.18 & 45.2 \\
            & 3996.9655 & 75.38 & 0.74 & 49.7 \\
            & 3996.9817 & 87.61 & 0.76 & 48.2 \\
            & 3996.9979 & 96.45 & 0.98 & 45.5 \\
            & 3997.0141 & 103.78 & 0.62 & 46.5 \\
            & 3997.0304 & 109.10 & 0.46 & 53.9 \\
            & 3997.0466 & 110.11 & 1.34 & 41.6 \\
            & 3997.0629 & 108.36 & 1.17 & 35.3 \\
\hline
\hline
\end{tabular}
\tablefoot{The BJD id is from 2454833.}
\end{table}

\end{appendix}

\end{document}